\documentclass[a4paper,12pt]{article}
\usepackage{a4}
\usepackage{graphics}
\usepackage{epsfig}
\usepackage{amssymb}
\usepackage{color}
\usepackage{cite}
\usepackage{epsf}
\usepackage{axodraw}

\definecolor{green3}{rgb}{0.,0.7,0.0}
\definecolor{red1}{rgb}{0.9,0,0}

\newcommand{\Pp}{{\rm P}_+}
\newcommand{\Pm}{{\rm P}_-}
\newcommand{\Pmp}{{\rm P}_\mp}
\newcommand{\Ppm}{{\rm P}_\pm}
\newcommand{\Cp}{{\rm C}_+}
\newcommand{\Cm}{{\rm C}_-}
\newcommand{\Cpm}{{\rm C}_\pm}

\newcommand{\la}{\lambda}

\newcommand{\eps}{\epsilon}

\newcommand{\charginopm}{\tilde{\chi}^\pm}
\newcommand{\charginomp}{\tilde{\chi}^\mp}

\newcommand{\chargino}{\tilde{\chi}^+}
\newcommand{\charginoplus}{\tilde{\chi}^+}
\newcommand{\charginominus}{\tilde{\chi}^-}

\newcommand{\neutralino}{\tilde{\chi}^0}

\newcommand{\sneutrino}{\tilde{\nu}}

\def\lsim{\raise0.3ex\hbox{$\;<$\kern-0.75em\raise-1.1ex\hbox{$\sim\;$}}}
\def\gsim{\raise0.3ex\hbox{$\;>$\kern-0.75em\raise-1.1ex\hbox{$\sim\;$}}}
\DeclareMathAlphabet{\scr}{U}{rsfs}{m}{n}

\begin{document}
\def\thefootnote{\fnsymbol{footnote}}

\hspace*{\fill} BONN-TH-2008-06\\
\begin{center}
{\large \bf 
CP-violating Higgs boson mixing in chargino production at the muon collider
}                                              
\end{center}
\vspace{0.5cm}
\begin{center}
{
\sc Olaf~Kittel%
\footnote{email: kittel@th.physik.uni-bonn.de}%
}
\end{center}
\begin{center}
{\small \it  Physikalisches Institut der Universit\"at Bonn,
                Nussallee 12, D-53115 Bonn, Germany
}
\vspace*{.5cm}
{
\sc Federico~von~der~Pahlen%
\footnote{email: pahlen@th.physik.uni-bonn.de}%
}
\end{center}
\begin{center}
{\small \it Instituto de F\'isica de Cantabria (CSIC-UC), E-39005 Santander, Spain
}
\end{center}

\begin{abstract}

We study the pair-production of charginos 
in the CP-violating Minimal Supersymmetric Standard Model
at center-of-mass energies around the heavy neutral Higgs boson resonances.
If these resonances are nearly degenerate,
as it can happen in the Higgs decoupling limit,
radiatively induced scalar-pseudoscalar transitions can be strongly enhanced.
The resulting mixing in the Higgs sector leads to large CP-violating effects,
and a change of their mass spectrum.
For longitudinally polarized muon beams, 
we analyze CP asymmetries %
which are sensitive to the interference
of the two heavy neutral Higgs bosons.
We present a detailed numerical analysis of the cross sections,
chargino branching ratios, and the CP observables.
We obtain 
sizable CP asymmetries, which would be accessible 
in future measurements at a muon collider.
Especially for intermediate values of the parameter $\tan\beta$, where the largest 
branching ratios of Higgs bosons into charginos are expected, 
this process allows
to analyze the Higgs sector properties and its interaction to supersymmetric fermions.

\end{abstract}

\def\thefootnote{\arabic{footnote}}
\setcounter{footnote}{0}

\newpage


\section{Introduction}

The CP-conserving Minimal Supersymmetric Standard Model~(MSSM) contains three neutral
Higgs bosons~\cite{HK,Gunion:1984yn,Gunion:1989we,Carena:2002es,Djouadi:2005gj},
the lighter and heavier CP-even scalars $h$ and $H$, respectively, and the 
CP-odd pseudoscalar $A$.
In the presence of CP phases, the MSSM Higgs sector is still CP-conserving
at Born level. However loop effects,  dominantly mediated by third generation
squarks, can generate significant CP-violating scalar-pseudoscalar transitions.
Thus the neutral CP-odd and CP-even Higgs states mix and form
the mass eigenstates $H_1$, $H_2$, $H_3$, with no definite CP
parities~\cite{Accomando:2006ga,CPNSH,PilaftsisAH,Pilaftsis:1997dr,Choi:2004kq,GunionetalHA,%
HAeffpotential,%
PW.explicitCP,%
Heinemeyer:2001qd}.
A detailed knowledge of their mixing pattern will be crucial for the
understanding of the MSSM Higgs sector in the presence of CP-violating phases.
The fundamental properties of the Higgs bosons have to be investigated in detail
to reveal the mechanism of electroweak symmetry breaking.

\medskip

The general theoretical formalism for Higgs mixing with CP phases is well 
developed~\cite{GunionetalHA,HAeffpotential,PW.explicitCP,Heinemeyer:2007aq}. 
The results have been implemented in sophisticated public computer
programmes~\cite{Frank:2006yh,Degrassi:2002fi,CPSuperH},
that allow for numerical higher order calculations of the Higgs masses,
widths and couplings. It was shown that CP-violating phases can lead to a 
lightest Higgs boson with mass of order 
$M_{H_1}=45$~GeV~\cite{Carena:2000ks},
which cannot be excluded by measurements at LEP~\cite{LEPbounds}.
CP phases can also change the predictions for the Higgs decay widths
drastically, see e.g. Ref.~\cite{Williams:2007dc} for recent
calculations for $H_2\to H_1H_1$ at higher order.
In general, angular distributions of Higgs decay products will be
crucial to probe their CP parities at colliders,
for recent studies see, e.g.,  Ref.~\cite{angulardist}.

\medskip

A particularly important feature of the CP-violating Higgs sector
is a possible resonance enhanced mixing of the states $H$ and $A$.
It is well known that states with equal conserved quantum numbers
can strongly mix if they are nearly degenerate, i.e., if 
their mass difference is of the order of their 
widths~\cite{Pilaftsis:1997dr,Choi:2004kq}.
This degeneracy occurs naturally in the Higgs decoupling limit of the MSSM,
where the lightest Higgs boson has Standard Model-like couplings and 
decouples from the significantly heavier Higgs bosons~\cite{decoupling}.
In the decoupling limit, a resonance enhanced mixing of the states
$H$ and $A$ can occur, which may result in nearly maximal CP-violating
effects~\cite{PilaftsisAH,Pilaftsis:1997dr,Choi:2004kq}.
Such effects can only be analyzed thoroughly in processes in which the heavy 
neutral Higgs bosons can interfere as nearby lying, intermediate 
resonances~\cite{Ellis:2004fs}. 
The relative phase information of the two Higgs states would be lost if only 
their masses, widths and branching ratios are calculated, e.g., by assuming 
that they are produced as single, on-shell resonances at colliders.

\medskip

In previous studies of the CP-conserving~\cite{Grzadkowski:1995rx,mumu90s,
Dittmaier:2002nd,Fraas:2003cx,Grzadkowski:2000hm,Asakawa:2000uj,Barger:1999tj,
Fraas:2004bq,Kittel:2005ma}
and CP-violating Higgs 
sector~\cite{Asakawa:2000es,Atwood:1995uc,Pilaftsis:1996ac,Babu:1998bf,
Choi:1999kn,Choi:2001ks,Bernabeu:2006zs,Hioki:2007jc,Dreiner:2007ay},
it was shown that the interferences of the heavy
neutral Higgs bosons can be ideally tested in $\mu^+\mu^-$ collisions.
Since the Higgs bosons are resonantly produced in the $s$-channel, the muon 
collider is known to be the ideal machine for measuring the neutral 
Higgs masses, widths, and couplings with high 
precision~\cite{Barger:1996vc,hefreports,Blochinger:2002hj}. 
For a systematic classification of CP observables
which test the Higgs interference, a preparation of initial muon 
polarizations \emph{and} the analysis of final fermion polarizations will be crucial.
The CP-even and CP-odd contributions of the interfering Higgs resonances to 
observables can be ideally studied if the beam polarizations are properly
adjusted~\cite{Grzadkowski:2000hm,Asakawa:2000uj,Barger:1999tj,Fraas:2004bq,
Kittel:2005ma,Asakawa:2000es,Atwood:1995uc,Pilaftsis:1996ac,Babu:1998bf,Choi:1999kn,Choi:2001ks}.
For example, for final SM fermions $f\bar f$, with $f=\tau, b,t$, 
polarization observables have been classified according to their CP 
transformation properties\cite{Asakawa:2000es}.
For non-diagonal chargino pair production, the C-odd 
observables are obtained by antisymmetrizing in the chargino indices of the 
cross sections, and complete the necessary set of 
observables~\cite{CPNSH,Kittel:2005ma,Osland:2007xw,Rolbiecki:2007se}.
Finally, the well controllable beam energy of the muon collider then allows to 
study the center-of-mass energy dependence of the observables around the 
Higgs resonances. 
This is an advantage over the photon collider,
where direct line-shape scans are not possible~\cite{gammacollider}.

\medskip

In this work, we extend the study of chargino production at the muon 
collider~\cite{Kittel:2005ma} to the CP-violating case. 
First results for chargino production
in the MSSM with explicit CP violation in the Higgs sector
have been reported in Ref.~\cite{CPNSH}.
We classify all CP-even and CP-odd observables for chargino production
$\mu^+\mu^- \to \charginominus_i\charginoplus_j,$ using longitudinally 
polarized muon beams. We analyze the longitudinal chargino polarizations by 
their subsequent leptonic two-body decays 
$\tilde\chi_j^+ \to \ell^+\tilde\nu_\ell,$ with $\ell=e,\mu,\tau$,
and the charge conjugated process
$\tilde\chi_j^- \to \ell^-\tilde\nu_\ell^{\ast}$ .
Asymmetries in the energy distributions of the decay leptons allow us to 
classify all CP-even and a CP-odd observables for chargino decay, 
which probe the chargino polarization.
Similar asymmetries can be defined for the energy distributions of $W$ 
bosons, stemming from the decays $\tilde\chi_j^\pm \to W^\pm\tilde\chi_k^0$.
We find that the muon collider provides the ideal testing ground for 
phenomenological studies of the Higgs interferences,
although it might only be built in the far future.
Our analysis gives a deeper understanding of the Higgs mixings,
and allows us to test the public codes regarding
the relative phase information of the Higgs states
in the presence of CP-violating phases.

\medskip

In Section~\ref{formalism}, we give our formalism for chargino production and 
decay with longitudinally polarized muon beams. In an effective Born-improved 
approach, we include the leading self-energy corrections into the Higgs 
couplings. We give analytical formulas for
the production and decay cross sections and distributions.
We show that also the energy distributions of the chargino
decay products depend sensitively on the Higgs interference.
In Section~\ref{AsymmetriesforPandD}, we classify
the asymmetries of the production cross section
and of the energy distributions according to their CP properties.
For the production of unequal charginos $\charginopm_1\charginomp_2$,
we also define a new set of C-odd asymmetries.
In Section~\ref{Numerical results}, we present
a detailed numerical study of the cross sections,
chargino branching ratios, and the CP observables.
We analyze their dependence on $\sqrt s$,
on the CP-violating phase $\phi_A$
of the common trilinear scalar coupling parameter $A$,
and on the gaugino and higgsino mass parameters $\mu$ and $M_2$.
We complete our
analysis by comparing the results with the CP observables obtained in
neutralino production~\cite{Fraas:2004bq,Dreiner:2007ay}.
We summarize and conclude in Section~\ref{Summary and conclusions}.

\section{Chargino production and decay formalism
  \label{formalism}}

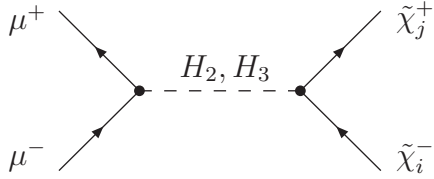
\begin{figure}[t]
\label{fig:charginoprod.res}
                \scalebox{1}{
\begin{picture}(10,2)(-3.4,0)
\DashLine(80,50)(140,50){5}
\Vertex(80,50){2}\Vertex(140,50){2}
\ArrowLine(140,50)(170,80)
\ArrowLine(170,20)(140,50)
\ArrowLine(50,20)(80,50)
\ArrowLine(80,50)(50,80)
\Text(1.6,0.9)[r]{$\mu^-$}
\Text(1.6,2.7)[r]{$\mu^+$}
\Text(6.2,0.9)[l]{$\charginominus_i$}
\Text(6.2,2.7)[l]{$\charginoplus_j$}
\put(3.35,1.95){$H_2,H_3$}
\end{picture}
}
\caption{Resonant Higgs exchange in chargino pair production
$\mu^+\mu^-\rightarrow\tilde\chi^-_i\tilde\chi^+_j$.}
\label{Fig:resHiggsMix}
\end{figure}

\begin{figure}[t]
                \scalebox{1}{
\begin{picture}(8,4.5)(0,0)
\Photon(80,50)(140,50){4}{3}
\Vertex(80,50){2}\Vertex(140,50){2}
\ArrowLine(140,50)(170,80)
\ArrowLine(170,20)(140,50)
\ArrowLine(50,20)(80,50)
\ArrowLine(80,50)(50,80)
\Text(1.6,0.9)[r]{$\mu^-$}
\Text(1.6,2.7)[r]{$\mu^+$}
\Text(6.2,0.9)[l]{$\tilde\chi_i^-$}
\Text(6.2,2.7)[l]{$\tilde\chi_j^+$}
\put(3.1,2){$\gamma ,Z,(H_1)$}
\end{picture}
\begin{picture}(8,2)(.4,0)
\Vertex(80,70){2}\Vertex(80,30){2}
\DashLine(80,30)(80,70){5}
\ArrowLine(50,0)(80,30)
\ArrowLine(80,70)(50,100)
\ArrowLine(110,0)(80,30)
\ArrowLine(80,70)(110,100)
\Text(1.6,0.3)[r]{$\mu^-$}
\Text(1.6,3.3)[r]{$\mu^+$}
\Text(4.1,0.3)[l]{$\tilde\chi_i^-$}
\Text(4.1,3.3)[l]{$\tilde\chi_j^+$}
\Text(3,1.75)[l]{$\tilde\nu_{\mu}$}
\end{picture}
}
\caption{\label{Fig:FeynProd} 
         Feynman diagrams for non-resonant chargino production 
         $\mu^+\mu^-\to\tilde\chi^-_i\tilde\chi^+_j$}
\end{figure}
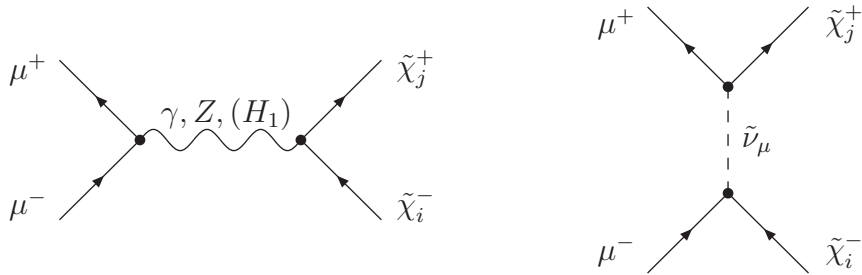

We study CP violation in the Higgs sector in pair production of charginos 
\begin{equation}
\mu^+  + \mu^- \to \charginominus_i + \charginoplus_j,
    \label{production}
\end{equation}
with longitudinally polarized muon beams.
We analyze the production of charginos at center-of-mass energies 
of the nearly mass degenerate heavy neutral Higgs bosons $H_2$ and $H_3$. 
They will be resonantly produced in the s-channel, see
the Feynman diagrams in Fig.~\ref{Fig:resHiggsMix}.
The significantly lighter Higgs boson $H_1$ is also
exchanged in the $s$-channel,
however is suppressed far from its resonance.
Together with sneutrino exchange $\tilde\nu_\mu$ in the $t$-channel,
and $Z$ and $\gamma$  exchange  in the $s$-channel,
they are the non-resonant continuum
contributions to chargino production, see
the Feynman diagrams in Fig.~\ref{Fig:FeynProd}.

To analyze the longitudinal chargino polarization,
we consider the subsequent CP-conserving but P-violating 
leptonic two-body decay of one of the charginos
\begin{equation}
        \tilde\chi_j^+ \to \ell^{+} + {\tilde\nu}_\ell,
   \label{decayell}
\end{equation}
and the charge conjugated process
$     \tilde\chi_j^- \to \ell^{-} + {\tilde\nu}_\ell^{\ast}$.
We will focus on the case $\ell =e,\mu$. However our results
can be extended to $\ell =\tau$, and for the chargino decay
into a $W$ boson
\begin{equation}
        \tilde\chi_j^\pm \to W^{\pm} + \tilde\chi_k^0,
   \label{decayW}
\end{equation}
for which we give the relevant formulas in Appendix~\ref{chardecay}.

\subsection{Higgs mixing \label{section:higgs-mixing}}

CP violation in the MSSM Higgs sector is induced
by scalar-pseudoscalar transitions at loop level.
The mixing is given by the symmetric and complex Higgs 
mass matrix~\cite{Frank:2006yh}
\begin{equation}
  {\rm \bf M}(p^2) =
\left(\begin{array}{ccc}
                        \ m_{h}^2 - \hat\Sigma_{hh}(p^2) \ &
                        - \hat\Sigma_{hH}(p^2)  &  - \hat\Sigma_{hA}(p^2)
\\[2mm]
                        - \hat\Sigma_{hH}(p^2)  &    \ m_{H}^2 -
\hat\Sigma_{HH}(p^2) \ &
                        - \hat\Sigma_{HA}(p^2)
\\[2mm]
                         - \hat\Sigma_{hA}(p^2)  &  - \hat\Sigma_{HA}(p^2) &
\ m_{A}^2 - \hat\Sigma_{AA}(p^2) \
      \end{array}
\right)
\label{eq:Weisskopf}
\end{equation}
at momentum squared $p^2$ 
in the tree-level basis of the CP eigenstates $h,H,A$.
Here $\hat\Sigma_{rs}(p^2)$, with $r,s = h,H,A$, are the renormalized self
energies of the Higgs bosons at one loop,
supplemented with higher-order contributions, see Ref.~\cite{Frank:2006yh}.
We neglect mixings with the Goldstone boson $G$ and the $Z$ boson.
Their contribution to the mass matrix is a two-loop effect, see discussion in Ref.~\cite{Frank:2006yh}.
Their effect on transition amplitudes, which enters already at the one-loop level,
is expected to be numerically small, see Ref.~\cite{Williams:2007dc}.
\footnote{
Our approach thus differs from Ref.~\cite{Ellis:2004fs},
where these self-energies have also been evaluated. 
Even if the contributions from Higgs mixing with $G$ and $Z$ turned out to be 
as large as those 
from the non-resonant $h$--$H$ and $h$--$A$ mixing contributions,
this would not result in a significant change of our numerical results.
}

The Higgs  propagator matrix
\begin{equation}
\label{eq:propagatorhHA}
   \Delta(p^2) = - i[ p^2 - {\rm \bf M}(p^2) ]^{-1},
\end{equation}
 has complex poles at 
$p^2=\mathcal M_{H_k}^2 =  M_{H_k}^2 - i M_{H_k} \Gamma_{H_k}$, $k=1,2,3$,
where $ M_{H_k} $ and $ \Gamma_{H_k} $ are the mass and width of the Higgs
boson mass eigenstate $H_k$, respectively, with convention
$ M_{H_1} \leq M_{H_2}  \leq M_{H_3} $.

At fixed $p^2=s$, we may diagonalize the Higgs mass 
matrix ${\rm \bf M}$, and consequently the propagator $\Delta$,
by a complex orthogonal matrix $C=C(s)$,
\begin{equation}
   {\rm \bf M}_{\rm D}(s) =  C \; {\rm \bf M}(s)\, C^{-1},
 \qquad \Delta_{\rm D}(s) =  C \, \Delta(s) \,C^{-1}. 
\label{eq:diagonalization}
\end{equation}
The diagonal elements of the propagator $ \Delta_{\rm D}$  
are given in the Breit-Wigner form by
\begin{eqnarray}
[\Delta_{D}(s)]_{kk}
  &=& \frac{-i}{s-   [{\rm \bf M_D}(s)]_{kk}}
  \equiv   \frac{-i}{s-M^{\prime 2}_{H_{k}}+
            iM^{\prime}_{H_{k}}\Gamma^{\prime}_{H_{k}}}
  \equiv   \Delta^\prime(H_{k}),
\label{propHiggs_D}
\end{eqnarray}
which defines the $s$-dependent real parameters $M^{\prime}_{H_{k}}$ and
$\Gamma^{\prime}_{H_{k}}$.
In the limit $s \to \mathcal M_{H_k}^2$,
these parameters are the on-shell mass and decay width of the Higgs boson $H_{k}$,
$M^\prime_{H_{k}}\to M_{H_{k}}$ and $\Gamma^\prime_{H_{k}} \to \Gamma_{H_{k}}$,
respectively.
If the $s$-dependence of the mass matrix  ${\rm \bf M}$ is weak
around the resonance $s \approx  M^2_{H_{k}}$ 
of a Higgs boson $H_k$, the $kk$-element of the  
propagator can be approximated by
\begin{eqnarray}
  \Delta^\prime(H_{k}) \approx  \Delta(H_{k})
  &=& \frac{-i}{s-M^2_{H_{k}}+iM_{H_{k}}\Gamma_{H_{k}}}.
 \label{propHiggs}
\end{eqnarray}
The approximations 
$M^\prime_{H_k} \approx M_{H_{k}}$
and
$\Gamma^\prime_{H_k} \approx \Gamma_{H_k}$
hold for instance, if no thresholds open around the Higgs resonance.

\medskip

In our numerical calculations, we  evaluate the mass matrix 
${\rm \bf M}(p^2)$
and the diagonalization matrix
$C(p^2)$
at fixed $p^2=M_{H_2}^2 \approx M_{H_3}^2$
with the program {\tt FeynHiggs 2.5.1}~\cite{Frank:2006yh,Degrassi:2002fi}.
We %
find
that the momentum dependence of ${\rm \bf M}(p^2)$ is weak 
around the resonances, and %
thus
$ \Delta^\prime(H_{k}) \approx \Delta(H_{k})$ to a very good
approximation.
For the following discussions, we %
set
$ \Delta^\prime(H_{k}) = \Delta(H_{k})$ for simplicity,
and drop the prime index of the propagator in our notation.

\subsection{Effective Higgs couplings and transition amplitudes
             \label{section:effcouplings}}

The amplitude %
for chargino production in muon-antimuon annihilation via Higgs exchange 
can be written in the general form
\begin{eqnarray}
T^P&=&
 \Gamma_{}^{(\chi)} \Delta_{}(s) \Gamma_{}^{(\mu)}
\label{eq:generalTPhHA}
\\[2mm]
&=&
 \Gamma_{}^{(\chi)} C^{-1}\;C \, \Delta(s) \,C^{-1} \; C \; \Gamma_{}^{(\mu)}
\\[2mm]
&=&
     \Gamma_{}^{(\chi)} C^{-1}_{}   \Delta_{D}(s) \,   C_{}\, \Gamma_{}^{(\mu)}
     =   \Gamma_{\rm eff}^{(\chi)}   \Delta_{D}(s)  \Gamma_{\rm eff}^{(\mu)},
\label{eq:generalTP}
\end{eqnarray}
where $\Gamma_{}^{(\chi)}$ and $\Gamma_{}^{(\mu)}$ are the one-particle
irreducible (1PI) Higgs vertices to charginos and muons, respectively.
We have used the basis transformation of the Higgs 
propagator $\Delta(s)$ 
at fixed center-of-mass energy $s$, Eq.~(\ref{eq:diagonalization}), 
which leads to the definition of the effective 1PI vertices 
$\Gamma_{\rm eff}^{(\mu)}$ and $\Gamma_{\rm eff}^{(\chi)}$.
The Higgs mixing corrections are thus effectively included via the 
diagonalization matrix $C$ into the 1PI vertices. 
While the diagonalization of the Higgs boson propagator matrix in Eqs.~(\ref{eq:generalTPhHA})-(\ref{eq:generalTP}) is not strictly necessary, it
yields the transition amplitudes in a diagonalized form.\footnote{
Note, however, 
that 
 there could be specific points in parameter space 
for which the propagator matrix cannot be diagonalized with a similarity transformation~\cite{Pilaftsis:1997dr},
but for which it is still invertible.
These points correspond to the singular situation where both the pole-masses {\it and} the widths of two Higgs bosons are equal.
In such a case, 
the Higgs exchange amplitudes can still be retrieved in the general form of Eq.~(\ref{eq:generalTPhHA}), 
as in, e.g., Refs.~\cite{Ellis:2004fs,Asakawa:2000es}, 
for similar studies of Standard Model fermion pair production. 
   }
In addition of simplifying the technical evaluation, 
in the new basis the Higgs boson states are approximately the pole-mass eigenstates,
as discussed in the preceding section.
We thus obtain compact analytic formulas for the squared amplitudes
of the Higgs exchange contributions, which we explicitely give
in Appendix~\ref{ab_coefficients}.

\medskip

If the Higgs bosons are nearly mass degenerate,
the radiative corrections to the propagator matrix are strongly enhanced by the Higgs mixing, 
and constitute 
the dominant contributions to the transition
amplitude $T^P$~(\ref{eq:generalTP}).
Thus we only include these radiative corrections and neglect specific vertex 
corrections.
The effective Higgs couplings to the initial muons, $c_{L,R}^{H_k\mu\mu}$,
and the final charginos, $ c_{L,R}^{H_k\chi_i\chi_j}$,
are then obtained by transforming the tree level couplings
$c_{L,R}^{h_\alpha\mu\mu}$ and $ c_{L,R}^{h_\alpha\chi_i\chi_j}$,
respectively, 
with the matrix $C$, see Eq.~(\ref{eq:diagonalization}),
\begin{eqnarray}
         c_{L,R}^{H_k\mu\mu}  &=&
        C_{k\alpha}   c_{L,R}^{h_\alpha\mu\mu},
\label{eq.Hmu.eff.c}\\[2mm]
   c_{L,R}^{H_k\chi_i\chi_j}  &=&
  \tilde C_{k\alpha}   c_{L,R}^{h_\alpha\chi_i\chi_j},\quad h_\alpha=h,H,A.
\label{eq.Hchi.eff.c}
\end{eqnarray}
The tree level Higgs couplings 
are defined and discussed, e.g., in detail in
Refs.~\cite{Kittel:2005ma,Gunion:1989we}.
Note that the final state chargino couplings 
transform with $\tilde C = {C^{-1} }^T$.
If the non-physical phases of the Higgs bosons are chosen appropriately, 
the matrix $C$ can be made complex orthogonal, which implies
$\tilde C = C$~\cite{orthogonalC}. 

\medskip

With the Born-improved effective couplings of Eqs.~(\ref{eq.Hmu.eff.c}) and
(\ref{eq.Hchi.eff.c}),
we can write the Higgs exchange amplitude for chargino production
\begin{eqnarray}
T^P =
\Delta(H_k)
\left[
\bar{v}(p_{\mu^+})\left(c_L^{H_k\mu\mu}P_L+c_R^{H_k\mu\mu}P_R \right)
u(p_{\mu^-})
\right]\phantom{.}\nonumber\\[3mm]
\times
\left[
\bar u(p_{\chi^+_j})\left(c_L^{H_k\chi_i\chi_j}P_L
                                   +c_R^{H_k\chi_i\chi_j}P_R \right)
v (p_{\chi^-_i})
\right]
\label{THiggsNoHel}
\end{eqnarray}
in its Born-improved form.
The amplitudes and Lagrangians for non-resonant
$\gamma$, $Z$ and $\tilde\nu_\mu$ exchange are given
in Appendix~\ref{nonresamplitudes},
and the Lagrangians for the chargino decays,
Eqs.~(\ref{decayell}) or (\ref{decayW}), are given in
Appendix~\ref{chardecay}.

\subsection{Squared amplitude}

In order to calculate the squared amplitude for 
chargino production~(\ref{production}) 
and the subsequent decay of chargino $\tilde\chi_j^+$,
Eqs.~(\ref{decayell}) or (\ref{decayW}),
with initial beam polarizations and complete spin correlations,
we use the spin density matrix formalism~\cite{Haber94,MoortgatPick:1998sk}. 
Following the detailed steps as given in Appendix~\ref{Density matrix formalism}, 
the squared amplitude in this formalism can be  written as
\begin{eqnarray}
        |T|^2 &=&  2|\Delta(\tilde\chi_j^+)|^2 
        (P D + \sum_{a=1}^3 {\Sigma}_P^{a} {\Sigma}_D^{a}),
\label{eq:tsquare}
\end{eqnarray}
with the propagator $\Delta(\tilde\chi_j^+)$ of the decaying chargino, 
see Eq.~(\ref{eq:charginopropagator}).
Here $P$ denotes the unpolarized production of the charginos
and $D$ the unpolarized decay.
The corresponding polarized terms are
${\Sigma}_P^{a} $ and $ {\Sigma}_D^{a}$,
and their product in Eq.~(\ref{eq:tsquare}) describes
the chargino spin correlations between production and decay. 
With our choice of the spin vectors, see Eqs.~(\ref{rhoP}) and (\ref{rhoD}), 
${\Sigma}_P^{3}/P$ is the longitudinal polarization of $\tilde\chi_j^+$, 
${\Sigma}_P^{1}/P$ is the transverse polarization in the production plane, 
and ${\Sigma}_P^{2}/P$ is the polarization perpendicular to the production plane.
The terms $D$ and ${\Sigma}_D^{a}$ for chargino decay are given in Appendix~\ref{chardecay}. 

The expansion coefficients of the squared chargino production 
amplitude $|T^2|$~(\ref{eq:tsquare}) subdivide into contributions from the 
Higgs resonances~(${\rm res}$), the continuum~(${\rm cont}$), 
and the resonance-continuum interference~(${\rm int}$), respectively, with
\begin{equation}
         P = P_{\rm res}+P_{\rm cont}, \qquad 
{\Sigma}_P^{a} = {\Sigma}_{\rm res}^{a} +{\Sigma}_{\rm cont}^{a} +{\Sigma}_{\rm int}^{a}, \quad a=1,2,3.
\label{contributions}
\end{equation}
The continuum contributions $P_{\rm cont}$, ${\Sigma}_{\rm cont}^{a}$ are 
those from the non-resonant $\gamma$, $Z$ and $\tilde\nu_{\mu}$ 
exchange channels, which we give in Appendix~\ref{nonresamplitudes}.
The contributions from $H_1$ exchange 
are suppressed by $m_\mu^2/s$.

The dependence of the resonant contributions
$P_{\rm res}$ and $\Sigma_{\rm res}^3$
on the longitudinal $\mu^+$ and $\mu^-$ beam polarizations  
${\mathcal P}_+$ and ${\mathcal P}_-$, respectively,
is given by\footnote{
   The resonant contributions 
   $\Sigma_{\rm res}^1$ and $\Sigma_{\rm res}^2$
   to the transverse polarizations of the chargino vanish
   for scalar Higgs bosons exchange in the $s$-channel.
   Note that there is no interference contribution to the 
   coefficients $P$ and ${\Sigma}_P^{3} $.
} 
\begin{eqnarray}
P_{\rm res}&=&
        (1 + {\mathcal{P}}_+{\mathcal{P}}_-)a_0  + 
        ({\mathcal{P}}_+ + {\mathcal{P}}_-)a_1, 
\label{eq.Pr}
\\
\Sigma_{\rm res}^3&=&
         (1 + {\mathcal{P}}_+{\mathcal{P}}_-)b_0 + 
         ({\mathcal{P}}_+ + {\mathcal{P}}_-)b_1.
\label{eq.Sr}
\end{eqnarray}
This expression
is useful
for analyzing the CP properties of 
$P_{\rm res}$ and $\Sigma_{\rm res}^3$.
The coefficients $a_n$ and $b_n$,
given explicitly in Appendix~\ref{ab_coefficients},
are functions of products of the Higgs couplings to muons and charginos, 
Eqs.~(\ref{eq.Hmu.eff.c}) and (\ref{eq.Hchi.eff.c}).
The coefficients also include the product of Higgs boson propagators, Eq.~(\ref{propHiggs_D}),
which strongly depend on the center-of-mass energy, as well as on the CP phases, which enter 
via the diagonalization matrix $C$, see the transformation in Eq.~(\ref{eq:diagonalization}).

\begin{table}
\renewcommand{\arraystretch}{1.2}
\caption{\label{coeffs:CandP} 
       Charge (C) and parity (P) properties of the coefficients 
       $a^\pm_0$, $a^\pm_1$, $b^\pm_0$, and $b^\pm_1$,  
       see Eqs.~(\ref{eq.Pr}) and (\ref{eq.Sr}),
       and the corresponding observables %
       for their determination, as defined in Section~\ref{AsymmetriesforPandD}.
       The four C-odd observables 
       vanish for diagonal chargino production, $i=j$.
}
\begin{center}
     \begin{tabular}{|c|c|c|c|c|}
\hline
coefficient     & C     & P     & CP  & observables \\
\hline
\hline
$ a^+_0 $       & $+$   & $+$   & $+$ &   $\sigma_{ij}^{\Cp\Pp}$ \\
\hline
$ a^+_1 $       & $+$   & $-$   & $-$ &   $\mathcal{A}^{\Cp\Pm}_{ij}$ \\
\hline
$ a^-_0 $       & $-$   & $+$   & $-$ &   $\mathcal{A}^{\Cm\Pp}_{ij}$ \\
\hline
$ a^-_1 $       & $-$   & $-$   & $+$ &   $\mathcal{A}^{\Cm\Pm}_{ij}$ \\
\hline
\hline
$ b^+_0 $       & $+$   & $-$   & $-$ &   $\mathcal{A}^{\Cp\Pm}_{ij,\ell}$ \\
\hline
$ b^+_1 $       & $+$   & $+$   & $+$ &   $\mathcal{A}^{\Cp\Pp}_{ij,\ell}$ \\
\hline
$ b^-_0 $       & $-$   & $-$   & $+$ &   $\mathcal{A}^{\Cm\Pm}_{ij,\ell}$ \\
\hline
$ b^-_1 $       & $-$   & $+$   & $-$ &   $\mathcal{A}^{\Cm\Pp}_{ij,\ell}$ \\
\hline
\end{tabular}
\end{center}
\renewcommand{\arraystretch}{1.0}
\end{table}

\subsection{C and P properties of the Higgs exchange coefficients}
              \label{CPproperties}

We classify the Higgs exchange coefficients $a_n$ and $b_n$ according to their 
charge~(C) and parity~(P) properties. 
Our aim is then to define a complete set of 
CP-even and CP-odd asymmetries in chargino production and decay,
in order to determine the Higgs couplings.
The kinematical dependence of the asymmetries can be probed
by line-shape scans.
Since a muon collider provides a good beam energy resolution,
it will be the ideal tool to analyze the strong $\sqrt{s}$ 
dependence of the asymmetries.

\medskip

The factors $a_0$ and $b_1$ are P-even,
    whereas $a_1$ and $b_0$ are P-odd. 
Note that the coefficients $a_{1}$ and $b_{1}$
only contribute for polarized muon beams.
For non-diagonal chargino production, we separate these coefficients 
into their C-even and C-odd parts,
respectively, by symmetric and antisymmetric combinations 
in the chargino indices,
\begin{eqnarray}
a_n^\pm &=& \frac{1}{2}\left[a_n(\charginominus_i\charginoplus_j) 
                       \pm a_n(\charginominus_j\charginoplus_i)\right],
\\
b_n^\pm &=& \frac{1}{2}\left[b_n(\charginominus_i\charginoplus_j) 
                       \pm b_n(\charginominus_j\charginoplus_i)\right],
\quad n=0,1.
\end{eqnarray}
The coefficients $a_n(\charginominus_j\charginoplus_i)$ 
and $b_n(\charginominus_j\charginoplus_i)$
of chargino $\charginominus_j$,
for the production of the charge conjugated pair of charginos,
$\mu^+\mu^-\rightarrow\tilde\chi^-_j\tilde\chi^+_i$,
are obtained by interchanging the indices $i$ and $j$
in the formulas of the coefficients defined in
Eqs.~(\ref{eq.anbn})-(\ref{eq.b1}),
which are defined for $\mu^+\mu^-\rightarrow\tilde\chi^+_j\tilde\chi^-_i$.
We summarize the C, P and CP properties of the coefficients, and thus of our observables, 
in Table~\ref{coeffs:CandP}.
Note that, 
for these observables, CP and 
CP$\tilde{\rm{T}}$ are equivalent, 
since they are built using the longitudinal polarizations only.
Here  $\tilde{\rm{T}}$ is the naive time reversal $t\to-t$,
which inverts momenta and spins without exchanging initial and final particles.

\subsection{Energy distributions of the chargino decay products}
            \label{energydistr}

The energy distribution of the lepton or $W$ boson from
the chargino decay, Eqs.~(\ref{decayell}) or (\ref{decayW}), respectively,
depends on the longitudinal chargino polarization.
In the center-of-mass system, the kinematical limits of the energy 
of the decay particle $\lambda=e,\mu,\tau,W$ are
\begin{eqnarray}
E_\lambda^{\rm max(min)} &=& \hat{E}_\lambda \pm \Delta_\lambda,
\label{kinlimits}
\end{eqnarray}
which read for the leptonic $(\lambda=\ell)$ chargino decays
\begin{eqnarray}
\hat{E}_\ell &=& \frac{E_\ell^{\rm max}+E_\ell^{\rm min}}{2} = 
\frac{ m_{\chi^\pm_j}^2-m_{\tilde\nu_\ell}^2}{2 m_{\chi^\pm_j}^2} E_{\chi^\pm_j},
\label{ehalf}
\\
\Delta_\ell &=& \frac{E_\ell^{\rm max}-E_\ell^{\rm min}}{2} = 
\frac{ m_{\chi^\pm_j}^2-m_{\tilde\nu_\ell}^2}{2 m_{\chi^\pm_j}^2}
|\vec{p}_{\chi^\pm_j}|,
\qquad \ell=e,\mu,\tau.
\label{edif}
\end{eqnarray}
The energy limits $E_W^{\rm max(min)}$ for the $W$ boson are 
given in Appendix~\ref{chardecay}.

Using the definition of amplitude squared, the cross section, Eqs.~(\ref{eq:tsquare}), 
and the explicit form of $\Sigma_D^3$~(\ref{etal}),
the energy distribution of the decay particle~$\lambda^\pm$ 
is~\cite{Fraas:2004bq,Kittel:2005ma}
\begin{equation}
\frac{d\sigma_{ij,\lambda^\pm}}{dE_\lambda} =
        \frac{\sigma_{ij,\lambda}}
        {2\Delta_\lambda}\left[ 1 + \,
        \eta_{\lambda^\pm}
        \frac{\bar{\Sigma}^3_P}{\bar{P}} 
        \frac{(E_\lambda- \hat{E_\lambda})}{\Delta_\lambda} \right],
\label{edist2}
\end{equation}
The integrated cross section for chargino production and decay is denoted by
$\sigma_{ij,\lambda^\pm}$, e.g. for the decay 
$\chargino_j\to\ell^+\tilde\nu_\ell$ it is, 
using the narrow width approximation for the propagator of the decaying chargino,
\begin{eqnarray}
 \sigma_{ij,\ell} =
 \sigma_{ij} \times{\rm BR}(\tilde\chi_j^+\to\ell^+\tilde\nu_\ell),
\label{eq:sigmatot}
\end{eqnarray}
where $\sigma_{ij}$ is the cross section for $\charginominus_i\charginoplus_j$ production. 
Explicit expressions for the cross sections
 $\sigma_{ij}$,
$\sigma_{ij,\ell}$, and  $\sigma_{ij,W}$  are given in Appendix~\ref{crossSection}.
The factor $\eta_{\lambda^\pm}$ is a measure of parity violation 
in the chargino decay. It is maximal
$\eta_{e^\pm},\eta_{\mu^\pm}= \pm 1$
for the decay into leptons of the first two generations,
while for the decay 
$\tilde\chi_j^\pm\to\tau^\pm\tilde\nu_\tau^{(\ast)}$,
it is generally smaller $|\eta_{\tau^\pm}|<1$.
In Appendix~\ref{chardecay},
we give explicit expressions of the factors $\eta_{\tau^\pm}$,
as well as $\eta_{W^\pm}$
for the decay $\tilde\chi_j^\pm\to W^\pm\tilde\chi^0_k$.

Further, from Eq.~(\ref{edist2}),
we see that the energy distribution is proportional to the
chargino polarization $\bar \Sigma^3_P/\bar P$.
The coefficients $\bar \Sigma^3_P$  and $\bar P$
are averaged over the chargino production solid angle
\begin{eqnarray}
\bar\Sigma^3_P  = \frac{1}{4\pi}\int \Sigma^3_P \,  d\Omega_{\chi^\pm_j}
              =  \Sigma_{\rm res}^3+\bar\Sigma_{\rm cont}^3, \qquad
\bar P = \frac{1}{4\pi}\int P  d\Omega_{\chi^\pm_j}
       = P_{\rm res} + \bar P_{\rm cont},
\label{eq:sigmabar}
\end{eqnarray}
denoted by a bar in our notation.
Note that the resonant contributions 
from Higgs exchange are isotropic 
$\bar\Sigma^3_{\rm res}=\Sigma^3_{\rm res}$,
$ \bar P_{\rm res}= P_{\rm res}$.

\medskip 

\begin{figure}[t]
\begin{picture}(14,5.6)
\put(-0.3,-16.2){\includegraphics{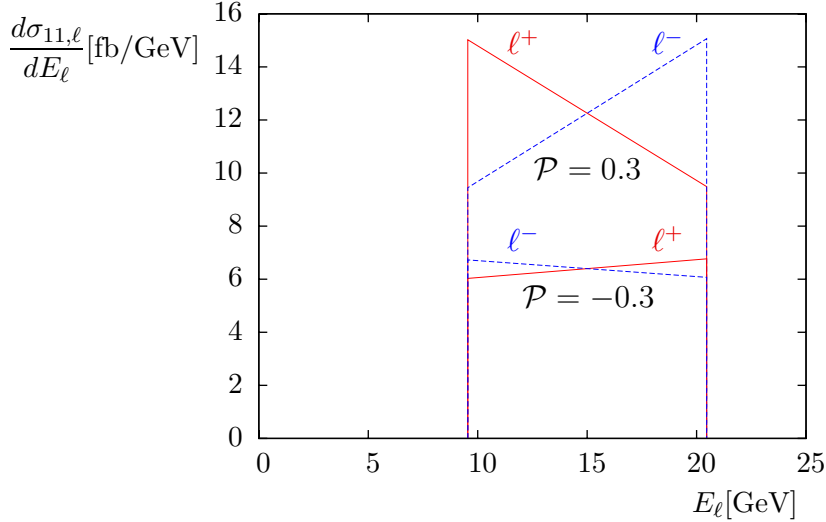}}

\put(9.1,6.2){\color{blue}$  \ell^-$}
\put(7.55,4.55){\color{black}$  {\mathcal P}=0.3$}
\put(7.2,6.2){\color{red1}$  \ell^+$}
\put(7.2,3.6){\color{blue}$  \ell^-$}
\put(7.4,2.85){\color{black}$  {\mathcal P}=-0.3$}
\put(9.1,3.6){\color{red1}$  \ell^+$}

\end{picture}
\caption{\small
        Energy distributions of the leptons $\ell^\pm$ for
        chargino production $\mu^+\mu^-\to\tilde\chi_1^+\tilde\chi_1^-$
        and decay $\tilde\chi_1^+\to\ell^+\tilde\nu_\ell$ (solid, red), and
        $\tilde\chi_1^-\to\ell^-\tilde\nu_\ell^\ast$ (dashed, blue)
        for $\ell = e$ or $\mu$,
        at $\sqrt{s}=(M_{H_2}+M_{H_3})/2$ 
        with longitudinal beam polarizations
        ${\mathcal P}_-={\mathcal P}_+\equiv {\mathcal P}=+0.3$
and        ${\mathcal P}=-0.3$.
        The MSSM parameters are given in Tables~\ref{scenarioA}
        and~\ref{scenarioAmasses}.
        The shown distributions have asymmetries (for ${\mathcal P}=0.3$)
        $\mathcal{A}_{11,\ell^+}=-11.3\%$ and
        $\mathcal{A}_{11,\ell^-}= 11.45\%$,
        (for ${\mathcal P}=-0.3$)
        $\mathcal{A}_{11,\ell^+}= 2.9\%$ and
        $\mathcal{A}_{11,\ell^-}= -2.6\%$,
        see Eq.~(\ref{asymmetryenergydist}).
}\label{fig:edist.l1l2}
\end{figure}

In Fig.~\ref{fig:edist.l1l2}, we show the energy distributions~(\ref{edist2}) 
of the leptons $\ell^\pm$ from the decays
$\tilde\chi_j^+\to\ell^+\tilde\nu_\ell$
and 
$\tilde\chi_j^-\to\ell^-\tilde\nu_\ell^*$,
for $\ell=e$ or $\mu$. 
The cutoffs in the energy distributions of the
leptons $\ell^+$ and $\ell^-$ correspond
to their kinematical limits, as given in Eq.~(\ref{kinlimits}).
We see the linear dependence of the distributions 
on the lepton energy.  Their slope is proportional 
to the longitudinal chargino polarization $\bar\Sigma_{P}^3/\bar P$,
see Eq.~(\ref{edist2}).
Note that the energy distribution might be difficult to
measure for a small chargino-sneutrino mass difference,
since the energy span of the observed lepton is proportional to
the difference of their squared masses, see Eq.~(\ref{edif}).

\section{Chargino production and decay observables
             \label{AsymmetriesforPandD}}

For chargino pair production 
$\mu^+\mu^-\rightarrow\tilde\chi^-_i\tilde\chi^+_j$,
we have expanded the resonant contributions 
$P_{\rm res}$~(\ref{eq.Pr}) and $\Sigma^3_{\rm res}$~(\ref{eq.Sr})
to the spin density matrix elements in terms of
the longitudinal muon beam polarizations ${\mathcal P}_\pm$.
We have classified the coefficients
$a_0^\pm$, $a_1^\pm$, $b_0^\pm$, and $b_1^\pm$ according to
their charge~(C) and parity~(P) transformation properties,
see Table~\ref{coeffs:CandP}.
In order to determine the coefficients $a_0^\pm$ and $a_1^\pm$,
we define one production cross section observable and
three asymmetries of the chargino production 
cross section in the following section.
To determine the coefficients $b_0^\pm$ and $b_1^\pm$ of the chargino
polarization,  the chargino decays have to be taken into account.
We therefore define an additional set of four asymmetries of the
energy distributions of the decay particle. 
In Table~\ref{coeffs:CandP}, we also list the observables which we define 
in the following sections. Non-diagonal chargino production $i\ne j$
leads to the four C-odd observables,
which vanish trivially
for diagonal chargino production $i=j$.

\subsection{Asymmetries of the chargino production cross section}
\label{sec:Asymmetries.chargino.prod}

In order to classify the CP observables 
in chargino production, we define in a first step
the symmetric and antisymmetric combinations 
\begin{eqnarray}
\sigma_{ij}^{\Cp}= \frac{1}{2}(\sigma_{ij} + \sigma_{ji}),
\qquad
\sigma_{ij}^{\Cm}= \frac{1}{2}(\sigma_{ij} - \sigma_{ji}),
\label{eq:sigmapm}
\end{eqnarray}
of the production cross section $\sigma_{ij}$~(\ref{crossectionProd}).
The combinations are relevant for non-diagonal chargino production, with $i\ne j$.
The indices ${\rm C}_+$ and ${\rm C}_-$ indicate the even and odd parity
under charge (C) conjugation, respectively.
In a second step, we obtain observables which have a definite parity (P)
by symmetrizing and antisymmetrizing in the common muon beam polarizations
${\mathcal P}_+ = {\mathcal P}_- \equiv {\mathcal P}$.

\medskip

The C- and P-even observable is obtained by symmetrizing 
$\sigma_{ij}^{\Cp}$~(\ref{eq:sigmapm}) in ${\mathcal P}$
\begin{equation}
\sigma_{ij}^{\Cp\Pp}=\frac{1}{2}
                      \left[\sigma_{ij}^{\Cp}(\mathcal{ P})
                          + \sigma_{ij}^{\Cp}(\mathcal{-P})
                      \right].
\label{eq:sigmacepe}
\end{equation}
Inserting the explicit form of $\sigma_{ij}$~(\ref{crossectionProd}) and $P_{\rm res}$~(\ref{eq.Pr}) 
the coefficient $a_0^+$ can be determined, 
\begin{equation}
\sigma_{ij}^{\Cp\Pp}=\frac{\sqrt{\lambda_{ij}}}{8\pi s^2}
\left[ (1+\mathcal{P}^2)a_0^+ +\bar P_{\rm cont}\right],
\end{equation}
if the continuum contributions
$\bar P_{\rm cont}$~(\ref{contributions}) 
can be subtracted, e.g, 
through a sideband analysis, 
where the cross section is extrapolated around the resonances~\cite{Fraas:2003cx}, 
and/or by chargino cross section measurements at the International Linear Collider (ILC)~\cite{TDR,Djouadi:2007ik}.

\medskip

The two CP-odd asymmetries are obtained by~\cite{Blochinger:2002hj,CPNSH}
\begin{eqnarray}
\mathcal{A}^{\Cp\Pm}_{ij}
&=& \frac
        {\sigma^{\Cp}_{ij}(\mathcal{P})-\sigma^{\Cp}_{ij}(\mathcal{-P})}
        {\sigma^{\Cp}_{ij}(\mathcal{P})+\sigma^{\Cp}_{ij}(\mathcal{-P})},
\label{eq:aprodCePo}
\end{eqnarray}
and
\begin{eqnarray}
\mathcal{A}^{\Cm\Pp}_{ij}
&=& \frac
        {\sigma^{\Cm}_{ij}(\mathcal{P})+\sigma^{\Cm}_{ij}(\mathcal{-P})}
        {\sigma^{\Cp}_{ij}(\mathcal{P})+\sigma^{\Cp}_{ij}(\mathcal{-P})}.
\label{eq:aprodCoPe}
\end{eqnarray}
The CP-even asymmetry is~\cite{CPNSH}
\begin{eqnarray}
\mathcal{A}^{\Cm\Pm}_{ij} 
&=& \frac
        {\sigma^{\Cm}_{ij}(\mathcal{P})-\sigma^{\Cm}_{ij}(\mathcal{-P})}
        {\sigma^{\Cp}_{ij}(\mathcal{P})+\sigma^{\Cp}_{ij}(\mathcal{-P})}.
\label{eq:aprodCoPo}
\end{eqnarray}
Using the definitions of the chargino production
cross section $\sigma_{ij}$~(\ref{crossectionProd}), 
and of the coefficient $P$~(\ref{eq.Pr}), we obtain
\begin{eqnarray}
\mathcal{A}^{\Cpm\Pm}_{ij}
= 
        \frac{ {2\mathcal P} a_1^\pm }{ (1 + {\mathcal P}^2)a_0^+ + \bar P_{\rm cont} }.
\label{eq:aprodCeoPodep}
\end{eqnarray}
These asymmetries thus allow to determine $a_1^\pm$.
The coefficient $a_0^-$ can be obtained from 
\begin{eqnarray}
\mathcal{A}^{\Cm\Pp}_{ij}
        =\frac{ (1 + {\mathcal P}^2) a_0^- }
              { (1 + {\mathcal P}^2) a_0^+ 
                + \bar P_{\rm cont} }.
\label{eq:aprodCoPedep}
\end{eqnarray}
Note that the continuum contributions $\bar P_{\rm cont}$
are P-even, $ P_{\rm cont}({\mathcal P}) = P_{\rm cont}(-{\mathcal P})$,
and C-even, $ P_{\rm cont}(\tilde\chi_i^-\tilde\chi_j^+) = 
P_{\rm cont}(\tilde\chi_i^+\tilde\chi_j^-)$,
and thus cancel in the numerator of the P-odd and C-odd asymmetries
$\mathcal{A}^{\Cpm\Pm}_{ij}$~(\ref{eq:aprodCeoPodep}) and
$\mathcal{A}^{\Cm\Pp}_{ij}$~(\ref{eq:aprodCoPedep}), respectively.

The maximum absolute values of the P-odd asymmetries depends on the beam
polarization~${\mathcal P}$
\begin{eqnarray}
\mathcal{A}^{\Cpm\Pm}_{ij\,\rm (max)}
= \frac{2\mathcal{P}}{1+\mathcal{P}^2},
\label{eq:amax}
\end{eqnarray}
which follows from Eq.~(\ref{eq:aprodCoPedep})
for vanishing continuum contributions $\bar P_{\rm cont}=0$.
The P-even asymmetry $\mathcal{A}^{\Cm\Pp}_{ij}$ can be
as large as $100\%$.

\medskip

The CP-odd asymmetries 
$\mathcal{A}^{\Cp\Pm}_{ij}$ and $\mathcal{A}^{\Cm\Pp}_{ij}$
vanish in the case of CP conservation, and
are sensitive to the CP phases of the Higgs boson 
couplings to the charginos and to the muons.
The CP-odd asymmetry $\mathcal{A}^{\Cp\Pm}_{ij}$ 
is CP$\tilde{\rm{T}}$-odd,
with $\tilde{\rm{T}}$ the naive time reversal $t\to-t$,
which inverts momenta and spins without exchanging initial and final particles.
Thus the asymmetry is due to the interference of CP phases with absorptive phases
from the transition amplitudes.
The absorptive phases are also called strong phases,
and can originate from intermediate particles in the Higgs self energies 
which go on-shell.
The asymmetry 
$\mathcal{A}^{\Cp\Pm}_{ij}$
is therefore sensitive 
to the CP phases of the Higgs boson couplings,
as well as to the phases of the Higgs propagators.

\medskip

In the Higgs decoupling limit~\cite{decoupling},
the heavy neutral Higgs bosons are 
nearly mass-degenerate.
Thus a mixing of the CP-even and CP-odd Higgs states $H$ and $A$ 
can be resonantly enhanced, and large CP-violating Higgs couplings can be
obtained~\cite{PilaftsisAH,Pilaftsis:1997dr}. 
In addition, if the Higgs bosons are nearly mass-degenerate, CP phases in the Higgs sector 
lead to a larger splitting of
the mass eigenstates $H_2$ and $H_3$. 
This, in general,
tends to increase the phase between the Higgs propagators,
giving rise to larger absorptive phases in the transition amplitudes.
On the contrary, the observables 
$\sigma_{ij}^{\Cp\Pp}$~(\ref{eq:sigmacepe}) and
$\mathcal{A}^{\Cm\Pm}_{ij}$~(\ref{eq:aprodCoPo}), 
being CP-even, will be reduced in general in the presence of 
CP-violating phases.

\subsection{Asymmetries of the lepton energy distribution}
\label{sec:asym.edis}

The longitudinal chargino polarization
is also sensitive to the Higgs interference
in the production $\mu^+\mu^-\to\tilde\chi^\mp_i\tilde\chi^\pm_j$,
and yields additional information on the Higgs couplings.
The chargino $\tilde\chi^\pm_j$ polarization can be analyzed by the
subsequent decays 
$\tilde\chi^\pm_j\to\ell^\pm\tilde\nu_{\ell}^{(\ast)}$, 
with $\ell=e,\mu,\tau$, and
$\tilde\chi^\pm_j\to W^\pm\tilde\chi_k^0$.
In Subsection~\ref{energydistr}, we have shown that the slope of the 
energy distribution of the decay particle $\lambda=\ell,W$, 
is proportional to the averaged longitudinal chargino polarization
$\bar\Sigma^3_P/\bar P$, see Eq.~(\ref{edist2}).
The polarization can be determined by 
the energy distribution asymmetry~\cite{Kittel:2005ma} 
\begin{eqnarray}
        {\mathcal A}_{ij,\lambda^\pm} &=& 
        \frac{\Delta \sigma_{ij,\lambda^\pm} }{\sigma_{ij,\lambda^\pm}}
 =  \frac{1}{2}
        \eta_{\lambda^\pm}
        \frac{\bar\Sigma^3_P}{\bar P}
\nonumber\\
 &=& 
 \frac{1}{2}
        \eta_{\lambda^\pm}
        \frac{(1 + {\mathcal{P}}_+{\mathcal{P}}_-)b_0 +
 ({\mathcal{P}}_+ + {\mathcal{P}}_-)b_1 +\bar\Sigma^3_{\rm cont} }
  {(1 + {\mathcal{P}}_+{\mathcal{P}}_-)a_0 +
    ({\mathcal{P}}_+ + {\mathcal{P}}_-)a_1 
 +\bar P_{\rm cont}},
\label{asymmetryenergydist}
\end{eqnarray}
with 
\begin{eqnarray}
  \Delta \sigma_{ij,\lambda^\pm}&=&
         \sigma_{ij,\lambda^\pm}(E_\lambda > \hat E_\lambda)
        -\sigma_{ij,\lambda^\pm}(E_\lambda < \hat E_\lambda),
\label{eq:deltasigma}
\end{eqnarray}
and $\lambda=e,\mu,\tau,W$.
Here we have used the explicit formula of the energy distribution
of the decay particle $\lambda^\pm=\ell^\pm,W^\pm$~(\ref{edist2}).

\medskip

An example of the energy distribution and the 
corresponding asymmetries ${\mathcal A}_{ij,\lambda^\pm}$
for the leptonic decay $\lambda=e,\mu$, 
for equal muon beam polarizations
${\mathcal P}_+ = {\mathcal P}_- \equiv {\mathcal P}$,
is given in~Fig.~\ref{fig:edist.l1l2}.
The area under each graph equals the corresponding
cross section of production and decay,
$\sigma_{11,\ell^\pm}=\sigma_{11}\times 
{\rm BR}(\tilde\chi^\pm_1\to\ell^\pm\tilde\nu_{\ell}^{(\ast)})$,
see Eq.~(\ref{eq:sigmatot}).
Here $\sigma_{11}$ is larger for ${\mathcal P}>0$ than for
${\mathcal P}<0$,
since the P-odd asymmetry $\mathcal{A}^{\Cp\Pm}_{11}$~(\ref{eq:aprodCePo}) 
is positive.
The slope of the curves is proportional to the 
averaged longitudinal chargino polarization 
$\bar\Sigma_P^3$,
and to the decay factor $\eta_{\ell^\pm}$, with
$\eta_{\ell^+}=-\eta_{\ell^-}=1$. Thus the slope of the energy
distribution of $\ell^+$ has the opposite sign than that of
$\ell^-$, each for  ${\mathcal P}>0$ and ${\mathcal P}<0$
separately.
However, the moduli of the slopes are not the
same for the two leptons for fixed ${\mathcal P}$, 
which can be seen in the different absolute values of
the asymmetries, e.g.,
$ {\mathcal A}_{11,\ell^+}=-11.3\%$ and
$ {\mathcal A}_{11,\ell^-}=11.45\%$
for ${\mathcal P}=+0.3$.
The difference of the absolute values is due to the small continuum contributions,  
which change sign, depending on the chargino charge, i.e.,
$ \bar\Sigma_{\rm cont}(\tilde\chi^+_1)$=
$-\bar\Sigma_{\rm cont}(\tilde\chi^-_1)$.

\medskip

The average chargino polarization depends on the continuum contributions 
$\bar\Sigma^3_{\rm cont}$, 
and is also proportional to $b_0$ and $b_1$, 
see Eq.~(\ref{asymmetryenergydist}). 
In order to separate these 
coefficients into their C-even and -odd parts
$b_n^+$, $b_n^-$, respectively,
we define the four generalized decay asymmetries of the energy 
distribution~\cite{CPNSH},
\begin{eqnarray}
\mathcal{A}^{\Cpm\Pp}_{ij,\lambda}&=&
        \frac{
               \Delta \sigma_{ij,\lambda^+}(\mathcal{ P})
               -
               \Delta \sigma_{ij,\lambda^+}(\mathcal{-P})
               \mp\Big[
               \Delta \sigma_{ij,\lambda^-}(\mathcal{ P})
               -
               \Delta \sigma_{ij,\lambda^-}(\mathcal{-P})\Big]
       }{
               \sigma_{ij,\lambda^+}(\mathcal{ P})
               +
               \sigma_{ij,\lambda^+}(\mathcal{-P})
               +
               \sigma_{ij,\lambda^-}(\mathcal{ P})
               +
               \sigma_{ij,\lambda^-}(\mathcal{-P})
        }
\label{eq:adecCeoPe}
\\[3mm]&=&
        \eta_{\lambda^+}
        \frac{   \mathcal{P}    b_1^\pm }
             {(1+\mathcal{P}^2) a_0^+ +  \bar{P}_{\rm cont}},
\label{eq:adecCeoPedep}
\\[3mm]
\mathcal{A}^{\Cp\Pm}_{ij,\lambda}&=&
        \frac{
               \Delta \sigma_{ij,\lambda^+}(\mathcal{ P})
               +
               \Delta \sigma_{ij,\lambda^+}(\mathcal{-P})
               -\Big[
               \Delta \sigma_{ij,\lambda^-}(\mathcal{ P})
               +
               \Delta \sigma_{ij,\lambda^-}(\mathcal{-P})\Big]
       }{
               \sigma_{ij,\lambda^+}(\mathcal{ P})
               +
               \sigma_{ij,\lambda^+}(\mathcal{-P})
               +
               \sigma_{ij,\lambda^-}(\mathcal{ P})
               +
               \sigma_{ij,\lambda^-}(\mathcal{-P})
        }
\label{eq:adecCePo}
\\[3mm]&=&
        \frac{ 1 }{2}
        \eta_{\lambda^+}
        \frac{   (1+ \mathcal{P}^2) b_0^+ }
                {(1+ \mathcal{P}^2) a_0^+ + \bar{P}_{\rm cont}},
\label{eq:adecCePodep}
\\[3mm]
\mathcal{A}^{\Cm\Pm}_{ij,\lambda}&=&
        \frac{
               \Delta \sigma_{ij,\lambda^+}(\mathcal{ P})
               +
               \Delta \sigma_{ij,\lambda^+}(\mathcal{-P})
               +
               \Delta \sigma_{ij,\lambda^-}(\mathcal{ P})
               +
               \Delta \sigma_{ij,\lambda^-}(\mathcal{-P})
       }{
               \sigma_{ij,\lambda^+}(\mathcal{ P})
               +
               \sigma_{ij,\lambda^+}(\mathcal{-P})
               +
               \sigma_{ij,\lambda^-}(\mathcal{ P})
               +
               \sigma_{ij,\lambda^-}(\mathcal{-P})
        }
\label{eq:adecCoPo}
\\[3mm]&=&
        \frac{ 1 }{2}
         \eta_{\lambda^+}
        \frac{  (1+ \mathcal{P}^2) b_0^- + \bar\Sigma^3_{\rm cont}}
                {(1+\mathcal{P}^2) a_0^+ +  \bar{P}_{\rm cont}},
\label{eq:adecCoPodep}
\end{eqnarray}
for equal muon beam polarizations 
${\mathcal P}_+ = {\mathcal P}_- \equiv {\mathcal P}$.
Note that the continuum contributions  $\bar\Sigma^3_{\rm cont}$ of the chargino 
$\tilde\chi^\pm_j$ polarization are C-odd,
$ \bar\Sigma^3_{\rm cont}(\tilde\chi_i^-\tilde\chi_j^+) =$ 
$-\bar\Sigma^3_{\rm cont}(\tilde\chi_i^+\tilde\chi_j^-)$,
and P-even, 
$ \bar\Sigma^3_{\rm cont}({\mathcal P}) = 
  \bar\Sigma^3_{\rm cont}(-{\mathcal P})$,
and thus cancel (only) in the numerator of the C-even and/or P-even asymmetries
$\mathcal{A}^{\Cpm\Pp}_{ij,\lambda}$~(\ref{eq:adecCeoPedep}) and
$\mathcal{A}^{\Cp\Pm}_{ij,\lambda}$~(\ref{eq:adecCePodep}), respectively.
The asymmetries for the decay into a $\tau$ or $W$ boson,
$\tilde\chi^\pm_j\to\tau^\pm\tilde\nu_\tau^{(\ast)}$, 
and
$\tilde\chi^\pm_j\to W^\pm\tilde\chi_k^0$,
respectively,
are generally smaller than those for the decays into an electron or muon,
due to $|\eta_{\tau^\pm}|,|\eta_{W^\pm}|\leq |\eta_{e^\pm}|=|\eta_{\mu^\pm}|=1$,
see Eqs.~(\ref{etatau}) and (\ref{etaW}).

\medskip

The CP-even asymmetry 
$\mathcal{A}^{\Cp\Pp}_{ij,\lambda}$~(\ref{eq:adecCeoPe})
is due to the correlation between the longitudinal polarizations 
of the initial muons and final charginos.
Large values of %
$\mathcal{A}^{\Cp\Pp}_{ij,\lambda}$, and also of
$\mathcal{A}^{\Cm\Pm}_{ij,\lambda}$~(\ref{eq:adecCoPo}),
can be obtained 
when both Higgs resonances are nearly degenerate,
and if their amplitudes are of the same magnitude.
However, a scalar-pseudoscalar mixing in the presence
of CP phases will in general
increase the mass splitting of the Higgs bosons,
and the reduced overlap of the Higgs resonances
also reduces the CP-even asymmetry 
$\mathcal{A}^{\Cp\Pp}_{ij,\lambda}$.

The CP-odd asymmetries
$\mathcal{A}^{\Cm\Pp}_{ij,\lambda}$~(\ref{eq:adecCeoPe}) and 
$\mathcal{A}^{\Cp\Pm}_{ij,\lambda}$~(\ref{eq:adecCePo})
are sensitive to CP-violating phases and thus
vanish for CP-conserving Higgs couplings.
Similarly to the CP-odd polarization asymmetry 
$\mathcal{A}^{\Cp\Pm}_{ij}$~(\ref{eq:aprodCePo})
for chargino production, the decay asymmetry
$\mathcal{A}^{\Cp\Pm}_{ij,\lambda}$
is approximately
maximal if the Higgs mixing is resonantly enhanced.
As pointed out earlier, this can happen naturally
in the Higgs decoupling limit.

\medskip

Finally we 
count the total number of observables %
which are available in chargino production and decay
with longitudinally polarized beams.
There are two production and two decay asymmetries/observables,
each for  $\tilde\chi_1^+\tilde\chi_1^-$ and $\tilde\chi_2^+\tilde\chi_2^-$
production and decay. 
Note that the C-odd observables vanish for diagonal chargino production.
For $\tilde\chi_1^\pm\tilde\chi_2^\mp$ production,
there are four production observables and four decay observables for
the decay of chargino $\tilde\chi_1^\pm$, as well as additional four decay
observables for the decay of $\tilde\chi_2^\pm$. 
These $20$ observables 
can be used to determine the Higgs couplings in chargino
production, see also Table~\ref{coeffs:CandP}.

\section{Numerical results
  \label{Numerical results}}

We analyze numerically the 
CP-even and CP-odd asymmetries 
for chargino production,
$\mu^+\mu^-\to\tilde\chi^+_1\tilde\chi^-_1$,
and 
$\mu^+\mu^-\to\tilde\chi^\pm_1\tilde\chi^\mp_2$.
For the chargino decays,
$\tilde\chi^\pm_{1,2} \to \ell^\pm\tilde\nu_\ell^{(\ast)}$,
we study the CP-even and CP-odd decay asymmetries
of the leptonic energy distributions, which allow
to probe the longitudinal chargino polarizations.
The feasibility of measuring the asymmetries depends also
on the chargino production cross section and decay branching ratios, 
which we discuss in detail.
We will identify regions of the parameter space
where a resonant enhanced mixing of the Higgs bosons
will lead to nearly maximal CP-violating effects.

We induce CP violation in the Higgs sector 
by a non-vanishing phase $\phi_A$ of the common 
trilinear scalar coupling parameter
$A_t=A_b=A_\tau\equiv|{A}|\exp(i\phi_A)$
for the third generation fermions.
This assignment is also compatible with the
bounds on CP-violating phases from experiments
on electric dipole moments 
(EDMs)~\cite{Yao:2006px,Harris:1999jx,Regan:2002ta,Romalis:2000mg}.
For simplicity a 
we keep the gaugino mass parameters $M_1$, $M_3$, and the Higgs mass parameter $\mu$ real.
For the calculation of the Higgs masses, widths 
and couplings, we use
the program {\tt FeynHiggs 2.5.1}~\cite{Frank:2006yh,Degrassi:2002fi}, 
see also~\cite{Dreiner:2007ay}.
We fix $\tan\beta=10$, since the Higgs boson decays into charginos
are most relevant for intermediate values of $\tan\beta$.
Smaller values of $\tan\beta$
favor the $t\bar t$ decay channel, while
larger values enhance decays into $b\bar b$ and $\tau\bar\tau$.
For the branching ratios and width
of the decaying chargino, we include the two-body decays~\cite{Kittel:2005rp}
\begin{eqnarray}
\tilde\chi^\pm_{1,2} &\to& W^\pm+\tilde\chi^0_k,\, e^\pm+\tilde\nu_e,
\,\mu^\pm+\tilde\nu_\mu,\, \tau^\pm+\tilde\nu_\tau,
\,\nu_e+\tilde{e}^\pm_L,\,\nu_\mu+\tilde\mu^\pm_L,
\,\nu_\tau+\tilde\tau^\pm_{1,2},\nonumber\\
\tilde\chi^\pm_2 &\to& Z^0+\tilde\chi^\pm_1,\,H^0_1+\tilde\chi^\pm_1.
\label{chardecaymodes}
\end{eqnarray}
and neglect three-body decays. 
We parametrize the slepton masses by $m_0$ and $M_2$,
which enter in
the approximate solutions 
to the renormalization group 
equations, see Appendix~\ref{chardecay}.
For the numerical discussion, we fix $m_0=70$~GeV.
Thus we obtain light sneutrino masses, which enable
the chargino decays $\tilde\chi^\pm_1 \to \ell^\pm\tilde\nu_\ell$.
We parametrize the diagonal entries of the
squark mass matrices by the common SUSY scale
parameter $M_{\rm SUSY}=M_{\tilde Q_3}=M_{\tilde U_3}=M_{\tilde D_3}$.
We fix $M_{\rm SUSY}=500$~GeV to suppress Higgs decays into the heavier squarks.
In order to reduce the number of parameters, we assume  
the GUT relation for the gaugino mass parameters
$M_1=5/3 \, M_2\tan^2\theta_W $.
Finally, we choose longitudinal muon beam polarizations of
${\mathcal P}_+={\mathcal P}_-={\mathcal P}=\pm0.3$,
as well as a luminosity of ${\mathcal L}=1~{\rm fb}^{-1}$. 
These values 
should be feasible at a muon collider running at $\sqrt{s}\sim 0.5~{\rm TeV}$\cite{hefreports}.

\medskip

\begin{table}
\renewcommand{\arraystretch}{1.2}
\caption{\label{scenarioA} SUSY parameters for the benchmark 
                           scenario~{CP$\chi$}.
The slepton masses are parametrized by $m_0$ and $M_2$,  
the squark masses by $M_{\rm SUSY}$.
}
\begin{center}
     \begin{tabular}{|c|c|c|c|}
\hline
$ M_{H^\pm} = 500~{\rm GeV}$ & 
$ \tan\beta = 10 $ & 
$       |{A}| = 1~{\rm TeV}$ & 
$    \phi_A = 0.2 \pi$ \\
\hline
$ M_{\rm SUSY} = 500~{\rm GeV}$ &
$          \mu = 400~{\rm GeV}$ & 
$          M_2 = 240~{\rm GeV}$ & 
$          m_0 = 70~{\rm GeV}$ \\
\hline
\end{tabular}
\end{center}
\renewcommand{\arraystretch}{1.0}
\end{table}

\begin{table}
\renewcommand{\arraystretch}{1.2}
\caption{SUSY masses, widths, branching ratios, and decay factors $\eta$
         for the benchmark scenario~{CP$\chi$},
        evaluated with  
       {\tt FeynHiggs 2.5.1}~\cite{Frank:2006yh,Degrassi:2002fi}.
         \label{scenarioAmasses}}
\begin{center}
      \begin{tabular}{|c|c|c|c|}
\hline
$        M_{H_1}  = 126.0~{\rm GeV}$ & 
$   m_{\chi^0_1}  =   118~{\rm GeV}$ &
$ m_{\tilde{e}_R} =   141~{\rm GeV}$ & 
${\rm BR}(\tilde\chi_1^+\to \ell^+\tilde \nu_\ell) =11  \%$ \\
\hline
$        M_{H_2}  = 492.7~{\rm GeV}$ & 
$   m_{\chi^0_2}  =   223~{\rm GeV}$ &
$ m_{\tilde{e}_L} =   230~{\rm GeV}$ & 
${\rm BR}(\tilde\chi_1^+\to W^+\tilde \chi_1^0) =20  \%$ \\
\hline
$        M_{H_3}  = 493.5~{\rm GeV}$ & 
$   m_{\chi^0_3}  =   405~{\rm GeV}$ &
$m_{\tilde\tau_1} =   138~{\rm GeV}$ & 
${\rm BR}(\tilde\chi_1^+\to \tilde\tau_1^+ \nu_\tau) = 48 \%$ \\
\hline
$   {\Gamma_{H_2} =  1.43~{\rm GeV}}$ & 
$ m_{\chi^\pm_1}  =  223~{\rm GeV}$ &
$m_{\tilde\tau_2} =  232~{\rm GeV}$ & 
$     \eta_{\tau^+} =0.99  $\\
\hline
$   {\Gamma_{H_3} =   1.30~{\rm GeV}}$ & 
$ m_{\chi^\pm_2}  =  425~{\rm GeV}$ &
$ m_{\tilde\nu}   =  215~{\rm GeV}$ & 
$     \eta_{W^+} =-0.30  $\\
\hline
\end{tabular}
\end{center}
\renewcommand{\arraystretch}{1.0}
\end{table}

\medskip

We center our numerical discussion around 
scenario~{CP$\chi$}, defined in Table~\ref{scenarioA}.
Inspired by the benchmark scenario 
CPX~\cite{Carena:2000ks} 
for studying enhanced CP-violating Higgs-mixing phenomena,
we set $|{A}|=2M_{\rm SUSY}=1~{\rm TeV}$, 
$M_3=800~{\rm GeV}$,
and a non-vanishing phase
$\phi_A = 0.2 \pi$.
We thus obtain large contributions from the trilinear 
coupling parameter ${A}$ of the third generation 
to the Higgs sector, both CP-conserving and CP-violating.

\subsection{Conditions for resonant enhanced Higgs mixing}
\label{sec:enhancement}

The CP-violating effects are maximized if the 
mixing of the Higgs states with different CP parities is resonantly 
enhanced.
This can happen naturally in the Higgs decoupling limit.
Resonant Higgs mixing occurs when the diagonal elements 
$m_{H}^2 - \hat\Sigma_{HH}(s)$ and $m_{A}^2 - \hat\Sigma_{AA}(s)$
of the Higgs mass matrix ${\rm \bf M}$, Eq.~(\ref{eq:Weisskopf}), 
are similar in size, 
and thus their difference is of the same order as the off-diagonal 
element $\hat\Sigma_{HA}(s)$.
When $m_{H}^2 - m_{A}^2 - \rm{Re}[ \hat\Sigma_{HH}(s)-\hat\Sigma_{AA}(s)]$
changes sign,
we interpret this condition as a level crossing of the CP eigenstates  $H$ 
and $A$~\cite{Choi:2004kq}.
The difference of the imaginary parts can be small,
$\rm{Im} \hat\Sigma_{HH}(s) - \rm{Im}\hat\Sigma_{AA}(s)\approx 0$,
if the Higgs channels into heavy squarks are closed,
as in our scenario~{CP$\chi$}.
Thus we obtain a large $H$--$A$ mixing even for 
moderate values of the CP-violating scalar-pseudoscalar 
self-energy transitions $\hat\Sigma_{HA}(s)$.
This also means that, in contrast to the benchmark scenario 
CPX~\cite{Carena:2000ks}, the Higgs mixing can be large
even for small values of $\mu$.
We thus choose 
$\mu = 400~{\rm GeV}$ and $M_2 = 240~{\rm GeV}$ 
of similar size in order
to enhance the branching ratios of the Higgs bosons into lighter charginos,
which are large only for mixed 
gaugino-higgsino charginos.
We give the masses of the Higgs bosons, charginos, sneutrinos,
light neutralinos, 
and the widths of the Higgs bosons
for scenario~{CP$\chi$}
in Table~\ref{scenarioAmasses}, 
where we also list the branching ratios for chargino~$\chargino_1$,
and the decay factors $\eta_{\tau^+}$~(\ref{etatau}) and 
$\eta_{W^+}$~(\ref{etaW}).
For non-diagonal chargino production 
we consider smaller values of $\mu$ and $M_2$ than in scenario~{CP$\chi$},
in order for this process to be kinematically allowed,
while keeping the same remaining parameters.


\subsection{ Production of $\tilde\chi_1^+\tilde\chi_1^-$}
\label{sec:prod.chi1.chi1}

\subsubsection{ $\sqrt{s}$ dependence}
\label{sec:sqrts.dependence}

For the scenario~{CP$\chi$},
we analyze the dependence of the asymmetries 
and the cross sections on the center-of-mass energy $\sqrt{s}$.
The CP-even and CP-odd observables exhibit a characteristic 
$\sqrt s$ dependence, mainly given
by the product of Higgs boson propagators $\Delta_{(kl)}$,
see Eq.~(\ref{deltadelta}).
A  muon collider will have a precise beam energy resolution,
and thus enables detailed line-shape scans.

\medskip

\begin{figure}[t]
\begin{picture}(16,13)
\put(-3,-10.3){\includegraphics{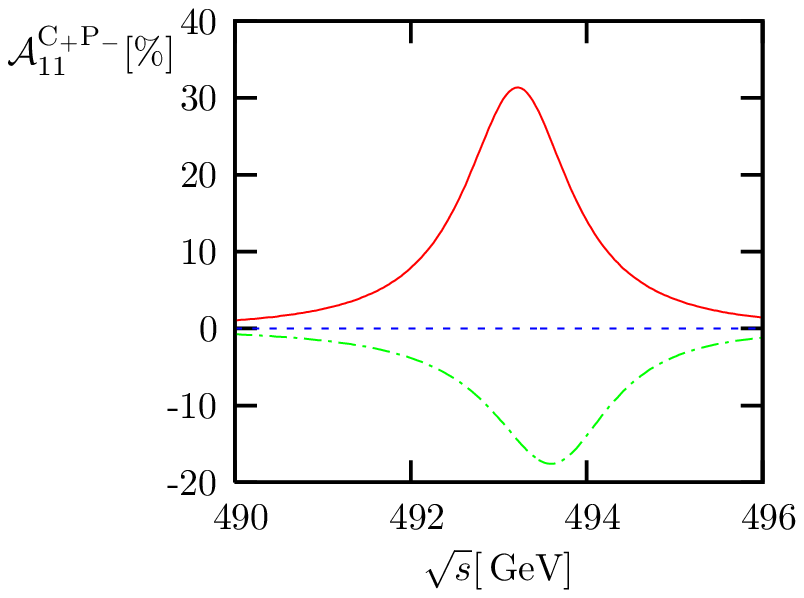}}
\put( 5.3,-10.3){\includegraphics{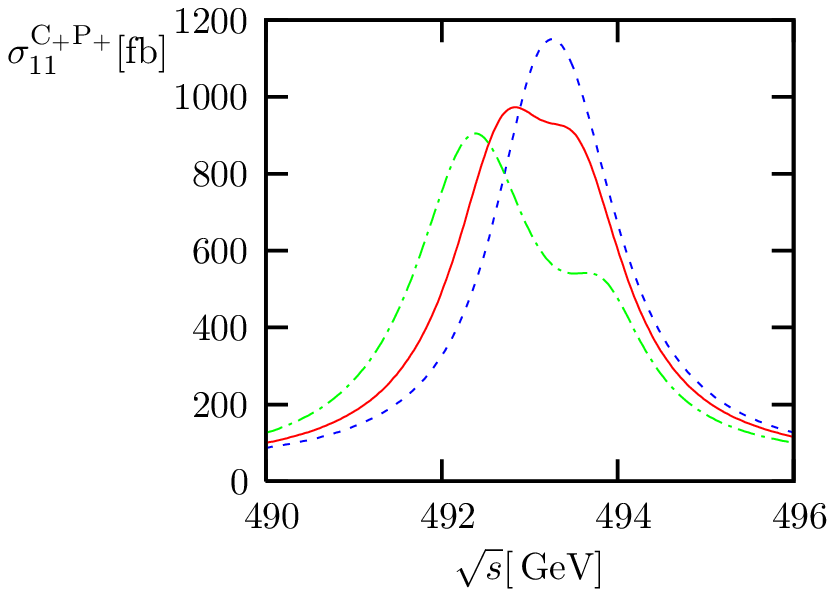}}
\put(0,7.){(a)}
\put(8,7.){(b)}
\put(-3,-17.3){\includegraphics{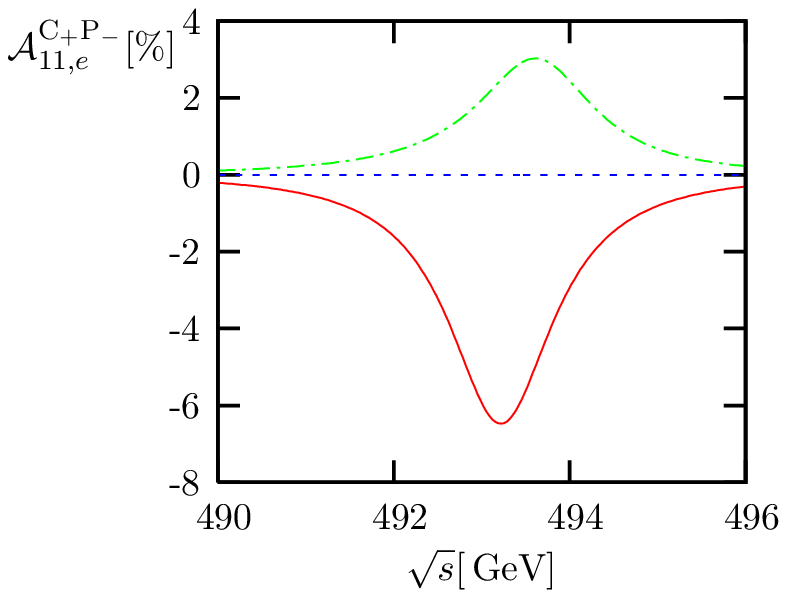}}
\put( 5.3,-17.3){\includegraphics{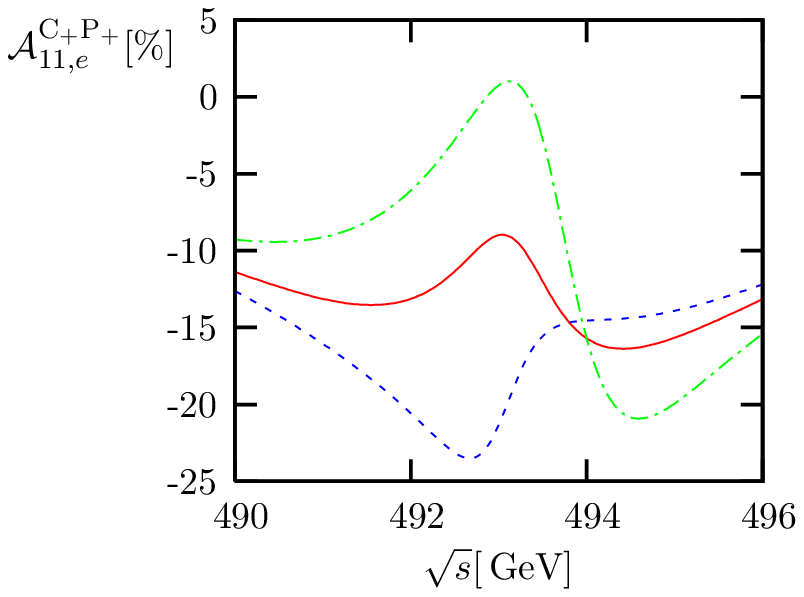}}
\put(0,0.15){(c)}
\put(8,0.15){(d)}
\end{picture}
\caption{\small  
        $\sqrt s$ dependence of 
        {\bf (a)}~the CP-odd production asymmetry 
        $\mathcal{A}^{\Cp\Pm}_{11}$, Eq.~(\ref{eq:aprodCePo}),
        and {\bf (b)}~the beam polarization averaged cross section
        $\sigma_{11}^{\Cp\Pp}$, Eq.~(\ref{eq:sigmacepe}),
        for chargino production
        $\mu^+\mu^- \to \tilde\chi_1^+\tilde\chi_1^-$,
        with common longitudinal beam polarization $|{\mathcal P}|=0.3$.
        For the subsequent decay
        $\tilde\chi_1^\pm\to e^\pm\tilde \nu_e^{(\ast)}$ 
        in {\bf (c)}~the CP-odd decay asymmetry
        $\mathcal{A}^{\Cp\Pm}_{11,e}$, Eq.~(\ref{eq:adecCePo}),
        and in {\bf (d)}~the CP-even decay asymmetry
        $\mathcal{A}^{\Cp\Pp}_{11,e}$, Eq.~(\ref{eq:adecCeoPe}).
        The  phase of the trilinear coupling $A$ is 
        $\phi_A=0$~(dotted, blue),
        $\phi_A=0.2 \pi$~(solid, red), and
        $\phi_A=0.6 \pi$~(dash-dotted, green). 
        The other SUSY parameters are given in Table~\ref{scenarioA}.
}
\label{fig:sqrts}
\end{figure}

In Fig.~\ref{fig:sqrts}(a), we show the CP-odd production asymmetry
$\mathcal{A}^{\Cp\Pm}_{11}$~(\ref{eq:aprodCePo})
for chargino production
$\mu^+\mu^- \to \tilde\chi_1^+\tilde\chi_1^-$
as a function of $\sqrt{s}$
around the heavy Higgs resonances $H_2$ and $H_3$.
At the peak value, $\sqrt{s}=(M_{H_2}+M_{H_3})/2\approx 493 $~GeV,
the interference of the two nearly degenerate Higgs bosons
is maximal, leading to an asymmetry of up to 
$\mathcal{A}^{\Cp\Pm}_{11}=30\%$.
The asymmetry measures the difference
of the chargino production cross section $\sigma_{11}({\mathcal P})$
for equal positive and negative muon beam polarizations
${\mathcal P}=\pm0.3$, see Eq.~(\ref{eq:aprodCePo}).
In Fig.~\ref{fig:sqrts}(b), we show the 
beam polarization averaged cross section
$\sigma_{11}^{\Cp\Pp}$, see Eq.~(\ref{eq:sigmacepe}),
for $\phi_A=0.2\pi$ (solid), $\phi_A=0$ (dotted), and $\phi_A=0.6\pi$ (dash-dotted).
We can observe that
the splitting of the two resonances is increased in the presence of
CP-violating phases in this scenario.
For $\phi_A=0.2\pi$ and $\phi_A=0.6\pi$ 
the two resonances are clearly visible in the line shape of the polarization averaged cross section 
$\sigma_{11}^{\Cp\Pp}$,
whereas it assumes the form of a single resonance for $\phi_A=0$,
where the Higgs bosons are extremely degenerate,
see Fig.~\ref{fig:sqrts}(b).

\medskip

The Higgs boson interference in chargino production also leads
to CP-odd and CP-even contributions to the average longitudinal 
chargino polarizations. In order to analyze the $\tilde\chi_1^\pm$
polarization, one can measure the CP-odd asymmetry 
$\mathcal{A}^{\Cp\Pm}_{11,\lambda}$~(\ref{eq:adecCePo}),
and the CP-even asymmetry
$\mathcal{A}^{\Cp\Pp}_{11,\lambda}$~(\ref{eq:adecCeoPe}),
of the energy distributions of the decay particle $\lambda=\ell,W$
in the chargino decay
$\tilde\chi_1^\pm \to \ell^\pm\tilde\nu_\ell^{(\ast)}$ or
$\tilde\chi_1^\pm \to W^\pm\tilde\chi_1^0$, respectively.
For simplicity, we discuss only the decay into an electron sneutrino,
$\tilde\chi_1^\pm\to e^\pm\tilde\nu_e^{(\ast)}$, i.e. $\lambda=e$.
With our choice of slepton sector parameters, i.e.,
approximate RGE relations for the masses, see Eqs.~(\ref{mslr})-(\ref{msn}), and a vanishing trilinear coupling for the first two slepton generations,
the same asymmetries are obtained for the decay 
into a muon sneutrino,
$\lambda=\mu$.
The asymmetries for the decay into a tau  sneutrino, $\lambda=\tau$,
or into a $W$ boson, $\lambda=W$,
are obtained by taking into account the decay factors
$\eta_\tau$~(\ref{etatau}) and $\eta_W$~(\ref{etaW}),
respectively, see Eqs.~(\ref{eq:adecCeoPedep}) and (\ref{eq:adecCePodep}).

\medskip

For the decay
$\tilde\chi_1^\pm\to e^\pm \tilde \nu_e^{(\ast)}$,
we show the $\sqrt{s}$ dependence of the
CP- and CP${\tilde{\rm T}}$-odd asymmetry
$\mathcal{A}^{\Cp\Pm}_{11,e}$ in Fig.~\ref{fig:sqrts}(c)
for $\phi_A=0.2\pi$.
The CP- and CP${\tilde{\rm T}}$-even asymmetry
$\mathcal{A}^{{\Cp\Pp}}_{11,e}$
is shown in Fig.~\ref{fig:sqrts}(d), both 
for $\phi_A=0$ and $\phi_A=0.2\pi$.
The phase $\phi_A$ tends to increase the mass splitting of the Higgs resonances. 
Their overlap is now reduced, leading in general to a suppression of the CP-even
asymmetry $\mathcal{A}^{\Cp\Pp}_{11,e}$, in particular
at the mean energy of the resonances $\sqrt{s}=(M_{H_2}+M_{H_3})/2$,
see the solid line in Fig.~\ref{fig:sqrts}(d). 
On the contrary, the larger Higgs splitting
increases the CP-odd asymmetries 
$\mathcal{A}^{\Cp\Pm}_{11,e}$ and $\mathcal{A}^{\Cp\Pm}_{11}$.

\medskip

All asymmetries for production and decay vanish asymptotically far from the 
resonance region. The continuum contributions from smuon, photon and $Z$ exchange
to the difference of the cross sections and to the average chargino
polarization cancel in the numerator, but contribute in the denominator
of the corresponding asymmetries, see 
their definitions in Section~\ref{AsymmetriesforPandD}.

\medskip

In the following Sections, we analyze 
the dependence of the 
production cross section 
and the asymmetries
on $|{A}|$ and $\phi_A$,
and finally on $M_2$ and $\mu$, 
fixing all remaining parameters to those of 
scenario~{CP$\chi$}. 
We fix the center-of-mass energy to
$\sqrt{s}=(M_{H_2}+M_{H_3})/2$, 
where we expect the largest CP-odd asymmetries
$\mathcal{A}^{\Cp\Pm}_{11}$ and
$\mathcal{A}^{\Cp\Pm}_{11,e}$,
see Figs.~\ref{fig:sqrts}(a) and (c), respectively.
For consistency, we also choose 
$\sqrt{s}=(M_{H_2}+M_{H_3})/2$
for the discussion of the CP-even decay asymmetry
$\mathcal{A}^{\Cp\Pp}_{11,e}$
although it is generally suppressed at this value if CP is violated.

\subsubsection{$|{A}|$ and $\phi_A$ dependence}
\label{Adependence}

\begin{figure}[t]
\begin{picture}(16,13.)
\put(-3,-10.3){\includegraphics{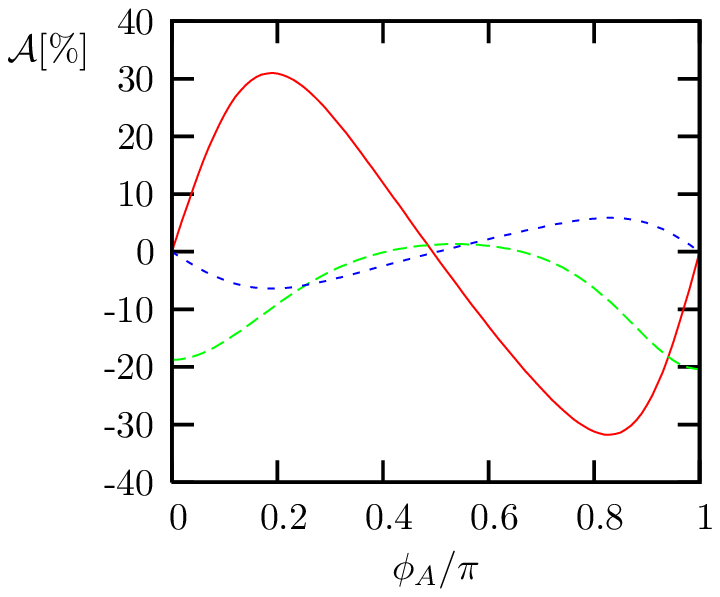}}
\put( 5.4,-10.3){\includegraphics{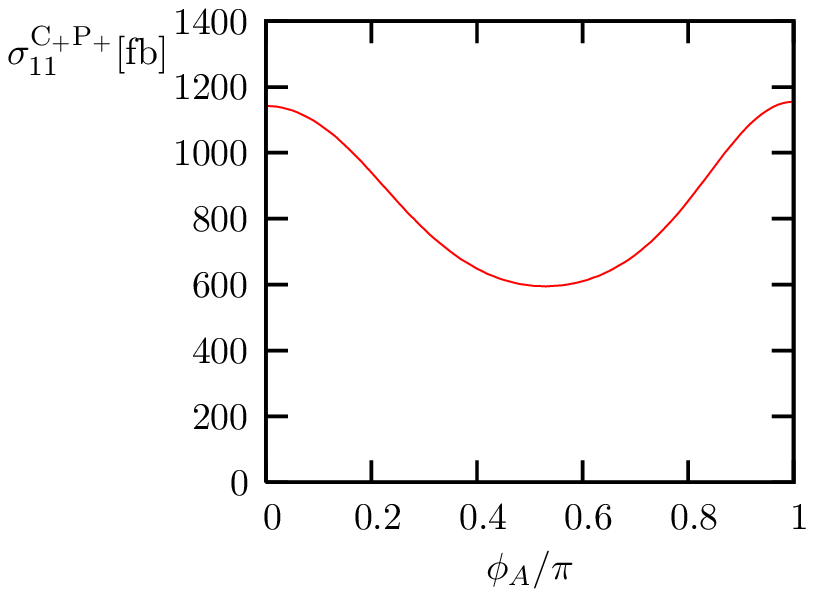}}
\put(.0,7.15){(a)}
\put(8.,7.15){(b)}
\put(-3,-17.3){\includegraphics{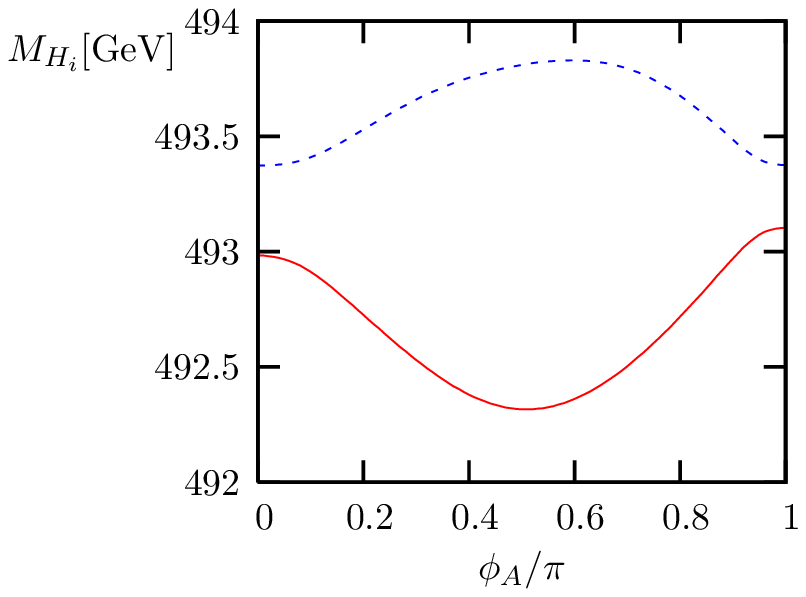}}
\put( 5.4,-17.3){\includegraphics{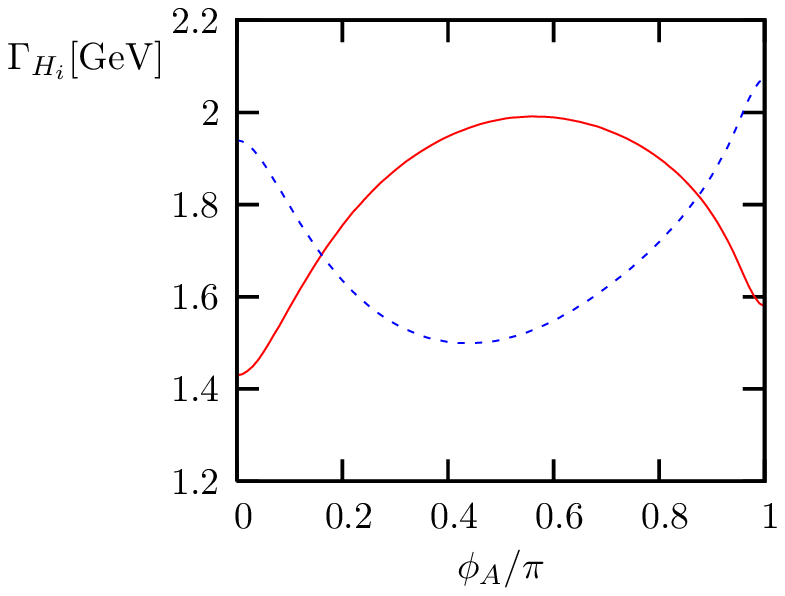}}
\put(.0,0.15){(c)}
\put(8.,0.15){(d)}

\put(2.8,11.8){$\mathcal{A}^{\Cp\Pm}_{11}$}
\put(4.9,11.0){$\mathcal{A}^{\Cp\Pm}_{11,e}$}
\put(1.5, 9.0){$\mathcal{A}^{\Cp\Pp}_{11,e}$}

\end{picture}
\caption{\small  
      Phase dependence of 
      {\bf (a)}~the CP-odd  asymmetry
      $\mathcal{A}^{\Cp\Pm}_{11}$ (solid, red), Eq.~(\ref{eq:aprodCePo}),
      for chargino production 
      $\mu^+\mu^- \to \tilde\chi_1^+\tilde\chi_1^-$
      at $\sqrt{s}=(M_{H_2}+M_{H_3})/2$,
      and for the subsequent decay 
      $\tilde\chi_1^\pm\to e^\pm\tilde \nu_e^{(\ast)}$
      the CP-even asymmetry 
      $\mathcal{A}^{\Cp\Pp}_{11,e}$ (dashed, green), Eq.~(\ref{eq:adecCeoPe}),
      and the CP-odd asymmetry
      $\mathcal{A}^{\Cp\Pm}_{11,e}$ (dotted, blue), Eq.~(\ref{eq:adecCePo}).
      In {\bf (b)}~the beam polarization averaged cross section 
      $\sigma_{11}^{\Cp\Pp}$, Eq.~(\ref{eq:sigmacepe}),
      with common longitudinal beam polarization $|{\mathcal P}|=0.3$.
      In {\bf (c)}~the Higgs masses $M_{H_i}$ and
      {\bf (d)}~the Higgs widths $\Gamma_{H_i}$,
      for $i=2$ (solid, red), and $i=3$ (dotted, blue).
      The other SUSY parameters are given in Table~\ref{scenarioA}.
}
\label{fig:phi}
\end{figure}

We analyze the dependence of the CP asymmetries on the phase $\phi_A$ of
the trilinear coupling ${A}$,
which is the only source of CP violation in our study.
The CP-odd asymmetries, 
$\mathcal{A}^{\Cp\Pm}_{11}$ and
$\mathcal{A}^{\Cp\Pm}_{11,e}$
see Fig.~\ref{fig:phi}(a),
are approximately maximal, if the 
mixing of the Higgs states is resonantly enhanced,
as discussed in Section~\ref{sec:enhancement}.
The degeneracy in the neutral Higgs bosons,
however, is lifted 
by the $H$--$A$ mixing, as can be observed
from Fig.~\ref{fig:phi}(c). A splitting of the order of the Higgs
widths $\Gamma_{H_{2,3}}$, shown in Fig.~\ref{fig:phi}(d),
leads to large absorptive phases, which are necessary for the presence of
CP$\tilde{\rm{T}}$-odd observables.
The 
increased Higgs mass splitting
leads, however, also to lower peak cross sections,
and thus to a lower polarization averaged cross section
$\sigma_{11}^{\Cp\Pp}$,
which we show in Fig.~\ref{fig:phi}(b) for common muon
beam polarizations $|{\mathcal P}|=0.3$.

\medskip

The asymmetries and cross sections for negative $\phi_A$
can be obtained from symmetry considerations.
Since the complex trilinear coupling ${A}$ is the only 
source of CP violation in our analysis, the CP-odd asymmetries
$\mathcal{A}^{\Cp\Pm}_{11}$ and
$\mathcal{A}^{\Cp\Pm}_{11,e}$
are odd with respect to the transformation $\phi_A \to -\phi_A$,
while the CP-even asymmetry $\mathcal{A}^{\Cp\Pp}_{11,e}$,
and the averaged cross section $\sigma_{11}^{\Cp\Pp}$, 
are even.

\medskip

\begin{figure}[t]
\centering
\begin{picture}(16,15.5)
\put(-2.15,-5.9){\includegraphics{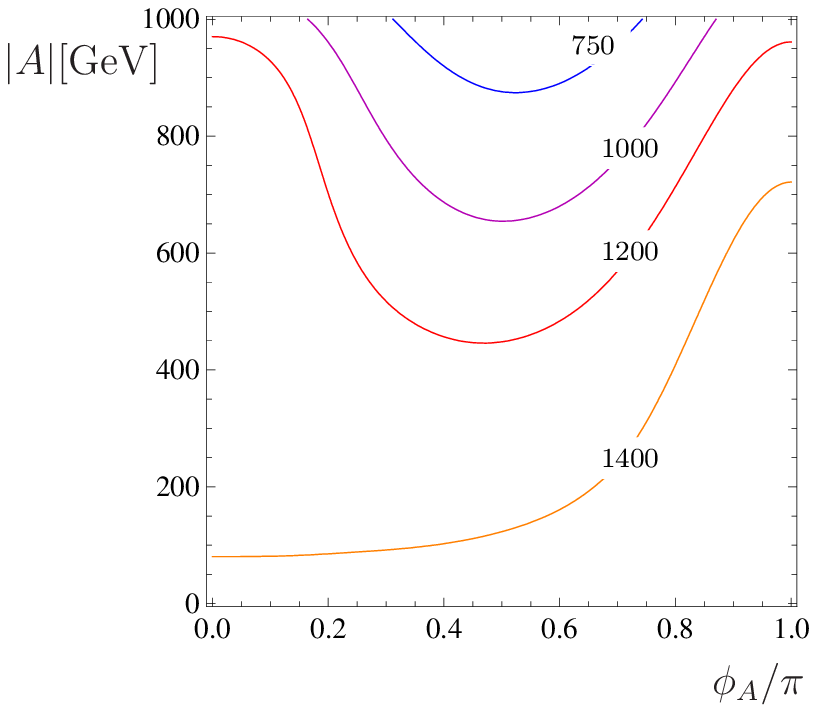}}
\put(6,-5.9){\includegraphics{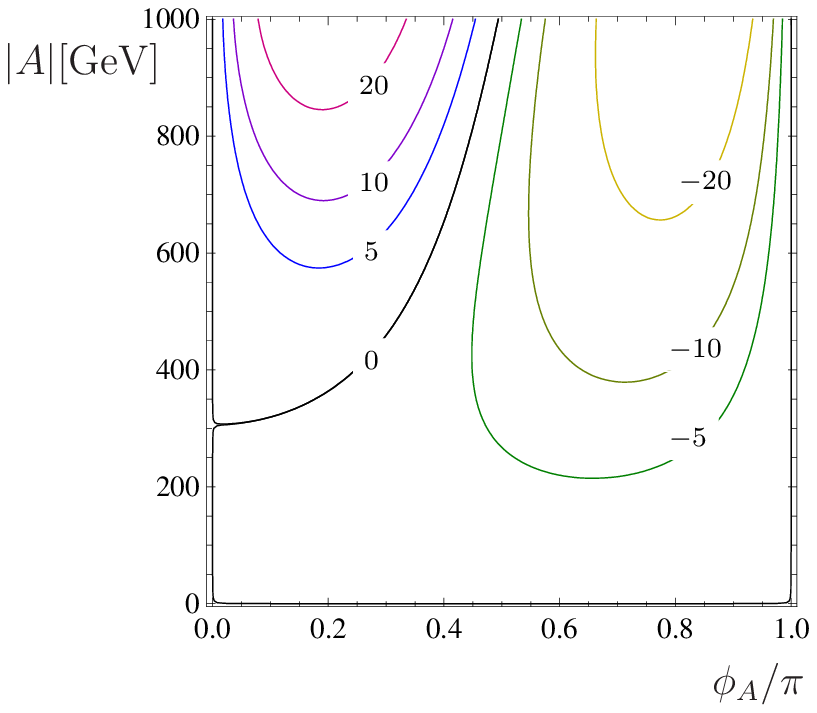}}
\put(1,8.15){(a)}
\put(9,8.15){(b)}
\put(-2.15,-13.9){\includegraphics{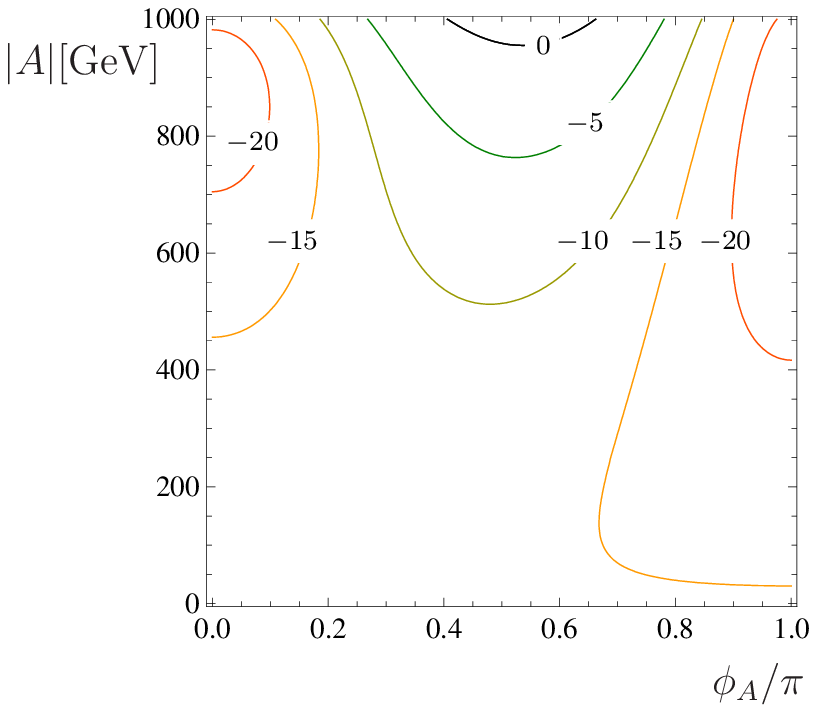}}
\put(6,-13.9){\includegraphics{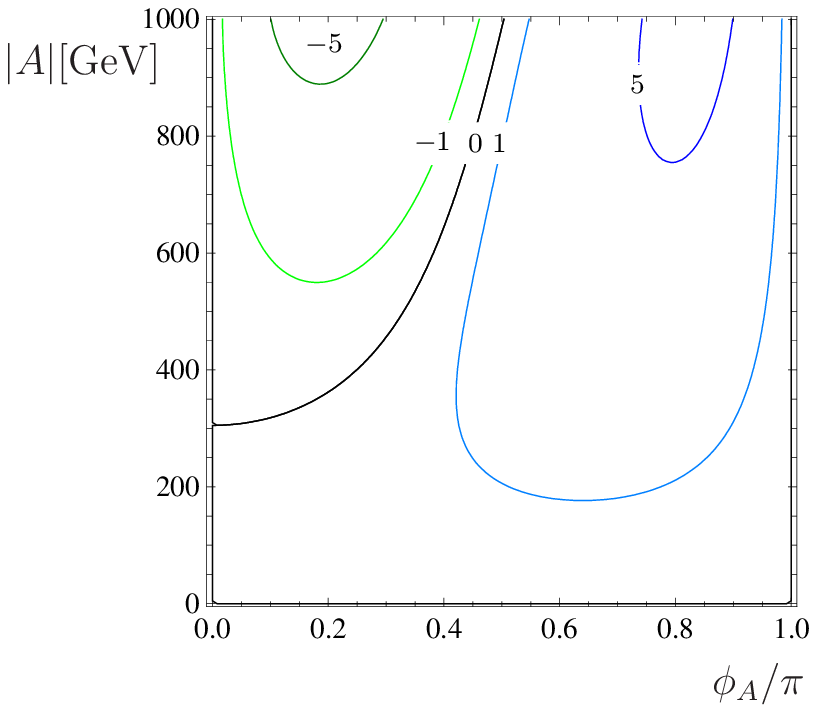}}
\put(1,0.15){(c)}
\put(9,0.15){(d)}
\put(0.7,14.9){averaged cross section $\sigma_{11}^{\Cp\Pp}$ in $\rm{fb}$ }
\put(9.8,14.9){Asymmetry $\mathcal{A}^{\Cp\Pm}_{11}$   in \%}
\put(1.7, 6.9){Asymmetry $\mathcal{A}^{\Cp\Pp}_{11,e}$ in \%}
\put(9.8, 6.9){Asymmetry $\mathcal{A}^{\Cp\Pm}_{11,e}$ in \%}
\end{picture}
\caption{\small  
      Contour lines in the $\phi_A$--$|{A}|$ plane
      for chargino production
      $\mu^+\mu^- \to \tilde\chi_1^+\tilde\chi_1^-$,
      for {\bf (a)} 
      the beam polarization averaged cross section 
      $\sigma_{11}^{\Cp\Pp}$, Eq.~(\ref{eq:sigmacepe}),
      and {\bf (b)} the CP-odd  asymmetry
      $\mathcal{A}^{\Cp\Pm}_{11}$, Eq.~(\ref{eq:aprodCePo}),
      at $\sqrt{s}=(M_{H_2}+M_{H_3})/2$ and
      common longitudinal beam polarization $|{\mathcal P}|=0.3$.
      For the subsequent decay 
      $\tilde\chi_1^\pm\to e^\pm\tilde \nu_e^{(\ast)}$,
      in {\bf (c)} the CP-even asymmetry 
      $\mathcal{A}^{\Cp\Pp}_{11,e}$, Eq.~(\ref{eq:adecCeoPe}),
      and in {\bf (d)} 
      the CP-odd asymmetry
      $\mathcal{A}^{\Cp\Pm}_{11,e}$, Eq.~(\ref{eq:adecCePo}).
      The SUSY parameters are given in Table~\ref{scenarioA}.
 }
\label{fig:asymm.AphiA}
\end{figure}

In Fig.~\ref{fig:asymm.AphiA}, we show contour lines of the 
averaged cross section $\sigma_{11}^{\Cp\Pp}$
and the asymmetries in the $\phi_A$--$|{A}|$ plane. 
The largest CP-odd asymmetries 
$\mathcal{A}^{\Cp\Pm}_{11}$ and
$\mathcal{A}^{\Cp\Pm}_{11,e}$
are obtained for $|{A}|\approx 2 M_{\rm SUSY} = 1$~TeV.
For larger values of $|{A}|$,
the lighter stops become kinematically accessible and
$H_2$ decays dominantly into $\tilde{t}_1^+ \tilde{t}_1^-$ pairs.
This leads to a suppression of the chargino production cross section,
and also to a suppression of the Higgs mixing for small $\mu$.
We therefore restrict our discussion to $|{A}| \lsim 1$~TeV.

\medskip

As we have observed in Fig.~\ref{fig:phi}(a),
the CP-even asymmetry  $\mathcal{A}^{\Cp\Pp}_{11,e}$
is in general larger in the CP-conserving limit.
This can be also seen in 
Fig.~\ref{fig:asymm.AphiA}(c), 
where the maximum of the asymmetry
is obtained for $\phi_A =0,\pi$,
and $|A| \approx  800~{\rm GeV}$. 
However, this is rather coincidental, and is due to the exact 
degeneracy of the Higgs bosons $H$ and $A$.

\subsubsection{  $\mu$ and $M_2$ dependence}
\label{sec:mu.and.M_2.dependence}
\begin{figure}[t]
\centering
\begin{picture}(16,15.5)
\put(-2.15,-5.9){\includegraphics{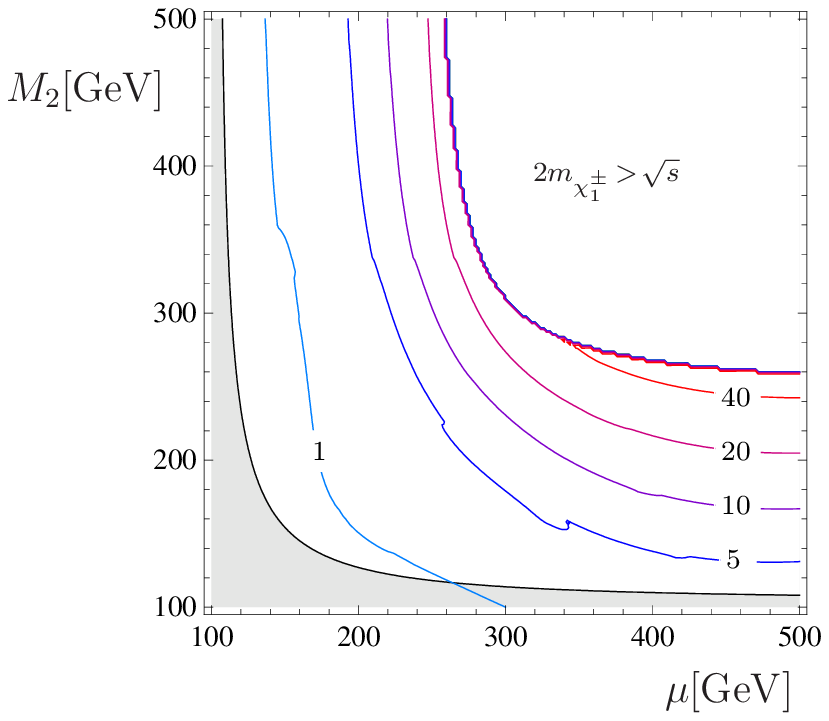}}
\put(6,-5.9){\includegraphics{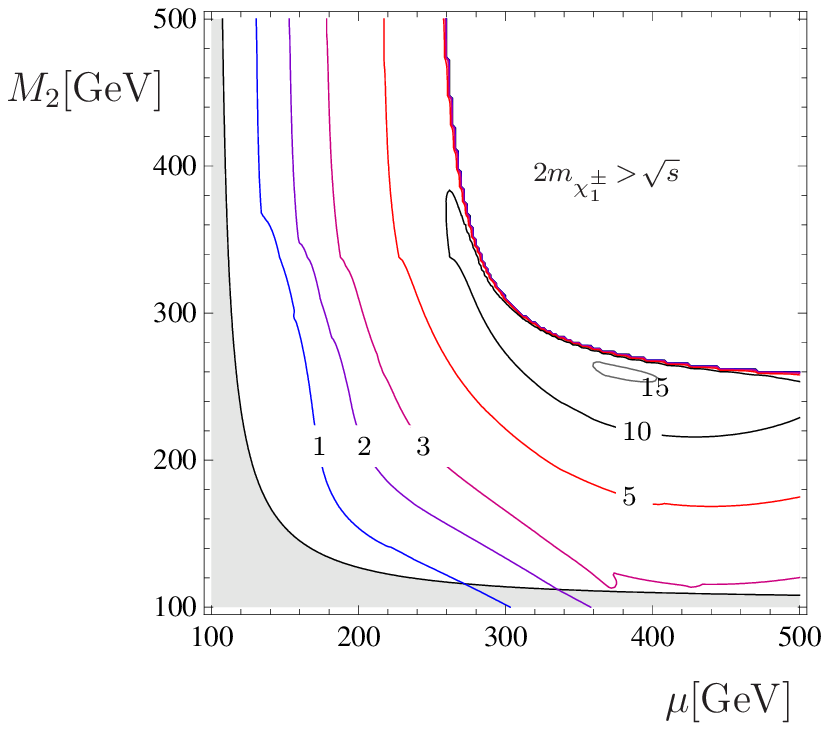}}
\put(1,8.15){(a)}
\put(9,8.15){(b)}
\put(-2.15,-13.9){\includegraphics{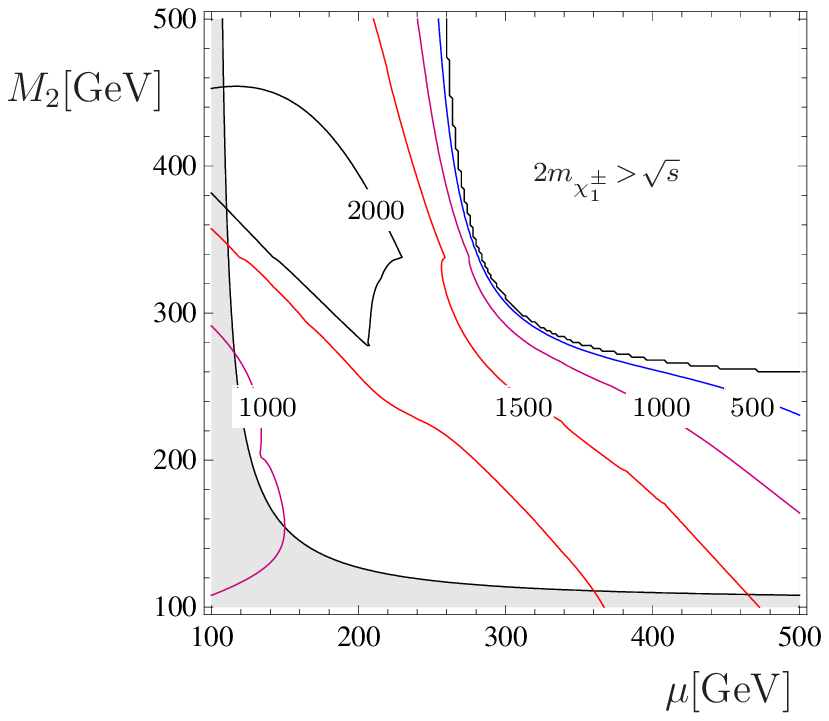}}
\put(6,-13.9){\includegraphics{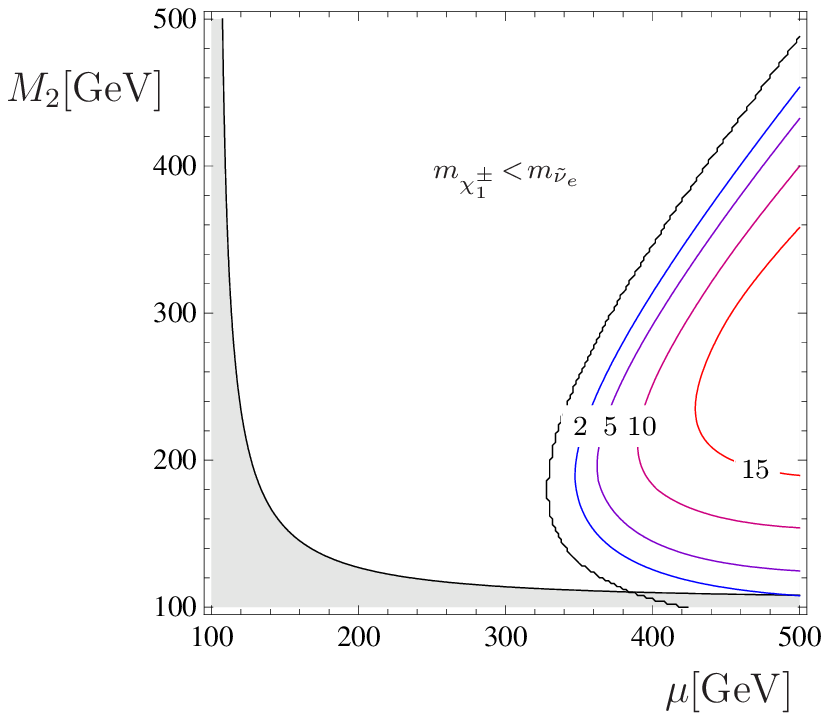}}
\put(1,0.15){(c)}
\put(9,0.15){(d)}
\put(1.4,14.9){    Asymmetry $\mathcal{A}^{\Cp\Pm}_{11}$ in \%}
\put(10.3,14.9){Significance  $\mathcal{S}^{\Cp\Pm}_{11}$}
\put(0.7, 6.9){averaged cross section $\sigma_{11}^{\Cp\Pp}$ in $\rm{fb}$ }
\put(10., 6.9){${\rm BR}(\tilde\chi_1^+\to e^+\tilde \nu_e)$ in \%}
\end{picture}
\caption{\small  
       {\bf (a)-(c)}
       Chargino production 
       $\mu^+\mu^-\to\tilde\chi_1^+\tilde\chi_1^-$
       at $\sqrt{s}=(M_{H_2}+M_{H_3})/2$ with 
       common muon beam polarization 
       $|\mathcal{P}|=0.3$.
       Contour lines in the $\mu$--$M_2$ plane for
       {\bf (a)}~the CP-odd asymmetry
       $\mathcal{A}^{\Cp\Pm}_{11}$, Eq.~(\ref{eq:aprodCePo}),
       {\bf (b)}~the corresponding significance 
       $\mathcal{S}^{\Cp\Pm}_{11}$, Eq.~(\ref{eq:SaprodCePo}),
       with ${\mathcal L}=1~{\rm fb}^{-1}$, and
       {\bf (c)}~the
       beam polarization averaged cross section 
       $\sigma_{11}^{\Cp\Pp}$, Eq.~(\ref{eq:sigmacepe}),
       for the SUSY parameters as given in Table~\ref{scenarioA}.
       The chargino branching ratio
       ${\rm BR}(\tilde\chi_1^+\to e^+\tilde \nu_e)$
       is given in {\bf (d)}.
       The shaded area is excluded by $m_{\chi_1^\pm}<103$~GeV.
 }
\label{fig:prod11}
\end{figure}

%
The couplings of the Higgs bosons to the charginos
strongly depend on the gaugino-higgsino composition of the charginos,
which is mainly determined by the values of $\mu$ and $M_2$.
For chargino production $\mu^+\mu^-\to\tilde\chi_1^+\tilde\chi_1^-$,
we show the CP-odd asymmetry
$\mathcal{A}^{\Cp\Pm}_{11}$~(\ref{eq:aprodCePo})
in the $\mu$--$M_2$ plane in Fig.~\ref{fig:prod11}(a)
with $|{\mathcal P}|=0.3$.
The absolute maximal value of the asymmetry is constrained by
the degree of beam polarization.
For $|{\mathcal P}|=0.3$ the maximum would be
$\mathcal{A}^{\Cpm\Pm}_{11\,\rm (max)}\approx 55\%$,
as follows from Eq.~(\ref{eq:amax}).
We observe in Fig.~\ref{fig:prod11}(a)
that the asymmetry reaches more than $40\%$ 
near the chargino production threshold,
where the coefficient $a_1$~(\ref{eq.a1}) 
receives large spin-flip contributions.
For smaller $\mu$ and $M_2$,
the interplay between unitarity and CPT symmetry~\cite{Pilaftsis:1997dr}
suppresses $\mathcal{A}^{\Cp\Pm}_{11}$,
see Ref.~\cite{Ellis:2004fs}
for a discussion on the related process $b \bar b \to \tau^+\tau^- $.
In addition,
the Higgs boson widths are increased, since decay channels
into light neutralinos and charginos open.
The larger widths result in a larger overlap of the Higgs resonances,
which reduces the absorptive phases in our scenario, and consequently 
suppresses the CP$\tilde{\rm T}$-odd asymmetry
$\mathcal{A}^{\Cp\Pm}_{11}$.

\medskip

In Fig.~\ref{fig:prod11}(c), we show the 
beam polarization averaged cross section
$\sigma_{11}^{\Cp\Pp}$~(\ref{eq:sigmacepe})
for chargino production $\mu^+\mu^-\to\tilde\chi_1^+\tilde\chi_1^-$,
which reaches up to $\sigma_{11}^{\Cp\Pp}\approx 2$~pb.
In addition 
the cross section shows no $p$-wave suppression at threshold,
since $H_2$ and $H_3$ are mixed CP eigenstates.
The corresponding statistical significance, see Eq.~(\ref{eq:SaprodCePo}),
shown in Fig.~\ref{fig:prod11}(b) for an integrated luminosity of ${\mathcal L}=1~{\rm fb}^{-1}$,
is largest
near threshold.
In Fig.~\ref{fig:prod11}(d), we
show the branching ratio for the chargino decay
$\tilde\chi_1^+\to e^+\tilde \nu_e$, which reaches
more than $15\%$, as the leptonic decay modes
into $\mu$ and $\tau$.
The main competing channels
are the decay into the $W$ boson,  and those into sleptons 
$\tilde e_L$, $\tilde\mu_L$, and $\tilde\tau_{1,2}$,
see Eq.~(\ref{chardecaymodes}).
The chargino decay $\tilde\chi_1^+\to\tilde\tau_1^+\nu_\tau$
is dominating, see the branching ratios in Table~\ref{scenarioAmasses},
and is the only two-body decay channel 
for $m_{\chi_1^\pm}<m_{\tilde\nu_\ell}$, see the
contour of the kinematical threshold in Fig.~\ref{fig:prod11}(d).

\medskip

\begin{figure}[t]
\centering
\begin{picture}(16,7.5)
\put(-2.15,-13.9){\includegraphics{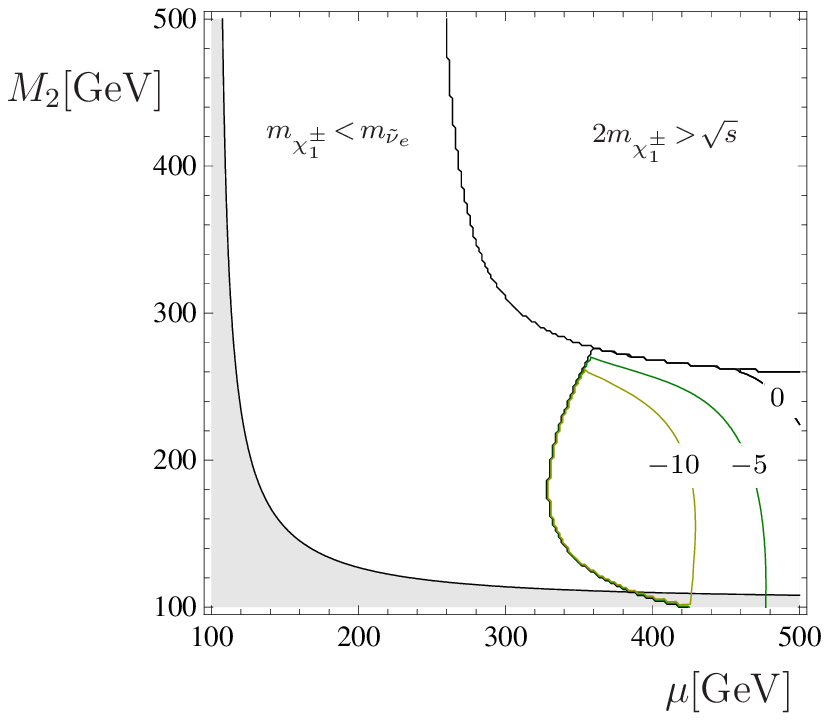}}
\put(6,-13.9){\includegraphics{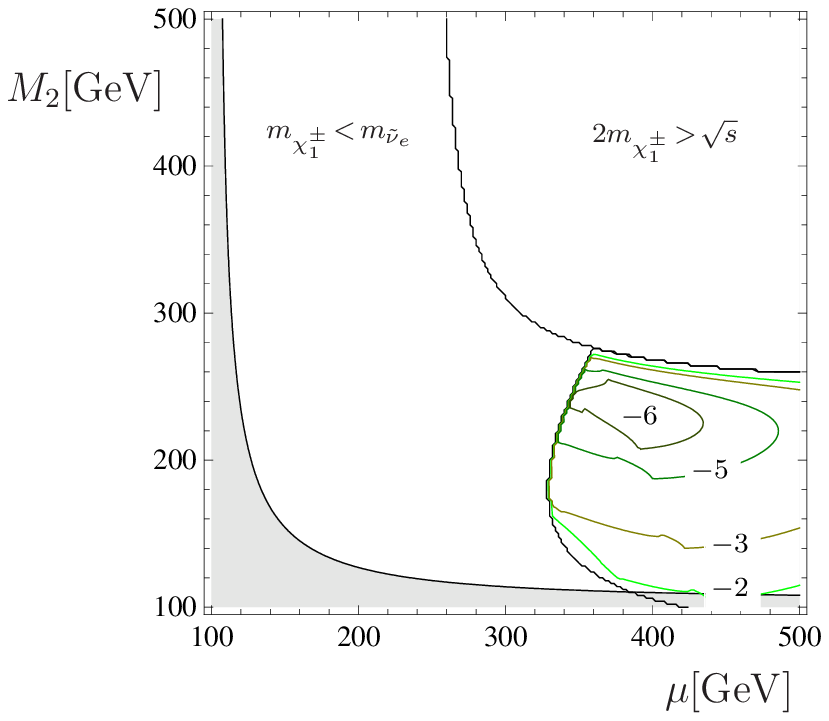}}
\put(1.0,0.15){(a)}
\put(9.0,0.15){(b)}
\put(1.7,6.9){Asymmetry $\mathcal{A}^{\Cp\Pm}_{11,e}$ in \%}
\put(9.8,6.9){Asymmetry $\mathcal{A}^{\Cp\Pp}_{11,e}$  in \%}
\end{picture}
\caption{\small 
        Contour lines in the $\mu$--$M_2$ plane for
        {\bf (a)}~the CP-odd asymmetry 
        $\mathcal{A}^{\Cp\Pm}_{11,e}$, Eq.~(\ref{eq:adecCePo}),
        and {\bf (b)}~the CP-even asymmetry
        $\mathcal{A}^{\Cp\Pp}_{11,e}$, Eq.~(\ref{eq:adecCeoPe}),
        for chargino production
        $\mu^+\mu^-\to\tilde\chi_1^+\tilde\chi_1^-$ 
        and decay $\tilde\chi_1^\pm\to e^\pm\tilde \nu_e^{(\ast)}$
        at $\sqrt{s}=(M_{H_2}+M_{H_3})/2$ with 
        longitudinally muon beam polarization
        $|\mathcal{P}|=0.3$,
        for the SUSY parameters as given in Table~\ref{scenarioA}.
        The corresponding chargino cross section and branching ratio 
        are shown in Fig.~\ref{fig:prod11}.
        The shaded area is excluded by $m_{\chi_1^\pm}<103$~GeV.
}
\label{fig:dec.asym.e}
\end{figure}


For the chargino decay
$\tilde\chi_1^\pm\to e^\pm\tilde \nu_e$,
we show the CP-odd and CP-even decay asymmetries 
$\mathcal{A}^{\Cp\Pm}_{11,e}$~(\ref{eq:adecCePo}),
and
$\mathcal{A}^{\Cp\Pp}_{11,e}$~(\ref{eq:adecCeoPe}),
in Figs.~\ref{fig:dec.asym.e}(a) and (b), respectively. 
As discussed before,
the CP-even asymmetry $\mathcal{A}^{\Cp\Pp}_{11,e}$ is suppressed 
by CP-violating effects due to the smaller overlap of the resonances 
at $\sqrt{s}=(M_{H_2}+M_{H_3})/2$.
Therefore we only find large values of 
$\mathcal{A}^{\Cp\Pp}_{11,e}$
for light neutralinos and charginos in the lower left corner of 
Fig.~\ref{fig:dec.asym.e}(b), where the larger Higgs widths counter 
the effect of the larger Higgs mass difference.
On the contrary, in that region the CP-odd asymmetry
$\mathcal{A}^{\Cp\Pm}_{11,e}$
is reduced due to smaller absorptive phases.
Finally, at threshold the longitudinal polarization of the chargino 
$\Sigma_{\rm res}^3$~(\ref{eq.Sr}) vanishes, 
and thus also both decay asymmetries, as follows from 
Eqs.~(\ref{eq.b0}) and~(\ref{eq.b1}).

\medskip

The statistical significances of the CP-odd and CP-even decay asymmetries
$\mathcal{A}^{\Cp\Pm}_{11,e}$,
and
$\mathcal{A}^{\Cp\Pp}_{11,e}$,
see Eqs.~(\ref{eq:SadecCeoPoe}), (\ref{eq:SadecCeoPeo}),
reach at most
$\mathcal{S}^{\Cp\Pm}_{11}\approx 2$
and
$\mathcal{S}^{\Cp\Pp}_{11}\approx 3$
respectively,
for an integrated luminosity of ${\mathcal L}=1~{\rm fb}^{-1}$.
Thus their measurement will be challenging.
Nonetheless, only these asymmetries 
allow to disentangle the contributions $b_0^+$ and $b_1^+$
to the longitudinal chargino polarization
$\Sigma_{\rm res}^3$~(\ref{eq.Sr}).
However these asymmetries can also be measured 
in the decays 
$\tilde\chi_1^\pm\to W^\pm\tilde\chi_1^0$,
and
$\tilde\chi_1^\pm\to \tau^\pm\tilde\nu_\tau^{(\ast)}$.
Note that the corresponding decay asymmetries are
smaller due to the decay factors $\eta_\lambda$, which are  
of the order
$|\eta_{W^\pm}|\approx 0.2-0.4$ and
$|\eta_{\tau^\pm}|\approx 1$.
A measurement of the asymmetries in these decay channels
is more involved,
since the $W$ and $\tau$ reconstruction efficiencies
have to be taken into account.

\subsubsection{Comparison with asymmetries in neutralino production}
              \label{sec:comparison}

The results
obtained for diagonal chargino production
 $\mu^+\mu^-\to\tilde\chi_1^+\tilde\chi_1^-$
in the previous section have  similarities with the corresponding asymmetries
for neutralino production, see Ref.~\cite{CPNSH}, and Ref.~\cite{Dreiner:2007ay} 
for a detailed review.
For instance, 
we have obtained similar line-shapes of the CP-odd production asymmetry,
$\mathcal{A}^{\Cp\Pm}_{11}$, Eq.~(\ref{eq:aprodCePo}),
as well as of the CP-odd and CP-even asymmetries for the chargino decays
$\mathcal{A}^{\Cp\Pm}_{11,e}$, Eq.~(\ref{eq:adecCePo}),
and 
$\mathcal{A}^{\Cp\Pp}_{11,e}$, Eq.~(\ref{eq:adecCeoPe}),
respectively, compared with the analogous  asymmetries for neutralino production.
The asymmetries also show a very similar dependence
on the common complex trilinear coupling parameter $A$, as well as
on the %
parameters $\mu$ and $M_2$.
In both processes, the production asymmetry is enhanced at threshold,
whereas the decay asymmetries show a distinct $p$-wave suppression. 
The similarities are due to similar kinematics, i.e., production of spin-half 
charginos or neutralinos, as well as  dynamics, i.e., Higgs interference in the production
of a C-even final state of massive gauginos and higgsinos.

Note that for neutralino or diagonal chargino production 
only C-even production asymmetries can be defined,
whereas after including the C-odd decay, only C-odd
decay asymmetries can be probed, see also the detailed discussions 
in Section~\ref{AsymmetriesforPandD}. For non-diagonal chargino production,
in addition two C-odd production asymmetries, as well as two C-even decay asymmetries,
are accessible, which we  discuss in the following section.

\subsection{ Production of $\tilde\chi_1^\pm\tilde\chi_2^\mp$}
\label{sec:prod.chi1.chi2}

\begin{figure}[t]
\centering
\begin{picture}(16,15.5)
\put(-2.15,-5.9){\includegraphics{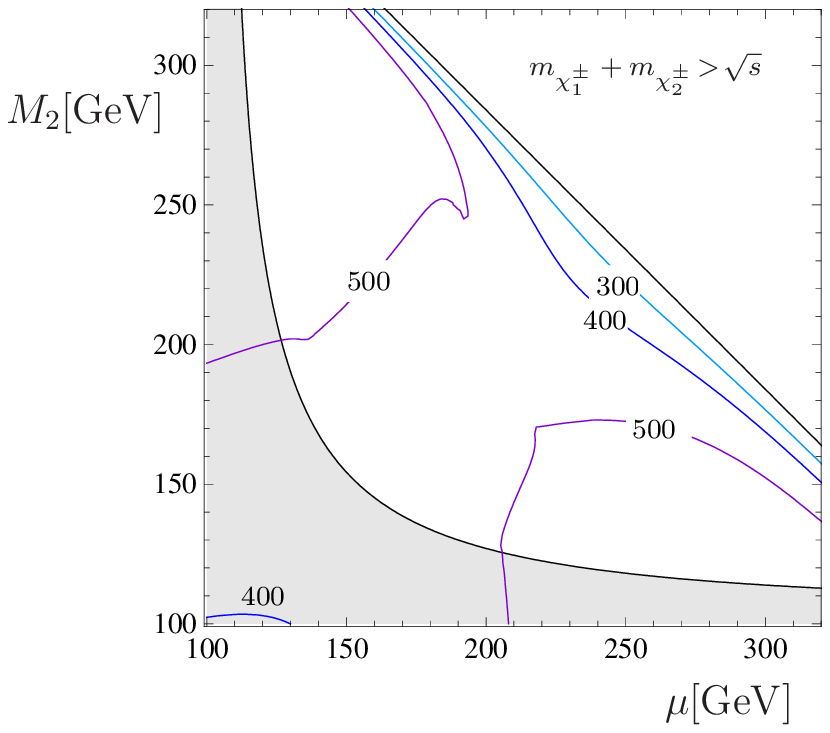}}
\put(6,-5.9){\includegraphics{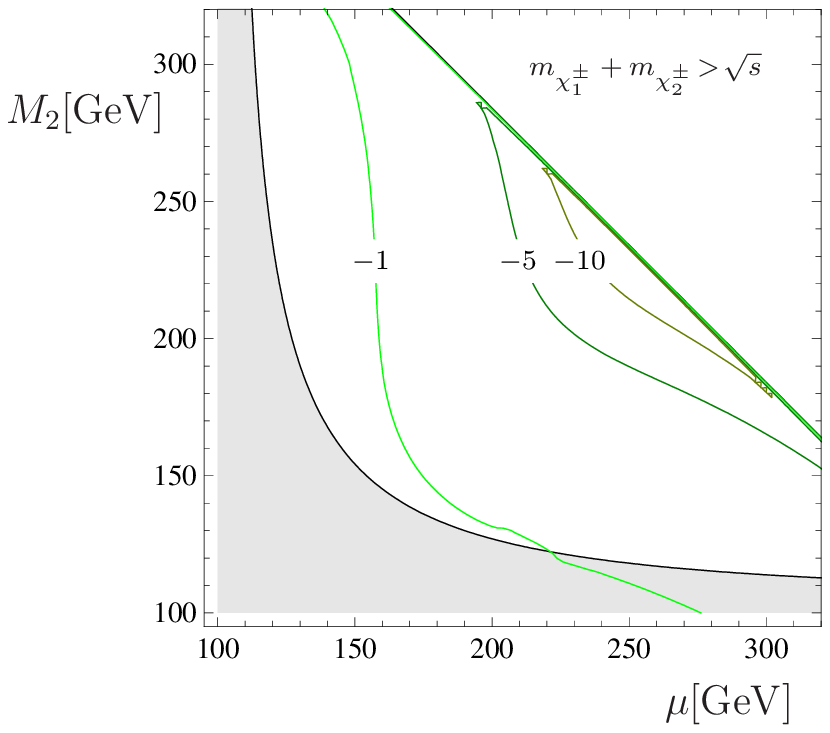}}
\put(1,8.15){(a)}
\put(9,8.15){(b)}
\put(-2.15,-13.9){\includegraphics{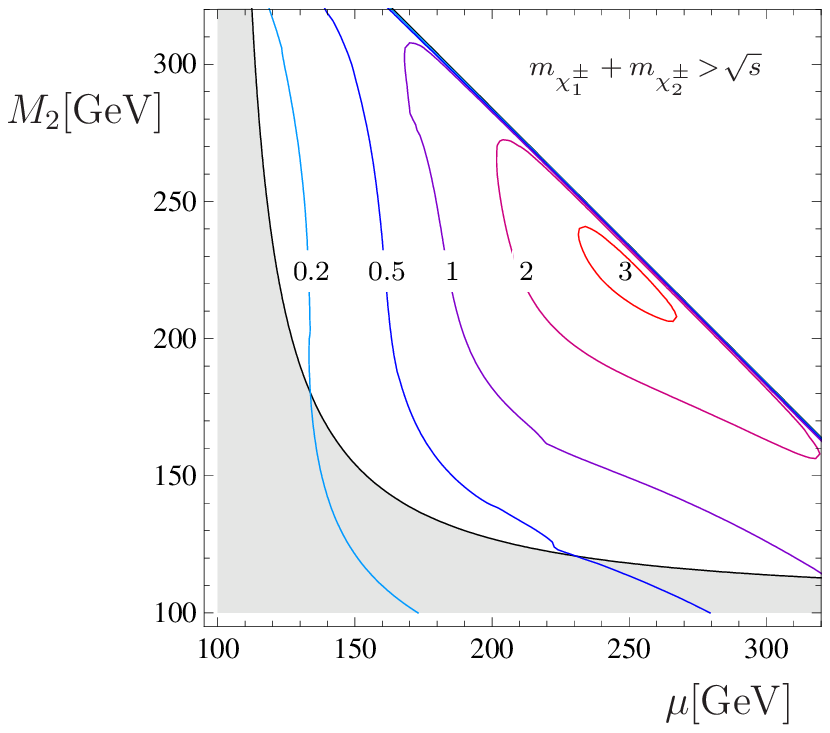}}
\put(6,-13.9){\includegraphics{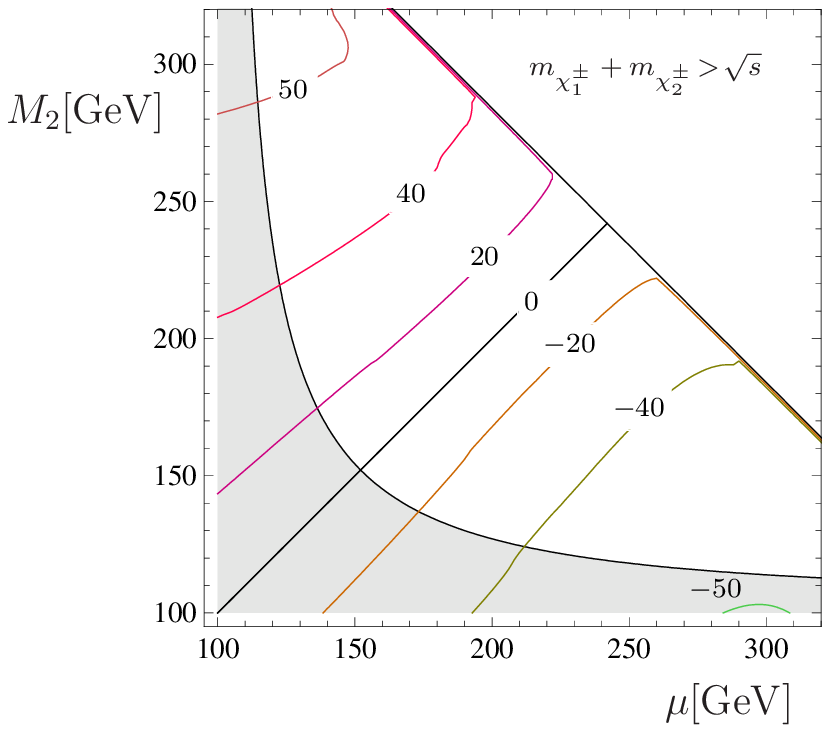}}
\put(1,0.15){(c)}
\put(9,0.15){(d)}
\put(0.7,14.9){averaged cross section $\sigma_{12}^{\Cp\Pp}$ in $\rm{fb}$ }
\put(9.8,14.9){   Asymmetry $\mathcal{A}^{\Cp\Pm}_{12}$ in \%}
\put(2.4, 6.9){Significance $\mathcal{S}^{\Cp\Pm}_{12}$}
\put(9.8, 6.9){   Asymmetry $\mathcal{A}^{\Cm\Pm}_{12}$ in \%}
\end{picture}
\caption{\small  
       Chargino production 
       $\mu^+\mu^-\to\tilde\chi_1^\pm\tilde\chi_2^\mp$
       at $\sqrt{s}=(M_{H_2}+M_{H_3})/2$ with 
       common muon beam polarization 
       $|\mathcal{P}|=0.3$.
       Contour lines in the $\mu$--$M_2$ plane for
       {\bf (a)}~the
       beam polarization averaged cross section 
       $\sigma_{12}^{\Cp\Pp}$, Eq.~(\ref{eq:sigmacepe}),
       {\bf (b)}~the CP-odd asymmetry
       $\mathcal{A}^{\Cp\Pm}_{12}$, Eq.~(\ref{eq:aprodCePo}),
       {\bf (c)}~the corresponding significance 
       $\mathcal{S}^{\Cp\Pm}_{12}$, Eq.~(\ref{eq:SaprodCePo}),
       with ${\mathcal L}=1~{\rm fb}^{-1}$,
       and {\bf (d)}~the CP-even asymmetry
       $\mathcal{A}^{\Cm\Pm}_{12}$, Eq.~(\ref{eq:aprodCoPo}).
       The SUSY parameters are given in Table~\ref{scenarioA}.
       The shaded area is excluded by $m_{\chi_1^\pm}<103$~GeV.
 }
\label{fig:prod12}
\end{figure}

For non-diagonal chargino production, 
$\mu^+\mu^-\to\tilde\chi_1^\pm\tilde\chi_2^\mp$,
we study the $\mu$--$M_2$ dependence of the cross section,
and of the CP-odd and CP-even production asymmetries
$\mathcal{A}^{\Cp\Pm}_{12}$, Eq.~(\ref{eq:aprodCePo}),
and 
$\mathcal{A}^{\Cm\Pm}_{12}$, Eq.~(\ref{eq:aprodCoPo}),
respectively.
The CP-odd asymmetry $\mathcal{A}^{\Cp\Pm}_{12}$
is largest at threshold, 
where it reaches up to $-10\%$, see Fig.~\ref{fig:prod12}(a).
Since the beam polarization averaged cross section 
reaches up to $400$~fb in that region,
we obtain a significance of $\mathcal{S}^{\Cp\Pm}_{12}<3$,
for an integrated luminosity of ${\mathcal L}=1~{\rm fb}^{-1}$,
see Fig.~\ref{fig:prod12}(c).
The CP-even production asymmetry exhibits a characteristic
$\mu\leftrightarrow M_2$ symmetry, see Fig.~\ref{fig:prod12}(d).
The Higgs chargino couplings transform as
$c_{L,R}^{H_k\chi_1\chi_2}\leftrightarrow c_{L,R}^{H_k\chi_2\chi_1}$
(\ref{eq.Hchi.eff.c}) under $\mu\leftrightarrow M_2$,
resulting in a sign change of $\mathcal{A}^{\Cm\Pm}_{12}$~\cite{Kittel:2005ma}.
Consequently, for $\mu= M_2$ the asymmetry 
vanishes, see the zero contour in 
Fig.~\ref{fig:prod12}(d).
On the other hand, the asymmetry is nearly maximal 
for $\mu\gg M_2$ or $\mu\ll M_2$, 
see the upper left and lower right corners
of Fig.~\ref{fig:prod12}(d).
The maximum absolute value would be
$\mathcal{A}^{\Cm\Pm}_{12\,\rm (max)}\approx 55\%$,
see Eq.~(\ref{eq:amax}) with  $|\mathcal{P}|=0.3$.
We do not show the CP-odd production asymmetry
$\mathcal{A}^{\Cm\Pp}_{12}$, Eq.~(\ref{eq:aprodCoPe}),
which  is only of order $1\%$ near threshold.

\medskip

\begin{figure}[t]
\centering
\begin{picture}(16,7.5)
\put(-2.15,-13.9){\includegraphics{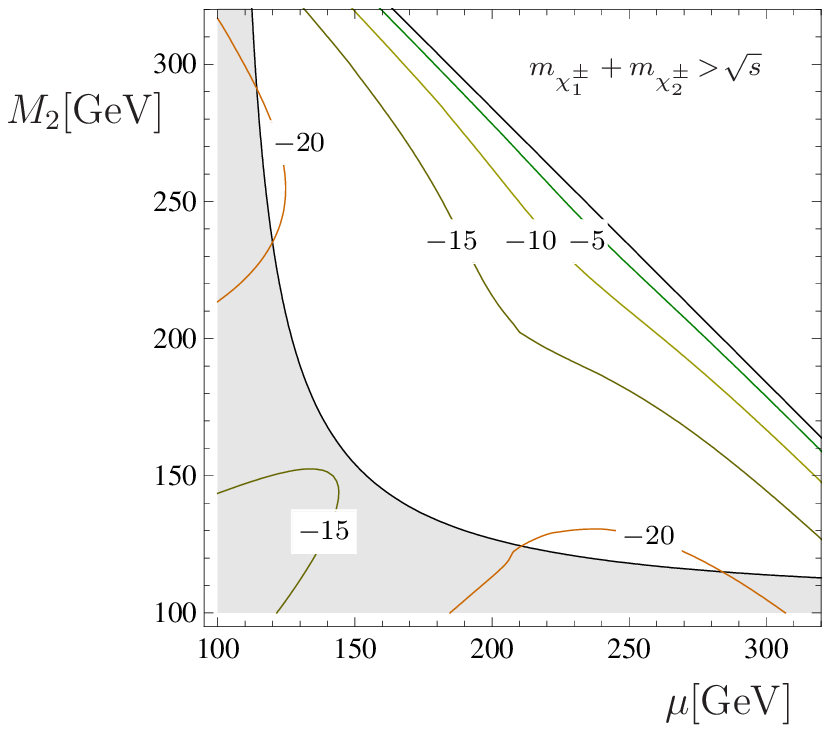}}
\put(6,-13.9){\includegraphics{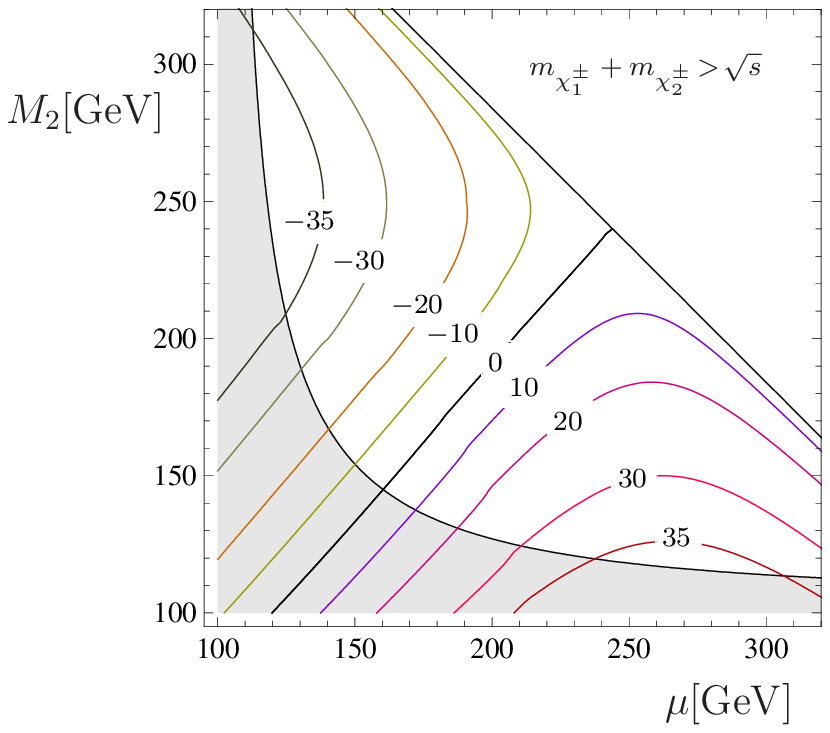}}
\put(1.0,0.15){(a)}
\put(9.0,0.15){(b)}
\put(1.7,6.9){Asymmetry $\mathcal{A}^{\Cp\Pp}_{12,e}$ in \%}
\put(9.8,6.9){Asymmetry $\mathcal{A}^{\Cm\Pm}_{12,e}$ in \%}
\end{picture}
\caption{\small 
        Chargino production 
        $\mu^+\mu^-\to\tilde\chi_1^\pm\tilde\chi_2^\mp$
        at $\sqrt{s}=(M_{H_2}+M_{H_3})/2$ with 
        common muon beam polarization 
        $|\mathcal{P}|=0.3$, and subsequent chargino 
        decay $\tilde\chi_2^\pm\to e^\pm\tilde \nu_e^{(\ast)}$.
        Contour lines in the $\mu$--$M_2$ plane for
        the CP-even decay asymmetries 
        {\bf (a)}~$\mathcal{A}^{\Cp\Pp}_{12,e}$, Eq.~(\ref{eq:adecCeoPe}), 
        and 
        {\bf (b)}~$\mathcal{A}^{\Cm\Pm}_{12,e}$, Eq.~(\ref{eq:adecCoPo}),
        for the SUSY parameters as given in Table~\ref{scenarioA}.
        The corresponding chargino production cross section 
        is shown in Fig.~\ref{fig:prod12}.
        The shaded area is excluded by $m_{\chi_1^\pm}<103$~GeV.
}
\label{fig:dec2.asym.e}
\end{figure}

For the subsequent chargino decay
$\tilde\chi_2^\pm\to e^\pm\tilde \nu_e^{(\ast)}$,
we show the CP-even decay asymmetries
$\mathcal{A}^{\Cp\Pp}_{12,e}$, Eq.~(\ref{eq:adecCeoPe}),
and
$\mathcal{A}^{\Cm\Pm}_{12,e}$, Eq.~(\ref{eq:adecCoPo}),
in Fig.~\ref{fig:dec2.asym.e}.
Both asymmetries are large in a wide region of the 
$\mu$--$M_2$ parameter space.
The asymmetry $\mathcal{A}^{\Cm\Pm}_{12,e}$ also
exhibits the characteristic
$\mu\leftrightarrow M_2$ symmetry,  see Fig.~\ref{fig:dec2.asym.e}(b),
as discussed above for the production asymmetry
$\mathcal{A}^{\Cm\Pm}_{12}$.
The decay asymmetry however has also contributions from
the continuum polarization
$\bar\Sigma^3_{\rm cont}$, see Eq.~(\ref{eq:adecCoPodep}).
Thus the zero contour in  Fig.~\ref{fig:dec2.asym.e}(b) gets slightly 
shifted from $\mu=M_2$. 
Since the continuum contributions are small compared to those from the resonance, 
the $\mu\leftrightarrow M_2$
symmetry of $\mathcal{A}^{\Cm\Pm}_{12,e}$, however, still holds approximately.

\medskip

The chargino $\tilde\chi_2^\pm$ two-body decays are open 
in the entire $\mu$--$M_2$ plane, 
with branching ratios up to 
BR$(\tilde\chi_2^+\to e^+\tilde \nu_e)= 10\%$.
Other competing channels are
$\tilde\chi_2^+\to W^+\tilde\chi_1^0$
and
$\tilde\chi_2^+\to\tau^+\tilde\nu_\tau $,
with branching ratios of also up to $10\%$.
For these decay channels the asymmetries
$\mathcal{A}^{\Cpm\Ppm}_{12,W}$
and
$\mathcal{A}^{\Cpm\Ppm}_{12,\tau}$
are accessible. Here, the 
appropriate decay factors $\eta_{W}$~(\ref{etaW}) and
$\eta_{\tau}$~(\ref{etatau}), respectively, have to
be taken into account.
The CP-odd decay asymmetries
$\mathcal{A}^{\Cp\Pm}_{12,e}$, Eq.~(\ref{eq:adecCePo}),
and
$\mathcal{A}^{\Cm\Pp}_{12,e}$, Eq.~(\ref{eq:adecCeoPe}),
do not exceed~$0.2\%$.

\medskip

\section{Summary and conclusions 
  \label{Summary and conclusions}}

In the MSSM with CP violation in the Higgs sector,
we have studied chargino production
at the muon collider around the resonances
of the two heavy neutral Higgs bosons.
For nearly degenerate neutral Higgs bosons,
as in the Higgs decoupling limit, the CP-violating Higgs mixing can be
resonantly enhanced,  which allows for large CP-violating effects.
We have defined the complete set of CP observables
to study the Higgs interference and their radiatively induced
Higgs mixings.
The muon collider with a precisely tunable beam energy
is an ideal tool to study the strong $\sqrt s$ dependence of the
CP asymmetries.

\medskip

Using longitudinally polarized muon beams,
we have classified the set of CP-even and CP-odd asymmetries
for chargino production, as well as for the chargino decays
into leptons. 
Appropriate observables can be defined for the decays into $W$ bosons.
In contrast to the production asymmetries, the asymmetries 
of the decay probe the longitudinal chargino polarization.
Due to spin correlations between production and decay,
the chargino polarization depends sensitively on the
Higgs interference.
Thus a different and independent set of asymmetries sensitive
to the Higgs couplings can be defined, that complement the 
production asymmetries.

\medskip
In a numerical study, we have parametrized CP violation in 
the Higgs sector by a common phase $\phi_A$
of the trilinear scalar coupling parameter.
We have analyzed the dependence of the asymmetries 
and cross sections on the complex parameter $A$, as well as on the
(real) gaugino parameters $\mu$ and $M_2$.
Large CP-violating asymmetries are obtained
if the $H$-$A$ Higgs mixing is resonantly enhanced.
We have discussed the conditions necessary for such a resonant
mixing. It occurs naturally in the Higgs decoupling limit,
when the Higgs mass difference is of the order of their widths.
We have identified the appropriate class of scenarios
which are adequate to study chargino production and decays.

\medskip

For $\tilde\chi_1^+\tilde\chi_1^-$ production,
the largest CP-odd production asymmetry can go up to $40\%$ 
for a degree of beam polarization of 
$\mathcal P = 0.3$.
The CP-odd decay asymmetry for the subsequent chargino decay
reaches up to  $10\%$.
By comparing the asymmetries
to the results which are known in the literature from non-diagonal
neutralino production,  we have
observed striking similarities. 
They are due to similar kinematics, i.e.,
production of spin-half charginos or neutralinos,
as well as to the dynamics, i.e., Higgs interference in the production
of a C-even final state of fermions.
We have completed the numerical study with the discussion of
non-diagonal chargino production.
Here, the C-odd and CP-even asymmetries are almost maximal
and reach more than $50\%$ for a degree of beam polarization of 
$\mathcal P = 0.3$.

\medskip

We find that chargino and neutralino pair production at a muon collider 
is complementary to similar studies for the production of Standard Model 
fermions.
While $t\bar t$ production is favored for small values of $\tan\beta$, 
third generation fermion production is largest for large $\tan\beta$.
The  strong effect  of the  CP phases on the Higgs mixing, and thus on their masses,
leads to a center-of-mass energy dependence of the observables
which can be ideally studied at the muon collider.

\medskip

We conclude that chargino production 
at the muon collider is the ideal testing ground to analyze and 
comprehend the phenomenology of MSSM Higgs mixing in the presence of CP-violating phases.
In particular, the crucial interference effects of the two nearly degenerated
heavy neutral Higgs resonances can well be analyzed and understood.
The proposed set of CP-sensitive observables
allows for a systematic test of the CP nature of the heavy 
neutral Higgs resonances, and their couplings to charginos.

\section{Acknowledgments}
We thank 
Sven Heinemeyer, Herbi Dreiner and Karina Williams
for helpful comments and discussions.
OK was supported by the SFB Transregio 33: The Dark Universe.
FP has been partially supported by the European
Community Marie-Curie Research Training Network under
contract MRTN-CT-2006-035505 ``Tools and Precision
Calculations for Physics Discoveries at Colliders''.
FP thanks the Grid Infrastructure of the EUFORIA project (FP7 Contract 211804).
OK thanks the Instituto de F\'isica de Cantabria (IFCA) for kind hospitality, 
where part of this work was completed.
%
%

\vspace*{1cm}

\begin{appendix}
\noindent{\Large\bf Appendix}
\setcounter{equation}{0}
\renewcommand{\thesubsection}{\Alph{section}.\arabic{subsection}}
\renewcommand{\theequation}{\Alph{section}.\arabic{equation}}

\setcounter{equation}{0}

\section{Density matrix formalism}
\label{Density matrix formalism}

We use the spin density matrix 
formalism~\cite{Haber94,MoortgatPick:1998sk,Dreiner:2007ay} 
for the calculation of the squared amplitudes for 
chargino production, Eq.~(\ref{production}), and 
decay channels, Eqs.~(\ref{decayell}) and~(\ref{decayW}).  
The amplitude for chargino production via resonant Higgs exchange, 
Eq.~(\ref{eq:tsquare}),
depends on the helicities $\lambda_\pm$ of the muons $\mu^\pm$ and 
the helicities $\lambda_i,\lambda_j$ of the produced charginos, 
$\mu^+\mu^-\rightarrow\tilde\chi^-_i\tilde\chi^+_j$,
\begin{eqnarray}
T^P_{\lambda_i\lambda_j\lambda_{+}\lambda_{-}} = 
\Delta(H_k)
\left[
\bar{v}(p_{\mu^+},\lambda_{+})\left(c_L^{H_k\mu\mu}P_L+c_R^{H_k\mu\mu}P_R \right)
u(p_{\mu^-},\lambda_{-})
\right]\phantom{.}\nonumber\\
\times
\left[
\bar u(p_{\chi^+_j},\lambda_j)\left(c_L^{H_k\chi_i\chi_j}P_L
                                   +c_R^{H_k\chi_i\chi_j}P_R \right)
v (p_{\chi^-_i},\lambda_i)
\right].
\label{THiggs}
\end{eqnarray}
We include the longitudinal beam polarizations of the muon-beams, ${\mathcal P}_{-}$ and 
${\mathcal P}_{+}$, with $ -1 \le {\mathcal P}_{\pm}\le +1$ in their density matrices
\begin{eqnarray}
\label{eq:density1}
\rho^{-}_{\lambda_{-} \lambda_{-}^\prime}  &=& 
     \frac{1}{2}\left(\delta_{\lambda_{-} \lambda_{-}^\prime} + 
      {\mathcal P}_{-}\tau^3_{\lambda_{-} \lambda_{-}^\prime}\right),\\[2mm]
\label{eq:density2}
\rho^{+}_{\lambda_{+} \lambda_{+}^\prime}  &=& 
     \frac{1}{2}\left(\delta_{\lambda_{+} \lambda_{+}^\prime} + 
   {\mathcal P}_{+}\tau^3_{\lambda_{+} \lambda_{+}^\prime}\right),
\end{eqnarray}
where  $\tau^3$ is the third Pauli matrix. 
The unnormalized spin density matrix 
of  $\charginominus_i\charginoplus_j$ production 
and $\tilde\chi_j^+$ decay are given by, 
\begin{eqnarray}
\rho^{P}_{\la_j\la^\prime_j} &=& 
\sum_{\la_i,\lambda_{+},\lambda_{+}^\prime,\lambda_{-} \lambda_{-}^\prime} 
\rho^{+}_{\lambda_{+} \lambda_{+}^\prime}
\rho^{-}_{\lambda_{-} \lambda_{-}^\prime}
{T}^{P}_{\la_i\la_j\lambda_{+}\lambda_{-} }
{T}^{P*}_{\la_i\la^\prime_j\lambda_{+}^\prime \lambda_{-}^\prime},
\label{rhop}
\\
        \rho^{D}_{\la^\prime_j\la_j} &=&                
                {{T}}^{D\ast}_{\la^\prime_j}{{T}}^{D}_{\la_j}.
\label{rhod}
\end{eqnarray} 
The amplitude squared for production and decay is then
\begin{eqnarray}
        |{{T}}|^2 &=& |\Delta(\tilde\chi_j^+)|^2 \sum_{\la_j \la_j^\prime} 
        \rho^P_{\la_j \la_j^\prime}\rho^D_{\la_j^\prime\la_j},
\label{tsquare}
\end{eqnarray}
with the chargino propagator
\begin{eqnarray}
\Delta(\tilde\chi_j^+)=\frac{i}{p^2_{\chi_j^+}-m_{\chi_j^+}^2+
        i m_{\chi_j^+}\Gamma_{\chi_j^+}}.
\label{eq:charginopropagator}
\end{eqnarray}
The spin density matrices, Eqs.~(\ref{rhop}) and~(\ref{rhod}),
can be expanded in terms of the Pauli matrices $\tau^a$
\begin{eqnarray}
\rho^P_{\la_j \la_j^\prime} &=& 
\delta_{\la_j \la_j^{\prime}}P + \sum_{a=1}^3 \tau_{\la_j \la_j^{\prime}}^a  {\Sigma}_P^{a},
\label{rhoP}
\\
\rho^D_{\la_j^\prime \la_j} &=& 
\delta_{\la_j^\prime \la_j}D + \sum_{a=1}^3 \tau_{\la_j^{\prime} \la_j}^a  {\Sigma}_D^{a},
\label{rhoD}
\end{eqnarray}
where we have defined a set of chargino spin vectors $s_{\chi_j^+}^a$.
In the center-of-mass system, they are 
\begin{equation}
        s_{\chi^+_j}^{1,\,\mu}= (0;1,0,0), \qquad       s_{\chi^+_j}^{2,\,\mu}= (0;0,1,0), 
        \qquad  s_{\chi^+_j}^{3,\,\mu}= 
        \frac{1}{m_{\chi^+_j}}(|\vec{p}_{\chi^+_j}|;0,0,E_{\chi^+_j}).
        \label{spinvec}
\end{equation}
We have chosen a coordinate frame 
such that the momentum of the chargino $\tilde\chi^+_j$ is
\begin{equation}
        p_{\chi^+_j}^{\,\mu}=(E_{\chi^+_j};0,0,|\vec{p}_{\chi^+_j}|),
\end{equation}
with
\begin{equation}
        E_{\chi_j^+} =\frac{s+m_{\chi^+_j}^2-m_{\chi^+_i}^2}{2 \sqrt{s}},\quad
    |\vec{p}_{\chi^+_j}|   =\frac{\sqrt{\lambda_{ij}}}{2 \sqrt{s}},
\end{equation}
and the triangle function
\begin{equation}
\lambda_{ij}= \lambda(s,m_{\chi^+_i}^2,m_{\chi^+_j}^2),
\label{triangle}
\end{equation}
with $\lambda(x,y,z)= x^2+y^2+z^2-2(xy+xz+yz)$.
Inserting the density matrices, Eqs.~(\ref{rhoP}) and~(\ref{rhoD}), 
into Eq.~(\ref{tsquare}), gives the amplitude squared
in the form of Eq.~(\ref{eq:tsquare}).

\section{Higgs exchange contributions}
\label{ab_coefficients}

We have expressed the resonant contributions to the 
spin-density matrix coefficients
as functions of the longitudinal $\mu^+$ and 
$\mu^-$ beam polarizations,
Eqs.~(\ref{eq.Pr}) and (\ref{eq.Sr}),
\begin{eqnarray}
P_{\rm res}&=&
        (1 + {\mathcal{P}}_+{\mathcal{P}}_-)a_0  + 
        ({\mathcal{P}}_+ + {\mathcal{P}}_-)a_1, 
\nonumber
\\
\Sigma_{\rm res}^3&=&
         (1 + {\mathcal{P}}_+{\mathcal{P}}_-)b_0 + 
         ({\mathcal{P}}_+ + {\mathcal{P}}_-)b_1.
\nonumber
\end{eqnarray}
The coefficients $a_n$ and $b_n$ on the r.h.s.\ are
\begin{eqnarray}
a_n  = \sum_{l,k\le l} (2-\delta_{kl}) a_n^{kl},\quad
b_n  = \sum_{l,k\le l} (2-\delta_{kl}) b_n^{kl},
\quad n=0,1; 
\label{eq.anbn}
\end{eqnarray}
with the sum over the contributions from the
Higgs bosons $H_{k}$, $H_{l}$ with $ k,l = 1,2,3$, 
respectively, and~\cite{CPNSH}
\begin{eqnarray}
a_0^{kl}~&=&
        \phantom{+}\!   \frac{s}{2}|\Delta_{(kl)}|
\Big[
                |c_\mu^+ | |c^+_\chi |
f_{ij}
        \cos(
                {\delta_\mu^+ }
                +
                {\delta_\chi^+ }
                +
                {\delta_\Delta }
        )
\nonumber\\ & & 
\phantom{
\!   \frac{s}{2}|\Delta_{(kl)}|
}
                 -|c_\mu^+ | |c_\chi^{RL} |
                m_i m_j
        \cos(
                {{\delta_\mu^+ }}
                +
                {{\delta_\chi^{RL} }}
                +
                {{\delta_\Delta }}
        )
\Big]_{(kl)},
\label{eq.a0}
\\
a_1^{kl}~&=& \!
        \phantom{+}\!   \frac{s}{2} |\Delta_{(kl)}|
\Big[
                |c_\mu^- | |c^+_\chi |
f_{ij}
        \cos(
                {{\delta_\mu^- }}
                +
                {{\delta_\chi^+ }}
                +
                {{\delta_\Delta }}
        )
\nonumber\\ & & 
\phantom{
\!   \frac{s}{2}|\Delta_{(kl)}|
}                -|c_\mu^- | |c_\chi^{RL} |
                m_i m_j
        \cos(
                {{\delta_\mu^- }}
                +
                {{\delta_\chi^{RL} }}
                +
                {{\delta_\Delta }}
        )
\Big]_{(kl)},
\label{eq.a1}
\\
b_0^{kl}~&=&\!
        -\frac{s}{4} |\Delta_{(kl)}|
\Big[
                |c_\mu^+ | |c^-_\chi |
\sqrt{\lambda_{ij}}
        \cos(
                {{\delta_\mu^+ }}
                +
                {{\delta_\chi^- }}
                +
                {{\delta_\Delta }}
        )
\Big]_{(kl)},
\label{eq.b0}
\\
b_1^{kl}~&=&\!
-
\frac{s}{4} |\Delta_{(kl)}|
\Big[
                |c_\mu^- | |c^-_\chi |
\sqrt{\lambda_{ij}}
        \cos(
                {{\delta_\mu^- }}
                +
                {{\delta_\chi^- }}
                +       
                {{\delta_\Delta }}
        )
\Big]_{(kl)}.
\label{eq.b1}
\end{eqnarray}
We have defined the products of couplings, suppressing the 
chargino indices~$i$ and~$j$,
\begin{eqnarray}
c_{\alpha(kl)}^\pm &=&  
c_R^{H_k\alpha\alpha} c_R^{H_l\alpha\alpha\ast}\pm
c_L^{H_k\alpha\alpha} c_L^{H_l\alpha\alpha\ast}
=
\Big[
        | c_\alpha^\pm | \exp({i {\delta_\alpha^\pm }})    
\Big]_{(kl)}  
,\quad \alpha=\mu, \chi,
\label{eq.clambda.mp}
\\
c_{\chi(kl)}^{RL} &=&  
c_R^{H_k\chi\chi} c_L^{H_l\chi\chi\ast}+
c_L^{H_k\chi\chi} c_R^{H_l\chi\chi\ast}
=
\Big[
|c_\chi^{RL}| \exp({i {\delta_\chi^{RL} }})
\Big]_{(kl)}  
,
\label{eq.cchi.RL}
\end{eqnarray}
the product of the Higgs boson propagators~(\ref{propHiggs}),
\begin{eqnarray}
      \Delta_{(kl)}&=&\Delta(H_k)\Delta(H_l)^\ast
=
\Big[
|\Delta|\exp({i {\delta_\Delta }})
\Big]_{(kl)}  
,\label{deltadelta}
\end{eqnarray}
and the kinematical functions 
$f_{ij}=(s-m_{\chi_i^-}^2- m_{\chi_j^+}^2)/2$, and 
${\lambda_{ij}}$, see Eq.~(\ref{triangle}).
We neglect interferences of the chirality violating Higgs
exchange amplitudes with the chirality conserving continuum amplitudes,
which are of order $m_\mu/\sqrt{s}$.
The contributions from $H_1$ exchange are small
far from its resonance.

Note that the longitudinal chargino polarization,
parametrized by the coefficients $b_0$ and $b_1$ of $\Sigma_{\rm res}^3$,
see Eqs.~(\ref{eq.b0}) and (\ref{eq.b1}), %
vanishes at threshold as $\sqrt{\lambda_{ij}}$. 
We have defined $b_0$ and $b_1$
for the longitudinal polarization of
the positively charged chargino $\tilde\chi^+_j$ in the reaction
$\mu^+\mu^-\rightarrow\tilde\chi^-_i\tilde\chi^+_j$.
The coefficients are the same for the other produced chargino
$\tilde\chi^-_i$, due to angular momentum conservation.

%
\section{Continuum contributions}
\label{nonresamplitudes}
%

The non-resonant 
chargino production~(\ref{production}) proceeds via $\gamma$ and $Z$ boson 
exchange in the $s$-channel, and muon-sneutrino $\tilde\nu_\mu$ exchange in
the $t$-channel, see the Feynman diagrams in Fig.~\ref{Fig:FeynProd}.
The MSSM interaction Lagrangians are~\cite{HK,MoortgatPick:1998sk}: 
\begin{eqnarray}
{\scr L}_{Z^0 \mu \bar \mu} &=&-\frac{g}{\cos\theta_W}
Z^0_{\nu}\bar \mu\gamma^{\nu}[L_\mu P_L+ R_\mu P_R]\mu,
\label{Zelel}\\[2mm]
{\scr L}_{\gamma \tilde\chi^+ \tilde\chi^-} &=&
- e A_{\nu} \bar{\tilde\chi}^+_i \gamma^{\nu} \tilde\chi_j^+
\delta_{ij},\quad e>0, \\[2mm]
{\scr L}_{Z^0\tilde\chi^+\tilde\chi^-} &=&
 \frac{g}{\cos\theta_W}Z_{\mu}\bar{\tilde\chi}^+_i\gamma^{\mu}
[O_{ij}^{'L} P_L+O_{ij}^{'R} P_R]\tilde\chi_j^+,\\[2mm]
{\scr L}_{\ell \tilde\nu\tilde\chi^+} &=&
- g V_{i1}^{*} \bar{\tilde\chi}_i^{+C} P_L \ell 
 \tilde{\nu}^{*}+\mbox{h.c.},\quad \ell=e,\mu,
\end{eqnarray}
with $i,j=1,2$ %
and $P_{L, R}=(1\mp \gamma_5)/2$.
The couplings are~\cite{HK,MoortgatPick:1998sk} 
\begin{eqnarray}
L_\mu&=&-\frac{1}{2}+\sin^2\theta_W, \quad
  R_\mu=\sin^2\theta_W \label{eq_5},\\[2mm]
 O_{ij}^{'L}&=&-V_{i1} V_{j1}^{*}-\frac{1}{2} V_{i2} V_{j2}^{*}+
\delta_{ij} \sin^2\theta_W,\\[2mm]
 O_{ij}^{'R}&=&-U_{i1}^{*} U_{j1}-\frac{1}{2} U_{i2}^{*} U_{j2}+
\delta_{ij} \sin^2\theta_W,
\end{eqnarray}
with the weak mixing angle $\theta_W$,
and the weak coupling constant $g=e/\sin\theta_W$,  $e>0$.
The complex unitary $2\times 2$ matrices
$U_{mn}$ and $V_{mn}$ diagonalize the chargino mass 
matrix $X_{\alpha\beta}$, $U_{m \alpha}^* X_{\alpha\beta}V_{\beta
n}^{-1}= m_{\chi^\pm_n}\delta_{mn}$, with $ m_{\chi^\pm_n}>0$.

\medskip

The helicity amplitudes for 
$\gamma$, $Z$ and $\tilde\nu_\mu$ exchange are
\begin{eqnarray}
T^P_{\lambda_i\lambda_j\lambda_{+}\lambda_{-}}(\gamma)&=&
-e^2 \Delta(\gamma)\delta_{ij} 
\left[\bar{v}(p_{\mu^+},\lambda_{+}) \gamma^{\nu} u(p_{\mu^-},\lambda_{-})\right]
         \nonumber\\& & \times
\left[ \bar{u}(p_{\chi^+_j}, \lambda_j) \gamma_{\nu}
            v (p_{\chi^-_i}, \lambda_i)\right],\label{char:T1}\\
T^P_{\lambda_i\lambda_j\lambda_{+}\lambda_{-}}(Z)&=&
- \frac{g^2}{\cos^2\theta_W}\Delta(Z)
  \left[\bar{v}(p_{\mu^+},\lambda_{+}) \gamma^{\nu} 
                  (L_\mu P_L+R_\mu P_R) u(p_{\mu^-},\lambda_{-}) \right]
                \nonumber\\ & & \times
\left[\bar{u}(p_{\chi^+_j}, \lambda_j) \gamma_{\nu} 
                (O^{'L}_{ji} P_L +O^{'R}_{ji} P_R)
                v(p_{\chi^-_i}, \lambda_i) \right], \label {char:T2}\\
T^P_{\lambda_i\lambda_j\lambda_{+}\lambda_{-}}(\tilde{\nu})&=&
-g^2 V_{j1} V_{i1}^{*}\Delta(\tilde{\nu})
        \left[\bar{v}(p_{\mu^+},\lambda_{+}) P_R v(p_{\chi^+_j}, \lambda_j) \right]
                \nonumber\\ & & \times
        \left[\bar{u}(p_{\chi^-_i}, \lambda_i) 
                P_L u(p_{\mu^-},\lambda_{-})\right],\label{char:T3}
        \end{eqnarray}
with the propagators
\begin{equation}\label{char:propagators}
        \Delta(\gamma) = \frac{i}{s },\quad
        \Delta(Z)      = \frac{i}{s-m^2_Z},\quad
        \Delta(\tilde \nu)  = \frac{i}{t-m^2_{\tilde \nu}},
\end{equation}
and  $s=(p_{\mu^+}+p_{\mu^-})^2$,
$t=(p_{\mu^+}-p_{\chi^+_j})^2$.
We neglect the $Z$-width in the propagator
$\Delta(Z)$ for energies beyond the resonance.
The Feynman diagrams are shown in 
Fig.~\ref{Fig:FeynProd}.
For $e^+e^-$ collisions, the amplitudes are given 
in Ref.~\cite{MoortgatPick:1998sk,gudidiss}.


\medskip

The continuum contributions $P_{\rm cont}$ are 
those from the non-resonant $\gamma$, $Z$ and $\tilde\nu_\mu$ 
exchange channels. 
The coefficient $P_{\rm cont}$ is independent of the chargino polarization. 
It can be decomposed into contributions from the 
different continuum channels
\begin{equation}
P_{\rm cont} =  
     P(\gamma \gamma)
   + P(\gamma Z) 
   + P(\gamma \tilde \nu)
   + P(Z Z)
   + P(Z\tilde \nu)
   + P(\tilde \nu\tilde \nu)
\end{equation}
with
\begin{eqnarray}
P(\gamma \gamma)&=&\delta_{ij}4e^4 |\Delta (\gamma)|^2
        (c_{L} + c_{R})
        E_b^2(E_{\chi_j^+} E_{\chi_i^-}+m_{\chi_j^+} m_{\chi_i^-}+
        q^2\cos^2\theta),\\
P(\gamma Z)&=&\delta_{ij}4\frac{e^2g^2}{\cos^2 \theta_W}E_b^2
        \Delta (\gamma)\Delta(Z)^{\ast}
        {\rm Re}\Big\{  
        (L_\mu c_{L} - R_\mu c_{R})
        (O^{'R\ast}_{ij} - O^{'L\ast}_{ij})
        2E_bq\cos\theta \,\quad
                \nonumber\\& & 
        +(L_\mu c_{L} + R_\mu c_{R})
        (O^{'L\ast}_{ij}+O^{'R\ast}_{ij})
        (E_{\chi_j^+} E_{\chi_i^-}+m_{\chi_j^+} m_{\chi_i^-}+q^2\cos^2\theta)
        \Big\},\\
P(\gamma \tilde \nu)&=&2\delta_{ij}e^2g^2E_b^2
        c_{L}\Delta (\gamma)\Delta (\tilde \nu)^{\ast}
        {\rm Re}\Big\{ V^{\ast}_{i1}V_{j1}\Big\}
         \nonumber\\& & \times
        (E_{\chi_j^+} E_{\chi_i^-}+m_{\chi_j^+} m_{\chi_i^-}-2E_bq\cos\theta
        +q^2\cos^2\theta),\\
P(Z Z)&=&\frac{2g^4}{\cos^4\theta_W}|\Delta (Z)|^2E_b^2\Big[
        (L_\mu^2 c_{L} - R_\mu^2  c_{R})
        (|O^{'R}_{ij}|^2-|O^{'L}_{ij}|^2)2E_bq\cos\theta
        \nonumber\\& &
        +(L_\mu^2c_{L} + R_\mu^2c_{R})
        (|O^{'L}_{ij}|^2+|O^{'R}_{ij}|^2)
        (E_{\chi_j^+} E_{\chi_i^-}+ q^2\cos^2\theta)
        \nonumber\\& &
        +(L_\mu^2 c_{L} + R_\mu^2 c_{R})
        2{\rm Re}\{O^{'L}_{ij}O^{'R\ast}_{ij}\}m_{\chi_j^+} m_{\chi_i^-}
        \Big], \\
P(Z\tilde \nu)&=&\frac{2g^4}{\cos^2\theta_W}
        L_\mu c_{L}E_b^2 \Delta (Z)\Delta (\tilde \nu)^{\ast}{\rm Re}\Big\{
        V^{\ast}_{i1}V_{j1} 
        \nonumber\\& & \times
        [O^{'L}_{ij}(E_{\chi_j^+} E_{\chi_i^-}-2E_bq\cos\theta+q^2\cos^2\theta) 
                +O^{'R}_{ij}m_{\chi_j^+} m_{\chi_i^-}]
        \Big\},\\
P(\tilde \nu\tilde \nu)&=& \frac{g^4}{2}c_{L}
|V_{i1}|^2|V_{j1}|^2 |\Delta (\tilde \nu)|^2
        E_b^2(E_{\chi_j^+} E_{\chi_i^-}-2E_bq\cos\theta+q^2\cos^2\theta ),
\end{eqnarray}
with the scattering angle $\theta \angle (\vec p_{\mu^-},p_{\tilde\chi_i^-})$. 
The longitudinal beam polarizations are included in the weighting factors  
\begin{equation}
c_L =(1-{\mathcal P}_-)(1+{\mathcal P}_+), \quad 
c_R= (1+{\mathcal P}_-)(1-{\mathcal P}_+).
\label{eq:cLcR}
\end{equation}
For equal beam polarizations 
${\mathcal P}_+ = {\mathcal P}_- \equiv {\mathcal P}$,
we thus have
$P_{\rm cont}({\mathcal P})=P_{\rm cont}(-{\mathcal P})$.
For $e^+e^-$ collisions, the terms for $P_{\rm cont}$ in the laboratory
system are also given 
in Refs.~\cite{gudidiss,Kittel:2005rp}, however they differ by
a factor of $2$ in our notation~(\ref{eq:tsquare}).
Note that the term $P_{\rm cont}$ is invariant under $i\leftrightarrow j$
exchange at tree level if CP is conserved.

\medskip

The continuum contributions 
$\Sigma_{\rm cont}^3$~(\ref{contributions}) to the
longitudinal $\tilde{\chi}^+_j$ polarization 
for chargino production 
$\mu^+\mu^-\rightarrow\tilde\chi^-_i\tilde\chi^+_j$
decompose into
   \begin{equation}
     \Sigma_{\rm cont}^3 =
     \Sigma_P^3(\gamma \gamma)
   + \Sigma_P^3(\gamma Z)
   + \Sigma_P^3(\gamma \tilde \nu)
   + \Sigma_P^3(Z Z)
   + \Sigma_P^3(Z\tilde \nu)
   + \Sigma_P^3(\tilde \nu\tilde \nu),\label{charg:sigmaP}
\end{equation}
with
\begin{eqnarray}
\Sigma_P^3(\gamma \gamma)&=&\delta_{ij}4e^4 |\Delta (\gamma)|^2
        (c_{L} - c_{R})
        E_b^2\cos\theta (q^2+  E_{\chi_j^+} E_{\chi_i^-}+m_{\chi_j^+} m_{\chi_i^-}),
       \label{eq:gammagamma}\\
\Sigma_P^3(\gamma Z)&=&\delta_{ij}4\frac{e^2g^2}{\cos^2 \theta_W}E_b^2
        \Delta (\gamma)\Delta (Z)^{\ast}
        \nonumber\\& &\times  
        {\rm Re}\Big\{(L_\mu c_{L} - R_\mu c_{R})
        (O^{'R\ast}_{ji} + O^{'L\ast}_{ji})
        (q^2+  E_{\chi_j^+} E_{\chi_i^-}+m_{\chi_j^+} m_{\chi_i^-})\cos\theta
                \nonumber\\& & 
        +(L_\mu c_{L} + R_\mu c_{R})
        (O^{'R\ast}_{ji}-O^{'L\ast}_{ji})
        q(E_{\chi_i^-}+E_{\chi_j^+}\cos^2\theta)
        \Big\},\\
\Sigma_P^3(\gamma \tilde \nu)&=&-\delta_{ij}2e^2g^2 c_{L}E_b^2
        \Delta (\gamma)\Delta (\tilde \nu)^{\ast}
         {\rm Re}\Big\{ V^{\ast}_{j1}V_{i1}
        \Big\} \nonumber\\& & \times
        [qE_{\chi_i^-} - (q^2+E_{\chi_j^+} E_{\chi_i^-})\cos\theta
        +qE_{\chi_j^+}\cos^2\theta- m_{\chi_j^+}  m_{\chi_i^-}\cos\theta],\\
\Sigma_P^3(Z Z)&=&\frac{2g^4}{\cos^4\theta_W}|\Delta (Z)|^2E_b^2\Big[
        (L_\mu^2 c_{L} + R_\mu^2 c_{R})
        (|O^{'R}_{ji}|^2-|O^{'L}_{ji}|^2)q(E_{\chi_i^-}+E_{\chi_j^+}\cos^2\theta)
        \nonumber\\& &
        +(L_\mu^2 c_{L} - R_\mu^2 c_{R})
        2{\rm Re}\Big\{O^{'L}_{ji}O^{'R\ast}_{ji}\Big\}
        m_{\chi_j^+} m_{\chi_i^-}\cos\theta
        \nonumber\\& &
        +(L_\mu^2 c_{L} - R_\mu^2c_{R})
        (|O^{'L}_{ji}|^2+|O^{'R}_{ji}|^2)
        (q^2+E_{\chi_j^+} E_{\chi_i^-})\cos\theta\Big], \\
\Sigma_P^3(Z\tilde \nu)&=&\frac{2g^4}{\cos^2\theta_W} L_\mu c_{L} E_b^2
        \Delta (Z)\Delta (\tilde \nu)^{\ast}
        {\rm Re}\Big\{V^{\ast}_{j1}V_{i1}
        [O^{'R}_{ji}m_{\chi_j^+} m_{\chi_i^-}\cos\theta
        \nonumber\\& & 
        -O^{'L}_{ji}(q E_{\chi_i^-}-(q^2+E_{\chi_j^+} E_{\chi_i^-})\cos\theta
        +qE_{\chi_j^+}\cos^2\theta)] 
        \Big\},\\
\Sigma_P^3(\tilde \nu\tilde \nu)&=& -\frac{g^4}{2}c_{L}
        |V_{j1}|^2|V_{i1}|^2 |\Delta (\tilde \nu)|^2 E_b^2%
        \nonumber\\& & \times
        [q E_{\chi_i^-}-(q^2+E_{\chi_j^+} E_{\chi_i^-})\cos\theta
        +qE_{\chi_j^+}\cos^2\theta]. \label{eq:snuetsneut}
\end{eqnarray}
For $e^+e^-$ collisions, the terms for $\Sigma_{\rm cont}^3$ 
for chargino $\tilde{\chi}^+_j$ in the laboratory
system are also given 
in Ref.~\cite{gudidiss,Kittel:2005rp}, however they differ by
a factor of $2$ in our notation~(\ref{eq:tsquare}).
In order to obtain the corresponding polarization terms for
the other chargino $\tilde{\chi}^-_i$, one has to substitute
$E_{\chi_j^+} \leftrightarrow E_{\chi_i^-}$ and
$m_{\chi_j^+} \leftrightarrow m_{\chi_i^-}$
in Eqs.~(\ref{eq:gammagamma})-(\ref{eq:snuetsneut});
in addition, one has to change the overall sign~\cite{gudidiss}.
Note that for the charge conjugated process
$\mu^+\mu^-\rightarrow\tilde\chi^+_i\tilde\chi^-_j$,
the averaged chargino $\tilde\chi^\pm_j$ polarization 
of the continuum, see Eq.~(\ref{eq:sigmabar}),
changes sign, i.e.,
$   \bar\Sigma_{\rm cont}^3(\tilde\chi^+_i\tilde\chi^-_j) 
= - \bar\Sigma_{\rm cont}^3(\tilde\chi^-_i\tilde\chi^+_j)$.
For equal beam polarizations 
${\mathcal P}_+ = {\mathcal P}_- \equiv {\mathcal P}$,
the terms obey
$\Sigma_{\rm cont}^3(-{\mathcal P})=\Sigma_{\rm cont}^3({\mathcal P})$,
see Eq.~(\ref{eq:cLcR}).

\section{Chargino decay into leptons and $W$ boson}
\label{chardecay}

The expansion coefficients of the chargino decay matrix~(\ref{rhoD}) 
for the chargino decay  $\tilde\chi_j^{+} \to \ell^{+}  \tilde\nu_\ell$, 
with  $\ell = e,\mu$, are
   \begin{eqnarray}
                D & = & \frac{g^2}{2} |V_{j1}|^2 
                (m_{\chi_j^{\pm}}^2 -m_{\tilde{\nu}_\ell}^2 ),
   \label{DR}   \\
      \Sigma^{a}_{D} &=&    
                 -g^2 |V_{j1}|^2 m_{\chi_j^{\pm}} (s^a_{\chi_j^{\pm}}
                        \cdot p_{\ell}).
   \label{SigmaD}
\end{eqnarray}
The coefficient $\Sigma^{a}_{D}$ for the charge conjugated process, 
$\tilde\chi_j^- \to \ell^-  \tilde\nu_\ell^\ast$, 
is obtained by inverting the sign of Eq.~(\ref{SigmaD}).
With these definitions, we can rewrite the factor 
$\Sigma^{3}_{D}$, that multiplies the 
longitudinal chargino polarization $\Sigma^{3}_{P}$ in Eq.~(\ref{eq:tsquare}), 
\begin{eqnarray}
        \Sigma^{3}_{D} &=& \eta_{\ell\pm} \frac{D}{\Delta_\ell}
        (E_\ell-\hat{E}_\ell),
\label{etal}
 \end{eqnarray}
where we have used 
\begin{eqnarray}
        m_{\chi_j^\pm} (s^3_{\chi_j^\pm}\cdot p_\ell)=
        -\frac{m_{\chi_j^\pm}^2}{|\vec{p}_{\chi^\pm_j} |}
        (E_\ell-\hat{E}_\ell).
\end{eqnarray}
Similar results can be obtained for the chargino
decay into a $\tau $, 
$\tilde\chi_j^{\pm} \to \tau^{\pm}  \tilde\nu_\tau^{(\ast)}$, and $W$
boson, $\tilde\chi_j^{\pm} \to W^{\pm}\tilde\chi_k^0$~\cite{Kittel:2005ma}. 
The corresponding interaction Lagrangians are~\cite{HK} 
\begin{eqnarray}
        {\scr L}_{\tau \sneutrino_\tau \tilde\chi^+} & = & 
                - g \bar{\tau}( V_{j1}  P_R -Y_{\tau}U_{j2}^*
                        P_L)\tilde{\chi}_j^{+C} \sneutrino_\tau + \mbox{h.c.},\\[2mm]
        {\scr L}_{W^-\tilde\chi^+\neutralino} & = & 
                g W^-_\mu  \bar{\tilde{\chi}}_k^{0} \gamma^\mu
                        (O_{kj}^L P_L+O_{kj}^R P_R) 
                \tilde{\chi}_j^{+} \sneutrino_\ell
        +       \mbox{h.c.},
\end{eqnarray}
respectively, with the couplings 
\begin{eqnarray}
        O_{kj}^L        &=&     -\frac{1}{\sqrt{2}}     N_{k4}V_{j2}^\ast
                                +                       (\sin\theta_W N_{k1}+\cos\theta_W N_{k2})       V_{j1}^\ast,
\\
        O_{kj}^R        &=&     +\frac{1}{\sqrt{2}}     N_{k3}^\ast U_{j2}
                                +                       (\sin\theta_W N_{k1}^\ast+\cos\theta_W
                                        N_{k2}^\ast)    U_{j1},
\end{eqnarray}
and $Y_{\tau}= m_{\tau}/(\sqrt{2}m_W\cos\beta)$.
The $4\times 4$ unitary matrix $N$ diagonalizes the neutralino mass
matrix $Y$ in the basis 
$\{ \tilde{\gamma},\tilde{Z},\tilde{h}_1,\tilde{h}_2 \}$ with 
$N_{il}^\ast Y_{lm}N_{mj}^\dagger = \delta_{ij} m_{\chi^0_j}$~\cite{HK}.

The expansion coefficients of the chargino decay matrix~(\ref{rhoD}) 
for $\tilde\chi_j^+ \to \tau^+  \tilde\nu_\tau$ are 
   \begin{eqnarray}
                D & = & \frac{g^2}{2} (|V_{j1}|^2 +Y_{\tau}^2|U_{j2}|^2)
                (m_{\chi_j^{\pm}}^2 -m_{\tilde{\nu}_\tau}^2 ),
   \label{DRtau} \\
      \Sigma^a_{D} &=&    
                 -g^2  (|V_{j1}|^2 -Y_{\tau}^2|U_{j2}|^2) 
                m_{\chi_j^{\pm}}(s^a_{\chi_j^{\pm}} \cdot p_{\tau}), 
                   \label{SigmaDtau}
\end{eqnarray}
and those for 
$\tilde\chi_j^+ \to W^+\tilde\chi_k^0$ are  
   \begin{eqnarray}
      D & = & \frac{g^2}{2}(|O^L_{kj}|^2+|O^R_{kj}|^2)
        \left[
                m_{\chi_j^{\pm}}^2+m_{\chi_k^0}^2-2m_W^2+\frac{(m_{\chi_j^{\pm}}^2-m_{\chi_k^0}^2)^2}{m_W^2}
        \right]
\nonumber\\ & &
        -6 g^2 {\rm Re}(O^L_{kj} O^{R\ast}_{kj})  m_{\chi_j^{\pm}} m_{\chi_k^0},
\label{DW}
\\[1mm]
      \Sigma^a_{D} &=& -g^2 (|O^L_{kj}|^2-|O^R_{kj}|^2)
\,
        \frac{  (m_{\chi_j^{\pm}}^2-m_{\chi_k^0}^2- 2 m_W^2)}{m_W^2}  
        \, m_{\chi_j^{\pm}} (s^a_{\chi_j^{\pm}}\cdot p_W).      
\label{SigmaDW}
   \end{eqnarray}
The coefficients $\Sigma^a_{D}$ for the charge conjugated processes, 
$\tilde\chi_j^- \to \tau^-  \tilde\nu_\tau^\ast$ and
$\tilde\chi_j^- \to W^-\tilde\chi_k^0$,
are obtained by inverting the signs of Eqs.~(\ref{SigmaDtau})
and~(\ref{SigmaDW}), respectively.

For the chargino decay $\tilde\chi_j^\pm \to W^\pm\tilde\chi_k^0$, 
the energy limits of the $W$ boson are 
$E_W^{{\rm max} ({\rm min})}= \hat{E}_W \pm \Delta_W$, see Eq.~(\ref{kinlimits}),
with
\begin{eqnarray}
        \hat{E}_W &=& \frac{E_W^{\rm max}+E_W^{\rm min}}{2} = 
        \frac{ m_{\chi^{\pm}_j}^2+m_W^2-m_{\chi_k^0}^2}{2 m_{\chi^{\pm}_j}^2} E_{\chi^{\pm}_j},
                \label{ehalfW}
\\
        \Delta_W &=& \frac{E_W^{\rm max}-E_W^{\rm min}}{2} = \frac{ \sqrt{
        \lambda(m_{\chi^{\pm}_j}^2,m_W^2,m_{\chi_k^0}^2)
        }}{2 m_{\chi^{\pm}_j}^2} |\vec{p}_{\chi^{\pm}_j}|.
\label{edifW}
\end{eqnarray}
The decay factor~(\ref{edist2}) is
\begin{eqnarray}
        \eta_{W^{\pm}}&=&   \pm   \frac{(|O^L_{kj}|^2-|O^R_{kj}|^2) f_1}
                                {(|O^L_{kj}|^2+|O^R_{kj}|^2) f_2  
                                        + {\rm Re}\{O^L_{kj} O^{R\ast}_{kj}\}  f_3},
\label{etaW}
\end{eqnarray}
with
\begin{eqnarray}
        f_1&=&  (m_{\chi_j^{\pm}}^2-m_{\chi_k^0}^2- 2 m_W^2) \sqrt{
        \lambda(m_{\chi^{\pm}_j}^2,m_W^2,m_{\chi^0_k}^2)
        }
, \nonumber\\
f_2&=&  (m_{\chi_j^{\pm}}^2+m_{\chi_k^0}^2- 2 m_W^2)~ m_W^2 + 
(m_{\chi_j^{\pm}}^2-m_{\chi_k^0}^2)^2, \nonumber\\
        f_3&=&  -12~ m_{\chi_j^{\pm}}~ m_{\chi_k^0}~ m_W^2.\nonumber
\end{eqnarray}
For the chargino decay
$\tilde\chi_j^\pm\to\tau^\pm\tilde\nu_\tau^{(\ast)}$,
the decay factor~(\ref{edist2}) is
obtained from Eqs.~(\ref{DRtau}) and (\ref{SigmaDtau})
\begin{eqnarray}
        \eta_{\tau^\pm}
  = \pm\frac{ |V_{j1}|^2 -Y_\tau^2|U_{j2}|^2}
        { |V_{j1}|^2 +Y_\tau^2|U_{j2}|^2}.
\label{etatau}
\end{eqnarray}
The coefficients $\eta_{W^{\pm}}$,
and also $\eta_{\tau^\pm}$~(\ref{etatau}), 
depend on the chargino couplings, as well as on
the chargino and neutralino masses, which could be measured at the 
international linear collider (ILC) with 
high precision~\cite{TDR,Djouadi:2007ik}.

\medskip

In order to reduce the free MSSM parameters,
we parametrize the slepton masses with
their approximate renormalization group equations (RGE)~\cite{Hall:zn}
\begin{eqnarray}
        m_{\tilde\ell_R}^2 &=& m_0^2 +m_\ell^2+0.23 M_2^2
        -m_Z^2\cos 2 \beta \sin^2 \theta_W,\label{mslr}\\
        m_{\tilde\ell_L  }^2 &=& m_0^2 +m_\ell^2+0.79 M_2^2
        +m_Z^2\cos 2 \beta(-\frac{1}{2}+ \sin^2 \theta_W),\label{msll} \\
       m_{\tilde\nu_{\ell}  }^2 &=& m_0^2 +m_\ell^2+0.79 M_2^2 +
       \frac{1}{2}m_Z^2\cos 2 \beta,
\label{msn}
\end{eqnarray}
with $m_0$ the common scalar mass parameter the GUT scale.

\section{Cross sections \label{crossSection}}

We obtain cross sections and distributions by integrating
the amplitude squared $|T|^2$~(\ref{eq:tsquare}) over
the Lorentz invariant phase space element $d{\rm Lips}$
\begin{equation}
        d\sigma=\frac{1}{2 s}|T|^2d{\rm Lips}.
\label{crossection}
\end{equation}
We use the narrow width approximation for the
propagator of the decaying chargino.
The approximation is justified for
$\Gamma_{\chi_j}/m_{\chi_j}\ll1$,
which holds in our case with
$\Gamma_{\chi_j}\lsim {\mathcal O}(1 {\rm GeV}) $.
Note, however, that the naive
${\mathcal O}(\Gamma/m)$-expectation of the error can easily receive
large off-shell corrections of an order of magnitude and more,
in particular at threshold or due to interferences
with other resonant or non-resonant processes.
For a recent discussion of these issues, 
 see~\cite{Hagiwara:2005wg,Berdine:2007uv}

Explicit formulas of the phase space for chargino
production~(\ref{production}) and decay~(\ref{decayell}),
can be found, e.g., in Ref.~\cite{Kittel:2005rp}.
The cross section for chargino production is
\begin{equation}
\sigma_{ij}=
             \frac{\sqrt{\lambda_{ij}}}{8\pi s^2}\bar P,
\label{crossectionProd}
\end{equation}
with the triangle function $\lambda_{ij}$~(\ref{triangle}),
and $\bar P$ given in Eq.~(\ref{eq:sigmabar}).
The integrated cross section for chargino production~(\ref{production})
and subsequent leptonic
decay $\chargino_j\to\ell^+\tilde\nu_\ell$~(\ref{decayell})
is given by
\begin{eqnarray}
 \sigma_{ij,\ell}=
                \frac{1 }{64\pi^2}
                \frac{\sqrt{\lambda_{ij}}}{s^2}   \,
\frac{(m_{\chi^+_j}^2-m_{\tilde{\nu_\ell}}^2)}
{m_{\chi^+_j}^3 {\Gamma_{\chi^+_j}}}
\,      \bar{P} D
= \sigma_{ij} \times{\rm BR}(\tilde\chi_j^+\to\ell^+\tilde\nu_\ell).
\label{eq:sigmatotell}
\end{eqnarray}
The integrated cross section for chargino production %
and subsequent decay $\tilde\chi_j^+\to W^+\neutralino_k$~(\ref{decayW}) is
\begin{eqnarray}
 \sigma_{ij,W}= 
                \frac{1 }{64\pi^2}
                \frac{\sqrt{\lambda_{ij}}}{s^2}   \,
                \frac
                {\sqrt{\lambda(m_{\chi^+_j}^2,m_{\chi^0_k}^2,m_W^2)}}
                {m_{\chi^+_j}^3 {\Gamma_{\chi^+_j}}}
\,      \bar{P} D
= \sigma_{ij} \times{\rm BR}(\chargino_j \to W^+\neutralino_k).
\label{eq:sigmatotW}
\end{eqnarray}

\section{Statistical significances}
\label{StatSignificances}

We define the statistical significances of the C-even cross section observable
$\mathcal{A}^{\Cp\Pm}_{ij}$ (\ref{eq:aprodCePo}) by~\cite{CPNSH} 
\begin{eqnarray}
{\mathcal S}^{\Cp\Pm}_{ij}
&=& |{\mathcal A}^{\Cp\Pm}_{ij}|
        \sqrt{ 2 (2-\delta_{ij}) \sigma_{ij}^{\Cp\Pp} { \mathcal L} },
\label{eq:SaprodCePo}
\end{eqnarray}
where ${\mathcal L}$ denotes the integrated luminosity for chargino production, 
and $\sigma_{ij}^{\Cp\Pp}$
is the C and P symmetrized chargino production cross section,
defined in Eq.~(\ref{eq:sigmacepe}).
There is a factor $2$ appearing in Eq.~(\ref{eq:SaprodCePo}),
since the asymmetries 
require two sets of equal beam polarizations $\pm{\mathcal P}$.
There is a factor $(2-\delta_{ij})$, since two 
independent cross section measurements are available for
$i\neq j$, but only one for $i=j$.
The significances for the C-odd asymmetries
$\mathcal{A}^{\Cm\Ppm}_{ij}$, Eqs.~(\ref{eq:aprodCoPe}),
and (\ref{eq:aprodCoPo}), are defined by
\begin{eqnarray}
{\mathcal S}^{\Cm\Ppm}_{ij}
&=& |{\mathcal A}^{\Cm\Ppm}_{ij}|
        \sqrt{4 \,\sigma_{ij}^{\Cp\Pp} { \mathcal L} }.
\label{significance1}
\end{eqnarray}
Note that the C-odd production asymmetries ${\mathcal A}^{\Cm\Ppm}_{ij}$
vanish trivially for $i=j$.

\medskip

The significances for the CP-odd decay asymmetries
$\mathcal{A}^{\Cpm\Pmp}_{ij,\lambda}$, 
Eqs.~(\ref{eq:adecCeoPe}) and (\ref{eq:adecCePo}),
are defined by~\cite{CPNSH}
\begin{eqnarray}
{\mathcal S}^{\Cpm\Pmp}_{ij,\lambda}
&=& |\mathcal{A}^{\Cpm\Pmp}_{ij,\lambda}|
        \sqrt{ 4 \,\sigma_{ij}^{\Cp\Pp} \,
{\rm BR}(\chargino_j\to\lambda^+\tilde N_\lambda)
{ \mathcal L}_{\rm eff} },
\label{eq:SadecCeoPoe}
\end{eqnarray}
and similarly for the CP-even asymmetries
$\mathcal{A}^{\Cpm\Ppm}_{ij,\lambda}$,
Eqs.~(\ref{eq:adecCeoPe}) and (\ref{eq:adecCoPo}),
\begin{eqnarray}
{\mathcal S}^{\Cpm\Ppm}_{ij,\lambda}
&=& |\mathcal{A}^{\Cpm\Ppm}_{ij,\lambda}|
        \sqrt{ 4 \,\sigma_{ij}^{\Cp\Pp} \,
{\rm BR}(\chargino_j\to\lambda^+\tilde N_\lambda)
{ \mathcal L}_{\rm eff} },
\label{eq:SadecCeoPeo}
\end{eqnarray}
with $\lambda=\ell$ or $W$, and  $\tilde N_\lambda$ the
associated sneutrino or neutralino, respectively.
The effective integrated luminosity is
$\mathcal{L}_{\rm eff} =\eps_\lambda \mathcal{L}$,
with the detection efficiency $\eps_\lambda$
of the leptons or $W$ bosons in the chargino decay
$\tilde\chi_j^\pm\to \ell^\pm\tilde\nu_\ell^{(\ast)}$ or
$\tilde\chi_j^\pm\to W^\pm\tilde\chi_k^0$, respectively.
There is a factor $4$ appearing in the significances,
Eqs.~(\ref{eq:SadecCeoPoe}) and (\ref{eq:SadecCeoPeo}),
since the asymmetries 
require two sets of equal beam polarizations $\pm{\mathcal P}$,
as well as two decay modes,
$\tilde\chi_j^+\to\lambda^+\tilde N_\lambda$
and the charge conjugated decay
$\tilde\chi_j^-\to\lambda^-\tilde N_\lambda^{(\ast)}$.

For an ideal detector a significance of, e.g., ${\mathcal S} = 1$ implies
that the asymmetries  
can be measured at the statistical 68\% confidence level.
In order to predict the absolute values of confidence levels,
clearly detailed Monte Carlo analysis
including detector and background 
simulations with particle identification and reconstruction
efficiencies would be required, which is however beyond the scope
of the present work.

\end{appendix}



\end{document}